\documentclass[fleqn,usenatbib]{mnras}

\usepackage{newtxtext,newtxmath}

\usepackage[T1]{fontenc}
\usepackage{ae,aecompl}

\usepackage{hyperref}
\usepackage{graphicx}	
\usepackage{blindtext}  
\usepackage{natbib}
\usepackage{rotating,times,graphicx,latexsym}
\usepackage{color}
\usepackage{longtable}
\usepackage{lscape}
\usepackage{lipsum} 
\usepackage{array}
\usepackage{flafter}

\usepackage{soul} 
\usepackage{xargs} 
\usepackage[pdftex,dvipsnames]{xcolor}  
\usepackage[colorinlistoftodos,prependcaption]{todonotes}
\newcommandx{\tobedone}[2][1=]{\todo[linecolor=red,backgroundcolor=red!25,bordercolor=red,inline,#1]{#2}}
\newcommandx{\changed}[2][1=]{\todo[linecolor=blue,backgroundcolor=blue!25,bordercolor=blue,inline,#1]{#2}\noindent}
\newcommandx{\thiswillnotshow}[2][1=]{\todo[disable,#1]{#2}}
\newcommandx{\lynne}[2][1=]{\todo[linecolor=Orchid,backgroundcolor=Orchid!25,bordercolor=Orchid,inline,#1]{#2}\noindent}
\newcommandx{\dirk}[2][1=]{\todo[linecolor=BurntOrange,backgroundcolor=BurntOrange!25,bordercolor=BurntOrange,inline,#1]{#2}\noindent}
\newcommandx{\jcw}[2][1=]{\todo[linecolor=SpringGreen,backgroundcolor=SpringGreen!40,bordercolor=SpringGreen,inline,#1]{#2}\noindent}

\newcommand{\kms}{kms$^{-1}$}

\newcommand{\hii}{H{\sc ii}~}
\graphicspath{ {images/} }

\title[Outbursting Be stars]{A survey for variable young stars with small telescopes: VI -  Analysis of the outbursting Be stars NSW\,284, Gaia\,19eyy, and VES\,263}
\author[Dirk Froebrich et al.]{Dirk Froebrich$^{1}$\thanks{E-mail: df@kent.ac.uk},
Lynne A. Hillenbrand$^{2}$, 
Carys Herbert$^{3}$, 
Kishalay De$^{4}$,
\newauthor
Jochen Eisl\"offel$^{5}\thanks{HOYS Observer}$, 
Justyn Campbell-White$^{6,7}$, 
Ruhee Kahar$^{3,7}$, 
\newauthor
Franz-Josef Hambsch$^{8,9,10}\dagger$, 
Thomas Urtly$^{11}\dagger$, 
Adam Popowicz$^{12}\dagger$,  
\newauthor
Krzysztof Bernacki$^{12}\dagger$, 
Andrzej Malcher$^{12}\dagger$, 
Slawomir Lasota$^{12}\dagger$, 
Jerzy Fiolka$^{12}\dagger$,   \newauthor
Piotr Jozwik-Wabik$^{12}\dagger$,
Franky Dubois$^{8,13}\dagger$, 
Ludwig Logie$^{8,13}\dagger$,
Steve Rau$^{8,13}\dagger$,  \newauthor
Mark\,Phillips$^{10,11,14}\dagger$, 
George Fleming$^{11}\dagger$, 
Rafael Gonzalez Farf\'{a}n$^{15,16}\dagger$, \newauthor
Francisco C. Sold\'{a}n Alfaro$^{10,15,17}\dagger$, 
Tim Nelson$^{18,19}\dagger$,
Stephen R.L. Futcher$^{11,18,20}\dagger$, \newauthor
Samantha M. Rolfe$^{21}\dagger$,
David A. Campbell$^{21}\dagger$, 
Tony Vale$^{10,11,22,23,24}\dagger$,
Pat Devine$^{14}\dagger$, \newauthor
Dawid Mo\'{z}dzierski$^{25}\dagger$,
Przemys{\l}aw J. Miko{\l}ajczyk$^{25,26}\dagger$,
Heinz-Bernd Eggenstein$^{27}\dagger$,  \newauthor
Diego Rodriguez$^{10}\dagger$,
Ivan L Walton$^{11}\dagger$,
Siegfried Vanaverbeke$^{8,13,28}$,  \newauthor 
Barry Merrikin$^{18}\dagger$, 
Yenal \"{O}{\u g}men$^{29}\dagger$, 
Alex Escartin Perez$^{15}\dagger$, \newauthor
Mario Morales Aimar$^{15,30}\dagger$,
Georg Piehler$^{31}\dagger$,
Lord Dover$^{3}\thanks{Observer Beacon Observatory}$, 
Aashini L. Patel$^{3}\ddagger$, 
\newauthor
Niall Miller$^{3,32}\ddagger$, 
Jack Finch$^{3}\ddagger$, 
Matt Hankins$^{33}$, 
Anna M. Moore$^{34}$, 
Tony Travouillon$^{34}$,  \newauthor 
Marek Szczepanski$^{35}\dagger$
\\
$^{1}$School of Physical Sciences, University of Kent, Canterbury CT2 7NH, UK\\
$^{2}$Department of Astronomy, MC 249-17, California Institute of Technology, Pasadena, CA 91125, USA\\
$^{3}$Centre for Astrophysics and Planetary Science, School of Physical Sciences, University of Kent, Canterbury CT2 7NH, UK\\
$^{4}$MIT-Kavli Institute for Astrophysics and Space Research, 77 Massachusetts Ave., Cambridge, MA 02139, USA\\   
$^{5}$Th\"{u}ringer Landessternwarte, Sternwarte 5, 07778 Tautenburg, Germany\\
$^{6}$European Southern Observatory, Karl-Schwarzschild-Stra\ss{}e 2, 85748 Garching, Germany\\
$^{7}$SUPA, School of Science and Engineering, University of Dundee, Nethergate, Dundee DD1 4HN, U.K.\\
$^{8}$Vereniging voor Sterrenkunde, werkgroep veranderlijke sterren, Oostmeers 122 C, 8000 Brugge, Belgium\\
$^{9}$Bundesdeutsche Arbeitsgemeinschaft f\"{u}r Ver\"{a}nderliche Sterne e.V. (BAV), Munsterdamm 90, D-12169 Berlin, Germany\\
$^{10}$AAVSO, 49 Bay State Road, Cambridge, MA 02138, USA\\
$^{11}$The British Astronomical Association, Variable Star Section, Burlington House Piccadilly, London W1J 0DU, UK\\
$^{12}$Department of Electronics, Electrical Engineering and Microelectronics, Silesian University of Technology, Akademicka 16, 44-100 Gliwice, Poland\\
$^{13}$Public Observatory AstroLAB IRIS, Provinciaal Domein De Palingbeek, Verbrandemolenstraat 5, B-8902 Zillebeke, Ieper, Belgium\\
$^{14}$Astronomical Society of Edinburgh, Edinburgh, UK\\
$^{15}$Observadores de Supernovas$^{\thanks{\tt \href{https://sites.google.com/view/sn2017eaw/}{Observadores de Supernovas}}}$, Spain\\
$^{16}$Uraniborg Obervatory, Calle Antequera 8, 41400, \'{E}cija, Sevilla, Spain\\ 
$^{17}$Science Department, Seville University, Av. de la Ciudad Jard\'{i}n, 20-22, 41005 Sevilla, Spain\\ 
$^{18}$Hampshire Astronomical Group, Clanfield, UK\\
$^{19}$Horndean Observatory, 6 Falcon Road, Horndean, Waterlooville, Hampshire, PO89BY, UK\\
$^{20}$Royal Astronomical Society, Burlington House, Piccadilly, London W1J 0BQ, UK\\
$^{21}$Bayfordbury Observatory, Department of Physics, Astronomy and Mathematics, University of Hertfordshire, UK\\
$^{22}$Wiltshire Astronomical Society, 2 Oathills, Corsham, SN13 9NL, UK\\
$^{23}$Bath Astronomers, 19 New King Street, Bath BA1 2BL, UK\\
$^{24}$The Herschel Society, The Herschel Museum of Astronomy, 19 New King Street, Bath BA1 2BL, UK\\
$^{25}$Astronomical Institute, University of Wroc{\l}aw, ul. M. Kopernika 11, 51-622 Wroc{\l}aw, Poland\\
$^{26}$Astronomical Observatory, University of Warsaw, Al. Ujazdowskie 4, 00-478 Warsaw, Poland\\
$^{27}$Volkssternwarte Paderborn, 33041 Paderborn, Germany\\
$^{28}$Center for Mathematical Plasma Astrophysics, University of Leuven, Belgium\\
$^{29}$Green Island Observatory, Karao{\u g}laono{\u g}lu Street 63A, Ge\c{c}itkale Ma{\u g}usa, North Cyprus\\ 
$^{30}$Observatorio de Sencelles, Sonfred Road 1, 07140 Sencelles. Mallorca, Spain\\
$^{31}$Selztal Observatory, D-55278 Friesenheim, Bechtolsheimer Weg 26, Germany\\
$^{32}$Centre for Astrophysics Research, University of Hertfordshire, Hatfield AL10 9AB, UK\\
$^{33}$Arkansas Tech University, Russellville, AR 72801, USA\\
$^{34}$Australian National University, Research School of Astronomy and Astrophysics, Mount Stromlo Observatory, Cotter Road, Weston Creek 2611, Australia\\
$^{35}$Department of Data Science and Engineering, Silesian University of Technology, Akademicka 16, 44-100 Gliwice, Poland\\
}
\date{Accepted XXX. Received YYY; in original form ZZZ}
\pubyear{2022}
\begin{document}
\label{firstpage}
\pagerange{\pageref{firstpage}--\pageref{lastpage}}
\maketitle

\begin{abstract}
This paper is one in a series reporting results from small telescope observations of variable young stars. Here, we study the repeating outbursts of three likely Be stars based on long-term optical, near-infrared, and mid-infrared photometry for all three objects, along with follow-up spectra for two of the three. The sources are characterised as rare, truly regularly outbursting Be stars. We interpret the photometric data within a framework for modelling light curve morphology, and find that the models correctly predict the burst shapes, including their larger amplitudes and later peaks towards longer wavelengths. We are thus able to infer the start and end times of mass loading into the circumstellar disks of these stars. The disk sizes are typically 3\,--\,6 times the areas of the central star. The disk temperatures are $\sim 40$\,\%, and the disk luminosities are $\sim 10$\,\% of those of the central Be star, respectively. The available spectroscopy is consistent with inside-out evolution of the disk. Higher excitation lines have larger velocity widths in their double-horned shaped emission profiles. Our observations and analysis support the decretion disk model for outbursting Be stars.
\end{abstract}




\section{Introduction}

Classical Be stars have long been recognized as objects in the later main sequence or early post-main sequence stage of evolution that are rapidly rotating \citep{Slettebak1982}.  By definition they are early-type stars exhibiting emission lines, typically H$\alpha$ that is double-peaked though a wide range of line profiles is presented, as well as some higher Balmer series lines and occasionally also \ion{He}{I} and/or \ion{Fe}{II} \citep{Slettebak1992}. Be stars generally have weak near-infrared excesses consistent with free-free emission \citep{Finkenzeller1984} and can be detected in the radio. The ``classical Be" nomenclature distinguishes them from Be phenomena arising due to binary interactions among evolved massive stars.  

Classical Be star spin rates are close enough to the breakup velocity \citep[$>$\,70\,\%;][]{Porter1996} that their observed properties are interpreted as being due to equatorial mass loss that produces a decretion disk\footnote{Older literature, including the original proposal by \cite{Lee1991} uses the term ``excretion disk".} (as opposed to an accretion disk). The disks are thought to be dust-free, and the gas emission is constrained from observed line widths to arise in the inner few stellar radii.  \cite{Rivinius2013} provides a general review of the Be phenomenon, and describes ``an outwardly diffusing gaseous Keplerian disk [...] fed by mass ejected from the central star [...] and governed by viscosity". 

A fraction of the classical Be population undergoes outburst behavior.  Although several mechanisms for putting material into the surrounding disk have been suggested, the currently favored hypothesis is a non-radial pulsation-driven means \citep[e.g.][and references therein]{Ressler2021}. The combination of rapid stellar rotation, presumably inherited from the main sequence, the radius inflation that occurs during post-main sequence evolution, and the existence of an outburst mechanism, leads to the star exceeding its breakup velocity, and hence to equatorial mass loss that replenishes the disk from the inside.  In the assumed viscous disk scenario, the addition of material at the inner disk edge implies transport of both mass and angular momentum outward.  According to \cite{Rivinius2013}, when the mass addition at the inner edge ceases, the sense of the mass flow reverses and the star can re-accrete any remaining disk material that has not otherwise escaped the system.

A well-known Be star database is described by \cite{Neiner2011}. Over the decades, various H$\alpha$ surveys, including IPHAS \citep{2005MNRAS.362..753D}, have continued to identify candidate Be stars. Recently, large-scale spectroscopic surveys such as APOGEE \citep{Chojnowski2015} and LAMOST \citep[e.g.][]{Wang2022} have contributed as well.  The observational properties of Be stars, and indeed the ratio of Be to non-Be is roughly dependent on spectral-type, with typical divisions into early (B0\,--\,B3), mid (B4\,--\,B7), and late (B8\,--\,B9) type Be stars.

The Be stars are also ubiquitous photometric variables.  They have been studied as such from the ground e.g. using OGLE \citep{Mennickent2000}, KELT \citep{Labadie-Bartz2017}, and ASAS \citep{Bernhard2018}. Studies at high precision and cadence have used CoRoT \citep{Seamann2018}, Kepler \citep{Rivinius2016,Labadie-Bartz2017}, and TESS \citep{Labadie-Bartz2022}. Variability types include stellar pulsation (both low-order g-modes and higher-frequency p-modes), other clustered periodic modes, stochastic photometric behavior that is attributed to disks, discrete outbursts, and finally, long-term trends. \cite{Labadie-Bartz2018} again working with KELT data, conducted a Be variability study focused on the outbursting category.

\cite{Mennickent2000} had earlier identified two families of outburst light curves: sharp and hump-like, and had found 13\% of the Be star sample to exhibit outbursts. \cite{Labadie-Bartz2017} found a higher 36\,\% of Be stars to exhibit outbursts, while \cite{Labadie-Bartz2018} from the same data set state 28\,\%. \cite{Labadie-Bartz2022}, using space-based TESS, find a similar 31\% of their Be star sample to show ``bursts", with sensitivity that allows identification of shorter duration and much lower amplitude bursts than would be detected as ``outbursts" in ground-based photometry. \cite{Bernhard2018}, however, find a much more sizable 3/4 of their Be star sample to show ``bursts". Both Labadie-Bartz and Bernhard demonstrate a spectral type dependence, with earlier type Be stars having a higher burst frequency.  The burst amplitude and duration are again correlated with the spectral class, in the sense of being larger and longer for the earlier type Be stars. Further, both authors quantify the burst rise times and decay times, with the latter being typically 2-3 times the former. Concerning the burst frequency, as these authors use different definitions of burst behavior (including negative departures from baseline brightness which are interpreted as outbursts seen through an edge-on disk and therefore dimming rather than brightening events), and may cover different parts of the B spectral type range differently, it is unclear what to make of the factors of several differences in the reported percentages (13\,\% vs about 30\,\% vs 73\,\%).

Our main interest is in the long duration, discrete outburst events. Such outbursting behavior is recurrent, and reported as somewhere between irregular \citep{Rivinius2013} and semi-regular \citep{Labadie-Bartz2017,Bernhard2018}, with reported rates of 0.5-5 per year per outbursting Be star. Recent examples of in-depth studies of individual outbursting Be sources are those of \cite{Richardson2021} focusing on spectroscopy \citep[see also the spectral time series of several objects shown in][]{Labadie-Bartz2018}, and \cite{Ghoreyshi2018} focusing on photometry. Detailed light curve modelling is also reported by \cite{Rimulo2018}. 

In this paper, we identify three newly recognised early-spectral class Be stars, as having repeating photometric and spectroscopic outbursts. The bursts are hump-like, occur on timescales of less than a year, and have visual amplitudes of up to 0.5 mag. We discuss the three stars in Sect.\,\ref{TheStars}, report the long-term photometric monitoring in Sect.\,\ref{photometry}, as well as spectroscopy in Sect.\,\ref{spectroscopy} of two of the three, that confirms the Be status and covers low and high states. In Sect.\,\ref{burstsection} we discuss our fit of the photometric outbursts with a simple model to determine the temperature and surface area of the emitting material. Finally, we discuss our findings in Sect.\,\ref{discussion}.

\begin{table}
\caption{\label{source_table} Adopted and determined properties of the three sources. We list the GaiaDR3 astrometry and stellar parameters, other literature values for the stellar parameters, average baseline magnitudes and colours from HOYS, and the typical burst properties. The references for the literature stellar parameters are as follows: (1) \citet{Carvalho2022}; (2) \citet{Munari2019}; (3) This work; (4) \citet{Comeron2012}.}
\centering
\setlength{\tabcolsep}{0.5pt}
\begin{tabular}{|c|c|c|c|}
\hline
Name & NSW\,284 & Gaia\,19eyy & VES\,263 \\ \hline
Aliases & [NSW2012]\,284 &   & SS\,447 \\ 
        & PTFS\,1821n  &   & Gaia18azl \\ \hline
\multicolumn{4}{|l|}{GaiaDR3 Astrometry} \\ \hline
RA (J2000)  & 21 38 39.81 & 08 30 42.50 & 20 31 48.85 \\
Dec (J2000) & +57 08 47.1 & -41 33 42.6  & +40 38 00.1 \\ \hline
p\,[mas] & 0.1880 [0.0153] & 0.1626	[0.0122] & 0.5733 [0.0105] \\
d\,[kpc] & 5.32 & 6.15 & 1.74 \\
RUWE & 1.155 & 0.956 & 1.113 \\
$\mu_\alpha$\,[mas/yr] & -2.549 [0.019] & -2.773 [0.013] & -2.951 [0.011] \\
$\mu_\delta$\,[mas/yr] & -2.093 [0.018] & 3.411	[0.013] & -5.336 [0.012] \\ \hline
\multicolumn{4}{|l|}{GaiaDR3 Stellar Parameters} \\ \hline
T\,[K] & 22,666 [1,000] & 15,415 [200] & -- \\
log g\,[cm s$^{-2}$] & 4.24 [0.12] & 3.70 [0.03] & -- \\
$A_V$\,[mag] & 2.9 & 2.6 & -- \\
M$_G$ [mag] & -1.66 & -1.62 & -- \\ \hline
\multicolumn{4}{|l|}{Literature Stellar Parameters} \\ \hline
$A_V$\,[mag] & 3.6 [1.7]$^{(1)}$ & & 4.1 [1.9] / 5.6 [0.15]$^{(1,2)}$ \\
SpT & $\sim$B3e$^{(3)}$ &  & B1\,II$^{(4)}$ \\
T\,[K] & 17,500$^{(3)}$ & 17,500$^{(3)}$ & 20,666$^{(4)}$ \\
L\,[L$_\odot$] &  &  & 13,000 [1000]$^{(2)}$ \\
M\,[M$_\odot$] &  &  & 9.1 / 12$^{(2,4)}$ \\ \hline
\multicolumn{4}{|l|}{Average baseline magnitudes from this work} \\ \hline 
B\,[mag] & 15.50 & 14.29 & 14.92 \\
V\,[mag] & 14.76 & 13.53 & 13.10 \\
R\,[mag] & 14.57 & 13.31 & 12.07 \\
I\,[mag] & 14.29 & 13.05 & 11.06 \\ \hline
\multicolumn{4}{|l|}{Average baseline colours from this work} \\ \hline 
B-V\,[mag] & 0.74 & 0.76 & 1.82 \\
V-R\,[mag] & 0.19 & 0.22 & 1.03 \\
R-I\,[mag] & 0.28 & 0.26 & 1.01 \\ \hline
\multicolumn{4}{|l|}{Typical burst characteristics from this work} \\ \hline
cadence [d] & 280-390 & 390-490 & 140-190 \\
duration [d] & 200 & 150 & 120 \\
R amp [mag] & $<$0.6 & $<$0.6 & $<$0.5 \\
\hline
\end{tabular}
\end{table}

\section{The Stars}\label{TheStars}

The three sources under study here are NSW\,284, Gaia\,19eyy, and VES\,263, each of which is described below. Table\,\ref{source_table} provides other identifiers, Gaia\,DR3 astrometry, photometry, and estimated stellar parameters including distance, extinction, spectral type, etc. \citep{2016A&A...595A...1G, 2022yCat.1355....0G}. Each of these three sources has a spectral energy distribution consistent with an early spectral type stellar photosphere, with no or only minor infrared excess.

\subsection{NSW~284}

This source is in Cepheus, projected into the south of the IC\,1396 \hii\ region, but a far background object according to the GaiaDR3 distance of about 5.3\,kpc. It was identified as an H$\alpha$ emission-line star by \cite{Nakano2012} and first presented as a variable star by \cite{2018MNRAS.478.5091F} who assumed it was a YSO. Here, we recognize the source as a Be star rather than a YSO.  Table\,\ref{source_table} provides its parameters. Reddening of $E(B-V)=1.16 \pm 0.56$\,mag \citep{Carvalho2022} combined with the Gaia G-band magnitude suggests an early B spectral type, consistent with our spectroscopic determination below that it is a B3 type star.  However, the measured B-V colour outside of the photometric bursts is about 0.70-0.75\,mag, while APASS reports $B-V = (15.043\pm0.003) - (14.206\pm0.007) = 0.8$\,mag; in either case a negative intrinsic color results, $(B-V)_0< -0.3$\,mag, suggesting an early O-type star. An alternate to this scenario is that strong Balmer continuum emission originating in hot circumstellar gas causes a blue-ing, even in the out-of-burst light curve.

\subsection{Gaia~19eyy}

Located in Puppis, this source has no previous literature despite being characterized on the  Gaia Alerts page\footnote{\url{http://gsaweb.ast.cam.ac.uk/alerts/alert/Gaia\,19eyy/}} as a YSO. Here, we analyze the source as a Be star outburster rather than a YSO based on the light curve similarity to\,NSW 284 and VES\,263.  The GaiaDR3 distance is about 6\,kpc and the extinction is unknown. As can be seen in Table\,\ref{source_table}, the apparent optical colours of Gaia\,19eyy during quiescence are almost identical to NSW\,284. Together with the similar distance, absolute Gaia magnitude and light curve behaviour, we infer that the source closely resembles a Be type source, despite the absence of spectroscopic confirmation.

\subsection{VES~263}

This source is in Cygnus and is a strong emission-line object that has appeared in many H$\alpha$ catalogs over the decades \citep[e.g.][but see \cite{Munari2019} for older references]{KW1997}. The star was determined by \cite{Berlanas2019} to be a kinematic member of Cyg~OB2, at mean distance 1760 pc, a result independently confirmed by \cite{Munari2019}. It was previously suggested as such by \cite{Comeron2012} who assigned a spectral type of B1~II and derived $A_V=4.4$ mag, as well as the other stellar parameters reported in Table\,\ref{source_table}. \cite{Munari2019} find $E(B-V)=1.80 \pm 0.05$ mag from spectral energy distribution (SED) fitting. From DIBs analysis, \cite{Carvalho2022} found an extinction value of $E(B-V)=1.33 \pm 0.61$ mag, consistent with the \cite{Comeron2012} value of $A_V$ for an $R_V=3.1$ as well as the DIBs measurements in \cite{Munari2019} but marginally lower than the SED fitting value in \cite{Munari2019}. The only paper discussing VES\,263 in any depth is \cite{Munari2019}, who present the source based on a Gaia Alert\footnote{\url{http://gsaweb.ast.cam.ac.uk/alerts/alert/Gaia18azl/}} as an eruptive Herbig~Ae/Be star. These authors assembled the historical record, which includes an approximately 10-year high state from $\sim1955-1965$ and their follow-up photometry and spectroscopy of the 2018 brightening of the photometric baseline that was reported by Gaia. Rather than a pre-main sequence star, here we suggest that VES\,263 is instead an evolved Be star undergoing outbursts.

\section{Photometry and long term light curves}\label{photometry}

We have assembled a comprehensive set of photometric data for each of the three sources, with Fig.\,\ref{hoys_lc} displaying the composite multi-wavelength light curves. In this section we discuss the data sets and describe the long term light curves.

\subsection{HOYS Photometry}

The vast majority of the optical photometry data of NSW\,284 and Gaia\,19eyy, used in our analysis, has been obtained as part of the Hunting Outbursting Young Stars (HOYS) citizen science project \citep{2018MNRAS.478.5091F}. This project observes a number of young, nearby clusters and star forming regions in optical filters with amateur telescopes. The HOYS observations for NSW\,284 are taken by a wide variety of observatories due to the nature of the project \citep{2018MNRAS.478.5091F,2020MNRAS.493..184E}. The southern position of the Gaia\,19eyy source means that all data are taken by the same observatory, the Remote Observatory Atacama Desert \citep[ROAD;][]{2012JAVSO..40.1003H}.

All HOYS target fields have deep images in all optical filters taken under photometric conditions as reference for relative photometry. The off-sets of the instrumental magnitudes in the reference images to the $B$, $V$, $R_{c}$, $I_{c}$ system (the filters used for the reference images) have been obtained with the Cambridge Photometric Calibration Server\footnote{\tt \href{http://gsaweb.ast.cam.ac.uk/followup}{http://gsaweb.ast.cam.ac.uk/followup}}. For simplicity we refer to the Cousins filters as $R$ and $I$, throughout the paper. For each HOYS target field we identify non-variable stars in our vast data set. Their colours and magnitudes are used to determine the colour terms in each image and correct them. The full details of this procedure can be found in \citet{2020MNRAS.493..184E}. It is those corrected magnitudes we use throughout for the analysis of the light curves.

\begin{figure*}
\centering
\includegraphics[angle=0,width=1.7\columnwidth]{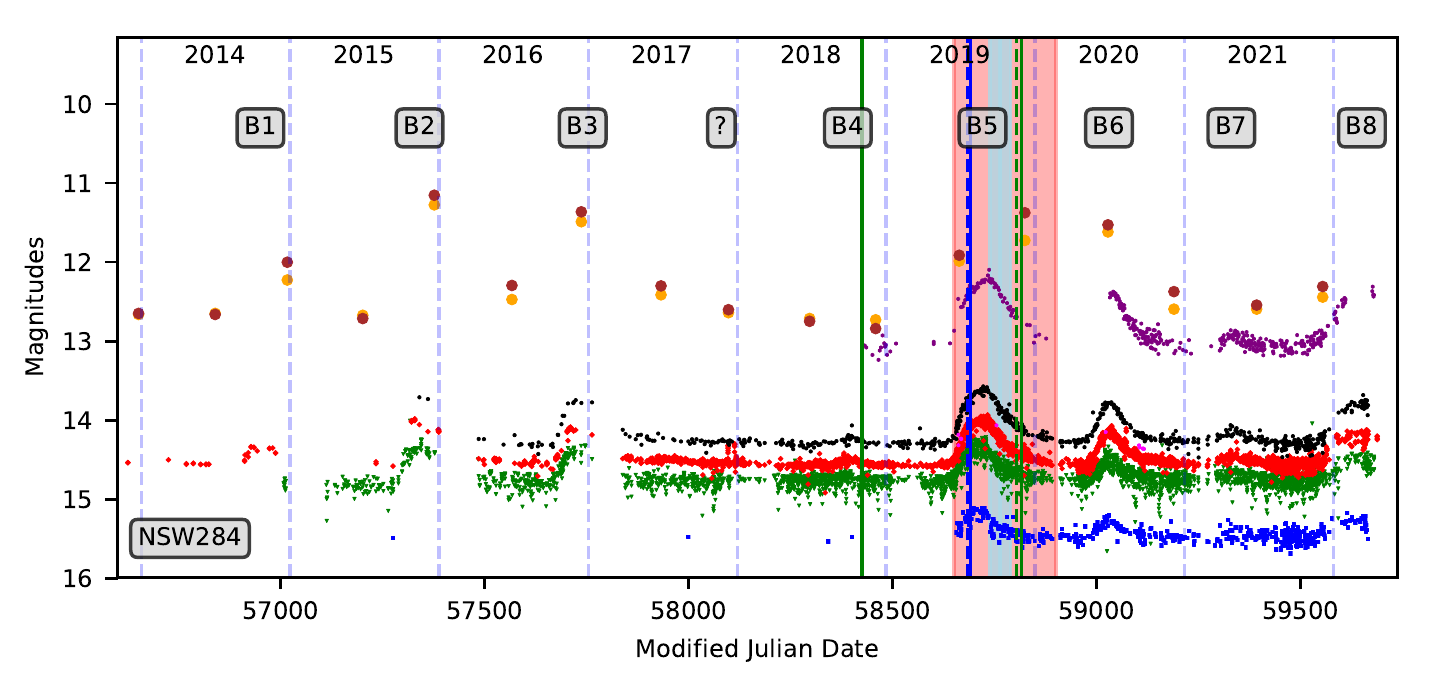} \\
\includegraphics[angle=0,width=1.7\columnwidth]{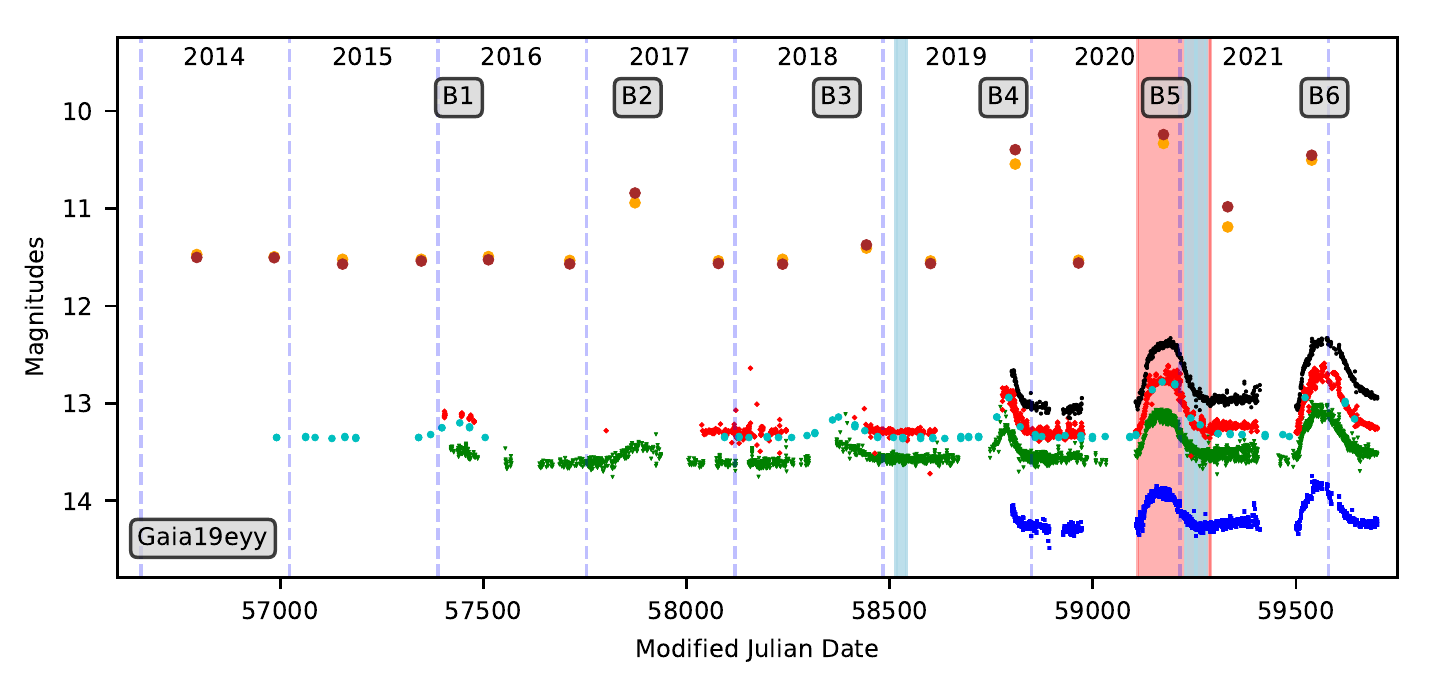} \\
\includegraphics[angle=0,width=1.7\columnwidth]{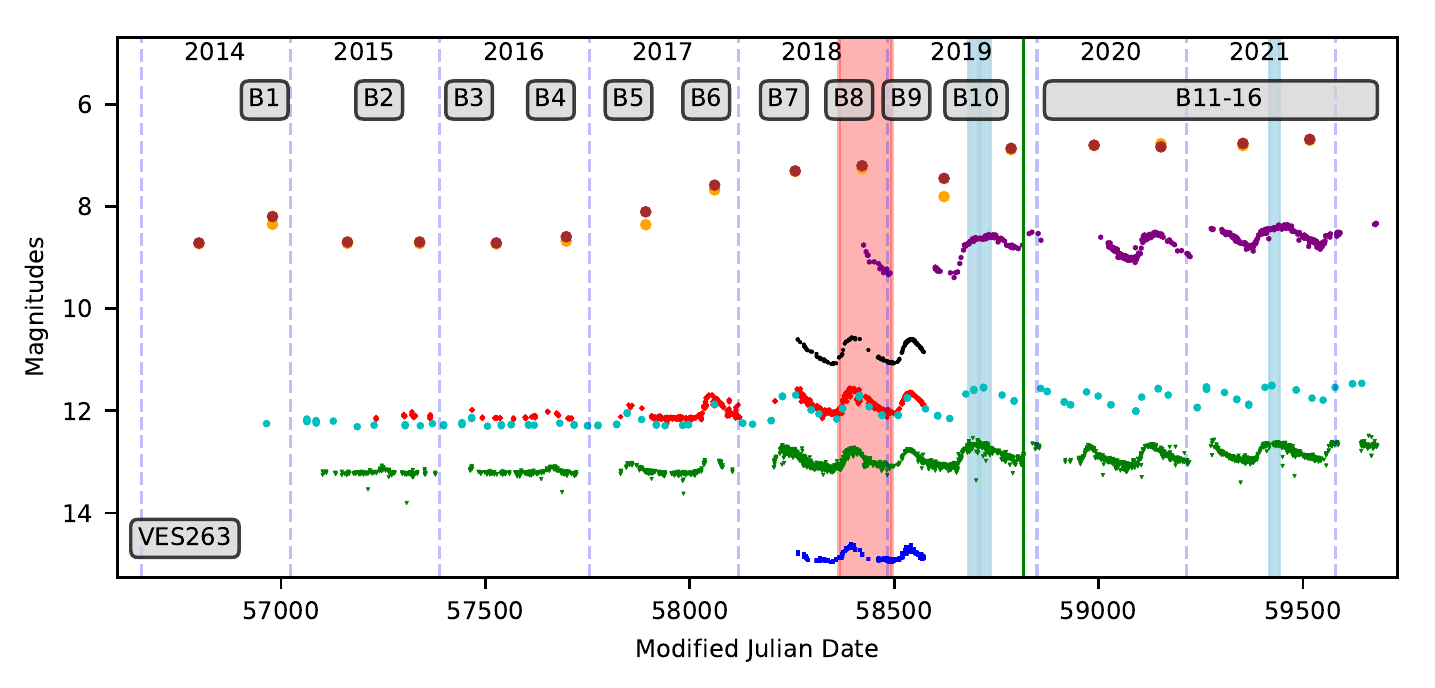} 
\caption{Long term light curves of NSW\,284 (top), Gaia\,19eyy (middle), and VES\,263 (bottom). Filters from bottom to top are colour coded: $B$ - blue, $V$ - green, $R$ - red, Gmag - turquoise, $I$ - black, $J$ - purple, $W1$ - orange, $W2$ - brown. The identified bursts are labeled and the "?" indicates a potentially failed burst. The red shaded bursts are discussed as examples in detail in Sect.\,\ref{burstsection}. The light blue shaded times indicate the TESS sectors with data available. Dashed vertical lines indicate January 1st each year. Green vertical lines indicate Keck spectra (solid - HIRES; dashed - NIRSPEC). Blue vertical lines indicate the Palomar 200\arcsec\ spectra (solid - DoubleSpec; dashed - TripleSpec). The optical ($BVRI$) photometry combines data from HOYS, ASAS-SN, ATLAS, PTF, ZTF, and from \citet{Munari2019} for VES\,263. If taken in different filters they are manually shifted into the nearest appropriate filter, e.g. g into V (see Sect.\,\ref{aux}) for details. \label{hoys_lc}}
\end{figure*}

In Fig.\,\ref{hoys_lc} we show the long term light curves for our sources. They combine the $B$, $V$, $R$, $I$ data from HOYS with optical and infrared photometry from auxiliary datasets discussed in Sect.\,\ref{aux}.

\subsection{Auxiliary Photometry}\label{aux}

In addition to the HOYS monitoring photometry, our sources of interest have been included in a number of different time domain surveys. Here, we reference and discuss the additional data that enable us to have a longer term and broader wavelength look at the bursts. All of those are included in the long term light curves displayed in Fig.\,\ref{hoys_lc}. 

\begin{itemize}

\item The object VES\,263 is not situated in one of the HOYS target fields. Thus, as baseline photometry, we use the $B$, $V$, $R$, and $I$ data from \citet{Munari2019} taken in 2018/19.

\item All three sources have data in the All-Sky Automated Survey for Supernovae \citep[ASAS-SN,][]{2014ApJ...788...48S, 2017PASP..129j4502K}. Photometry is available in the $V$ and $g$ filters. We manually shifted the data in Fig.\,\ref{hoys_lc} to match our HOYS $V$-band photometry. The magnitude off-sets are slightly different for the three sources due to their different colours - notably VES\,263 (see Table\,\ref{source_table}).

\item Photometry from the Asteroid Terrestrial-impact Last Alert System \citep[ATLAS,][]{Tonry2018} is also available for all three sources. Data taken in the $c$ (cyan) and $o$ (orange) filters have been adopted and manually shifted to match the baseline $V$ and $R$ photometry, respectively. The data for VES\,263 are very noisy (0.5\,mag) after MJD\,=\,58500. These are hence not included in Fig.\,\ref{hoys_lc}.

\item The Gaia Alerts \citep{Hodgkin2021} contain light curves for VES\,263 (Gaia18azl) and Gaia\,19eyy, but not NSW\,284 since it did not trigger an alert. We note that the Gaia $Gmag$ filter approximates $R$ for red sources, which is evident in Fig.\,\ref{hoys_lc}.

\item The Palomar-Gattini-InfraRed survey \citep[PGIR,][]{de2020, 2019NatAs...3..109M} observes the sky every two nights in the near-infrared $J$-band to a median depth of $J$\,=\,15.7\,mag (AB). The $J$-band light curves are available for NSW\,284 and VES\,263 beginning on JD\,=\,2458410,
but not Gaia\,19eyy due to its southerly declination. 

\item The WISE mission \citep{Wright2010,Cutri2012} and the repurposed NEOWISE time domain survey \citep{Mainzer2011,2014yCat.2328....0C} provides all-sky mid-infrared survey data and has light curves in the $W1$ and $W2$ bands for all three of our sources. Although the NEOWISE $\sim$6-month cadence provides data that are under-sampled relative to the variability timescales, there is detectable brightening in the $W1$ and $W2$ bands that corresponds to the optical bursts. NEOWISE also shows evidence that the burst behavior occurred before the start of HOYS.

\item Palomar Transient Factory \citep[PTF,][]{Law2009} $r$-band data is available for NSW\,284. We include these data, shifted into our $R$-band, in the long term light curve. There are a handful of data points for VES\,263, but due to the small number, we have not added them to Fig.\,\ref{hoys_lc}.

\item The Zwicky Transient Factory survey \citep[ZTF,][]{ztf_overview} has recorded data for NSW\,284 in $g$ and $r$ (internal ID = ZTF18abksgkt). For VES\,263 only $g$ data are available. They were retrieved from IPAC \citep{ztf_data} and cover April 2018 through to the present. As for the other surveys, we manually shifted the $g$ and $r$ data into our $V$ and $R$ photometry.

\item The Transiting Exoplanet Survey Satellite \citep[TESS,][]{2015JATIS...1a4003R} conducted short timescale monitoring at 30 minute cadence of all three sources. At present, each source has two sequential sectors of data available, with a sector spanning 24\,--\,28\,days, and two of the three sources have a third sector that is separated by nearly two years. The photometry has not been added into our long term light curves in Fig.\,\ref{hoys_lc}, but the individual sectors are indicated as light blue shaded areas.

\end{itemize}

\subsection{Summary of light curve behavior}

The light curves shown in Fig.\,\ref{hoys_lc} have some common features, but the three sources are also somewhat distinct in their photometric behavior. All sources show repeated bursts since the start of our data in 2014, with approximately 1\,--\,2 bursts per year. We have visually identified all bursts (see labels 'B' in Fig.\,\ref{hoys_lc}). We define as a burst times where the average brightness deviates in a sustained manner by more than the typical photometric scatter from the flat baseline in at least one of the filters. The most salient feature of the light curves is that the burst amplitudes are larger towards redder wavelengths. Furthermore, they tend to peak later at longer wavelengths. Table\,\ref{source_table} provides the median burst amplitudes, which are typically several tenths of a mag, and burst durations, which are typically several months. Below we give a brief description of the long term light curve for each source. 

\subsubsection{NSW284}

The object shows a flat light curve with superimposed bursts (see top panel of Fig.\,\ref{hoys_lc}). There are in total eight bursts during the 8.5\,yr of data. There is a clear gap between B3 and B4 in 2017/18. Based on the behaviour prior and after that, one could have expected a burst to appear at that time. The WISE magnitudes in late 2017 are marginally higher (by 0.1\,mag, or one sigma) than in the subsequent mid-2018 data point, hinting at a weak burst, which was not detectable in the optical data. We have indicated that gap as '?'. Including the questionable burst, the average gap between bursts is about 345\,d, but they range from 280\,d to 390\,d. One can see that prior to the questionable burst the gaps are slightly larger than one year, while afterwards they are shorter and of the order of 330\,d. 

While B1 is detected only in $R$ and the WISE data, the other bursts are consistently covered in the optical data. Starting from 2019 (B5), we have also regular $B$ and $J$-band photometry. Throughout, the amplitudes in all bursts are higher at longer wavelengths. Similarly, the brightness increase is faster than the decrease and the peak brightness occurs slightly later at longer wavelengths. The burst amplitudes and duration vary from burst to burst. For the stronger bursts the amplitudes range from 0.2\,mag in $B$ to almost 2\,mag in the WISE filters. Lower amplitude bursts appear shorter, probably in part because the declining part of the burst merges into the noise of the data sooner. There is no discernible pattern (within the limited number of bursts detected) that indicates whether a burst is stronger or weaker. The TESS data available for this source coincides with one of the declines in brightness (burst B5). Other than a general drop in flux, there are no other discernible features visible in this high cadence data.

\subsubsection{Gaia\,19eyy}

The long term light curve of Gaia\,19eyy, shown in the middle panel of Fig.\,\ref{hoys_lc}, is similar to NSW\,284. There are distinct bursts on top of a low-state brightness. Coverage of the objects at timescales significantly below the burst duration starts in early 2015. However, we only have good multi-filter data since the end of 2019, when the object got flagged up as variable star in the Gaia alerts \citep{Hodgkin2021}. Similar to NSW\,284 the bursts are semi regular, with six detected bursts over 7.5\,yr. Thus, the average cadence is approximately 450\,d, with gaps ranging from 390\,d to 490\,d. Again, there is a slight trend that the gaps between bursts decrease over time. 

All bursts but B1 and B3 show very strong WISE amplitudes. This seems to be caused by timing of the data rather than real differences, as B1 and B3 peak in the optical between two of the WISE data points. In contrast to NSW\,284, the optical bursts seem to be slightly more symmetric. Hence, the increase and decrease in brightness are of about the same length. The latest three bursts (B4 -- B6) are clearly stronger than the first three (B1 -- B3), but we do not think this is a significant trend. Thus, in all aspects the burst and long term light curves of Gaia\,19eyy and NSW\,284 are very similar. This includes the TESS data available, which does not show any short term variability.

\subsubsection{VES\,283}

The long term light curve of VES\,263 (shown in the bottom panel of Fig.\,\ref{hoys_lc}) has aspects similar to the two other objects. However, it also differs significantly. The optical 2018/19 data has been discussed in \citet{Munari2019}. We only have long term coverage in $V$ and $Gmag$ for the entire duration of the data. The $R$ data only covers pre-2018/19 and $J$-band coverage is available from late 2018.

Since 2014, our long term data indicates a total of 16 bursts. Initially (up to 2018) the bursts appear on top of a flat baseline brightness. Their cadence is approximately 180\,d (slightly decreasing) and the amplitudes do appear to increase from burst to burst in $V$ and $R$. After 2019 the bursts increase in amplitude and their cadence decreases to about 145\,d. Furthermore, the brightness does not return to the pre-2018 baseline magnitudes in-between bursts. 

The WISE light curve suggests a general, almost 2\,mag brightening in the MIR from 2018. Judging by the general increase of the faint state at optical wavelengths discussed above, this might in part be the case. However, it is also evident that the WISE observing cadence almost matches the outbursts of this source. Most WISE data are taken just after the optical peaks. Given that in the other objects the peak at longer wavelengths occurs later, it is likely that VES\,263 still varies in the WISE bands, and the steady increase seen in the data is only in part real, and in part caused by the observing cadence.

The TESS data for this object show some short term variations. For about half of the TESS coverage, regular variations with a period of about 17\,hr can by found. The peak-to-peak amplitudes vary from at most 0.7 percent to zero, i.e. they disappear into the noise. These variations cannot be attributed to  surface features or orbiting disk material, as the periods would be much larger. Thus, the most likely source for the variability are pulsations. If these are part of the trigger mechanism for the bursts, or the termination of the mass loading is not clear, as they are not observed in any of the other sources.

\section{Spectroscopic Observations and Findings}\label{spectroscopy}

We have obtained a number of spectra for two of our sources, NSW\,284 and VES\,263, during different parts of the light curve. Below we describe the data and discuss the spectra in detail.

\subsection{NSW284}

\subsubsection{Palomar 200" / DoubleSpec}

An optical spectrum was taken on 2019-07-27 with the Palomar 200" telescope and facility optical spectrograph DoubleSpec \citep[][in original form]{OG1982}. At that time the source was approximately half way through the brightness increase of burst B5. We used the D68 dichroic to separate the spectrograph arms. The source was observed using the 600\,l/mm and 4000\,\AA\ blaze (blue), and the 1200\,l/mm and 7100\,\AA\ blaze (red) gratings, respectively. Data were processed using standard tasks in IRAF to produce a one-dimensional wavelength and flux-calibrated spectrum between $\approx 4000-9000$\,\AA.

The blue-side spectrum shows strong absorption in the \ion{Na}{I}~D doublet as well as in the 5780\,\AA\ and 6614\,\AA\ (weaker) diffuse interstellar band (DIB) features. There is weak H$\alpha$ emission having an equivalent width $W_\lambda < 2$\,\AA, but the other Balmer lines are in very weak absorption.  The red-side spectrum is essentially featureless, aside from telluric contributions, with no evidence of Paschen line absorption. Although we believe the source to be an early type star (see below), this is not unexpected given the weakness of the Balmer lines.

\begin{figure}
\centering
\includegraphics[angle=0,width=\columnwidth]{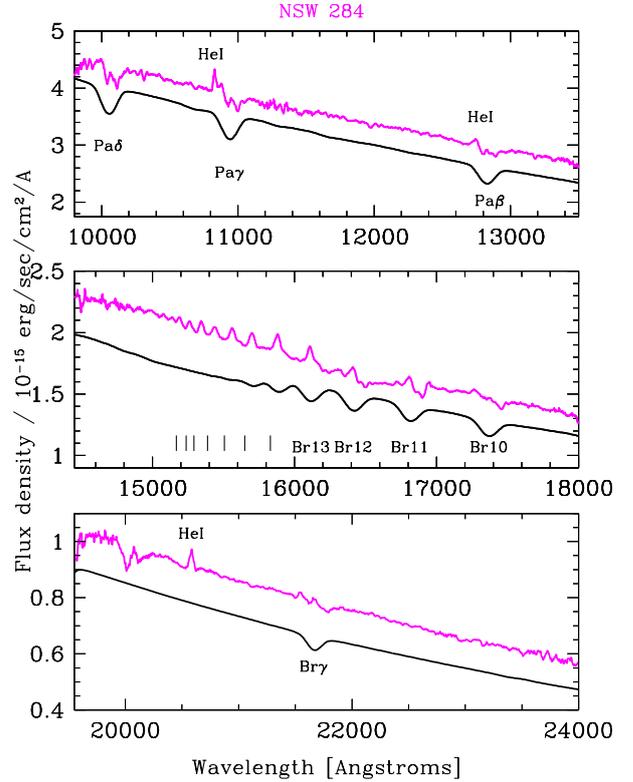} 
\caption{Near-infrared ($JHK$) spectrum of NSW\,284 (magenta) taken during the rise of burst B5, compared to a NextGen model spectrum that has been reddened by the extinction value in Table 1, and scaled (black). The spectrum temperature of 9800 K was chosen to highlight the expected locations of the near-infrared hydrogen lines. NSW\,284 exhibits a relatively featureless continuum, with only weak H and He emission lines. Paschen signatures at Pa$\delta$, Pa$\gamma$, and Pa$\beta$ are apparent, as is Brackett line emission, notably in the $H$-band lines as well as Br$\gamma$ in the $K$-band. There is also clear \ion{He}{I} emission at 10830 and 20581\,\AA. 
\label{fig:irspec}}
\end{figure}

\subsubsection{Palomar 200" / TripleSpec}\label{triplespec}

An infrared spectrum was taken on 2019-07-19 with the Palomar 200" telescope and facility infrared spectrograph TripleSpec \citep{Herter2008}. Similar to the DoubleSpec data, it was taken during the brightness increase of burst B5, specifically eight days earlier. The data was processed using a customized version of  the \texttt{spextool} package \citep{Cushing2004}, with telluric correction making use of the \texttt{xtellcorr} code \citep{Vacca2003}. Fig.\,\ref{fig:irspec} shows the final extracted and combined spectrum, which has $R\approx 2700$.

The spectrum has a blue continuum and is relatively featureless aside from hydrogen and helium signatures. \ion{H}{I} Brackett line emission, notably in Br$\gamma$ in the $K$-band and in the higher lines above Br13 in the $H$-band, as well as Paschen line emission at Pa$\beta$, Pa$\gamma$, and Pa$\delta$ in the $J$-band is apparent. The Paschen lines appear to have some structure, perhaps a mix of absorption and emission. The Brackett pattern has the peak emission strength at Br13 and Br14, with weaker lines above and below, indicating high optical depth. Several \ion{He}{I} emission lines are present throughout the spectral range, but there is no \ion{He}{II}. No metal lines can be identified, such as those exhibited by some Be stars \citep{Cochetti2022}.

\begin{figure*}
\centering
\includegraphics[angle=0,width=\columnwidth]{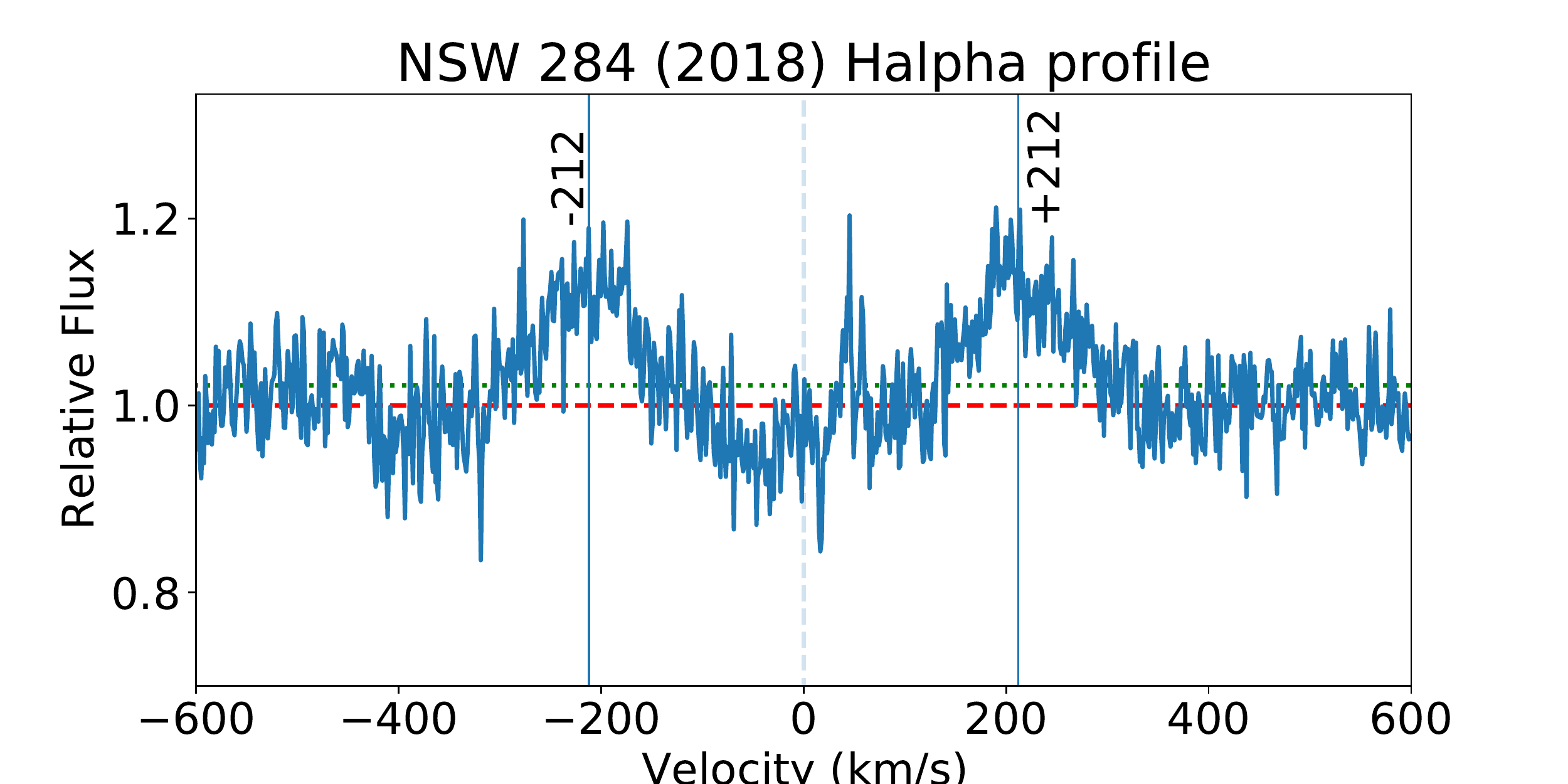} \hfill
\includegraphics[angle=0,width=\columnwidth]{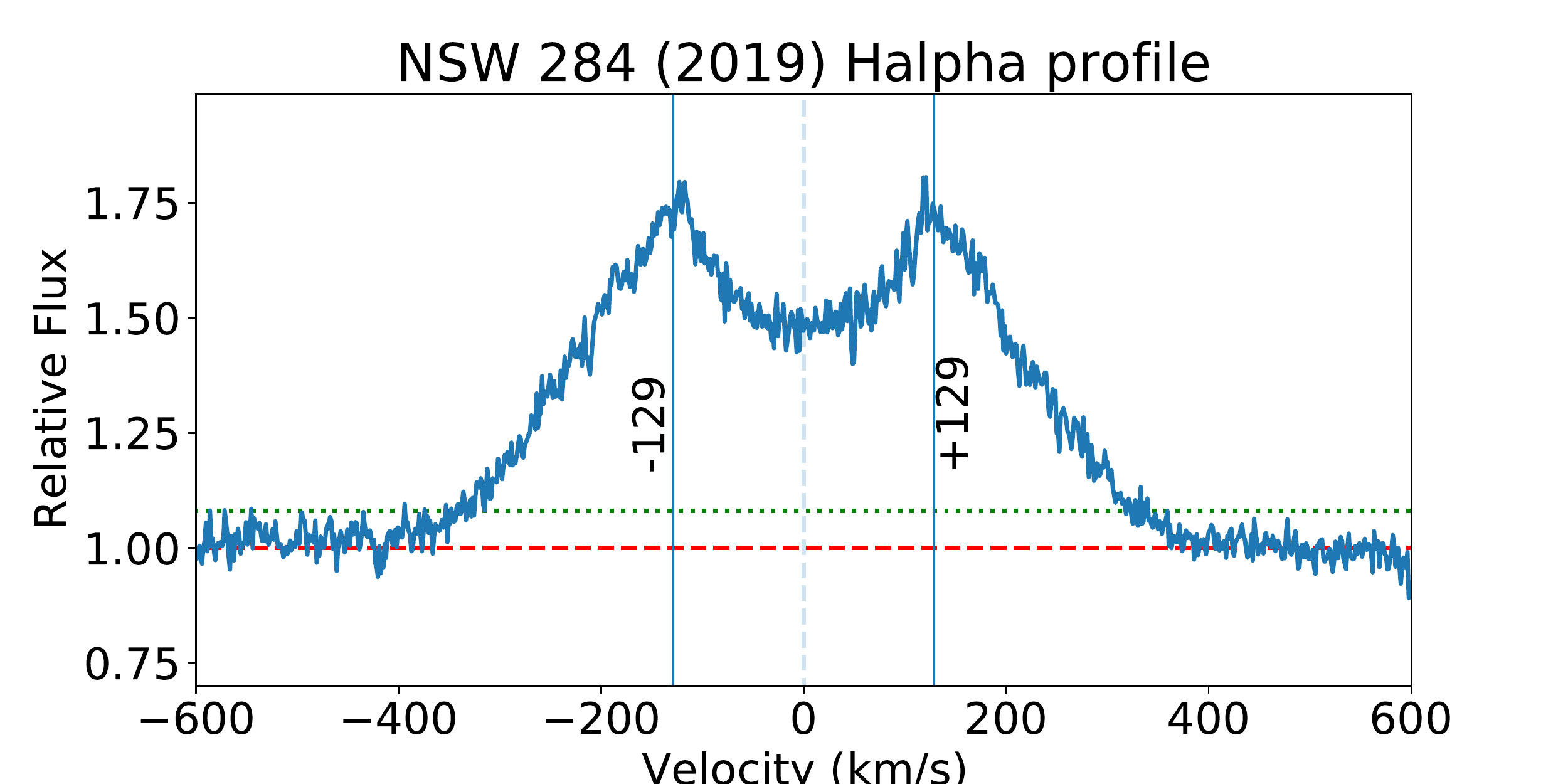} \\
\includegraphics[angle=0,width=\columnwidth]{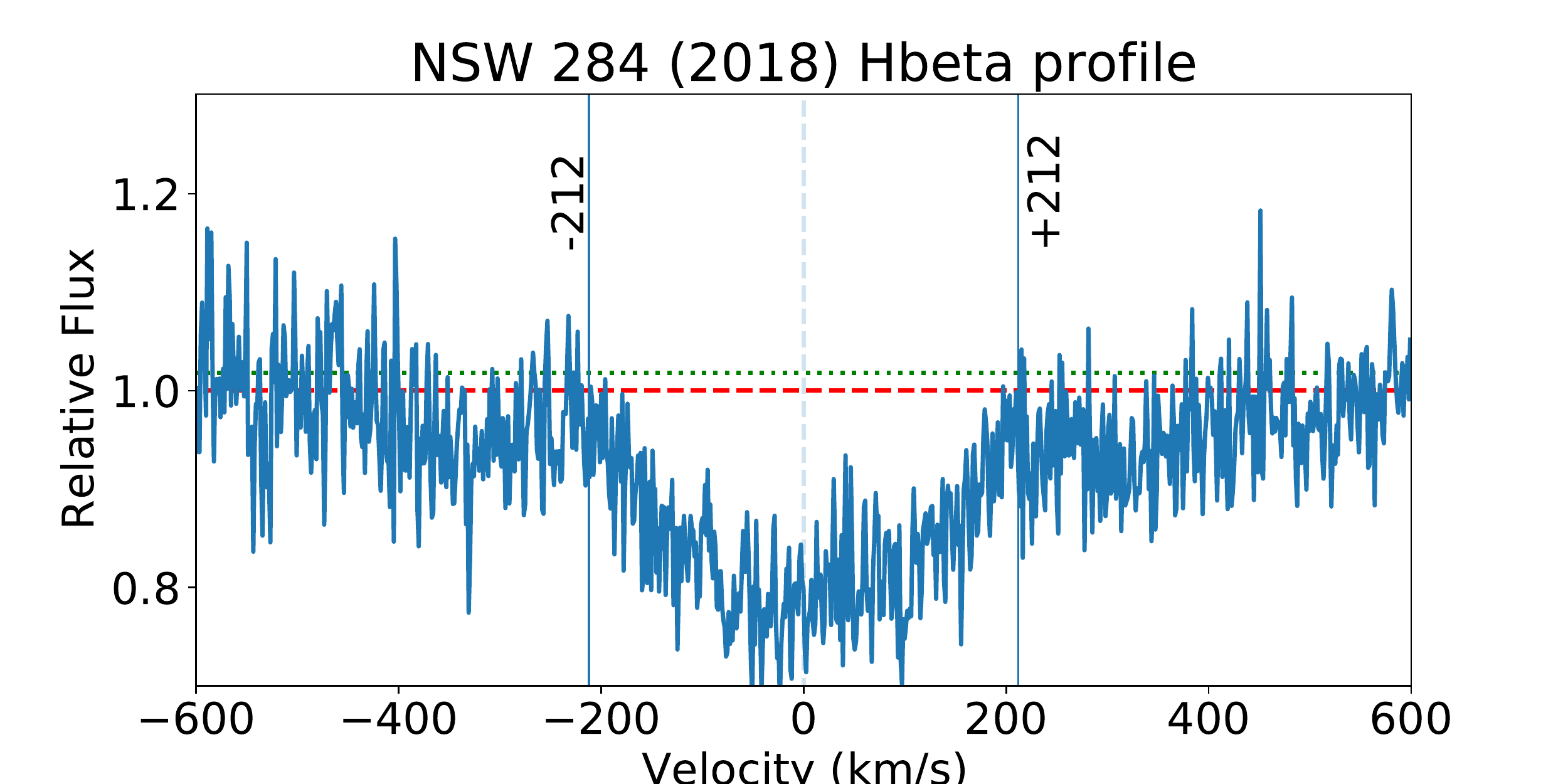} \hfill
\includegraphics[angle=0,width=\columnwidth]{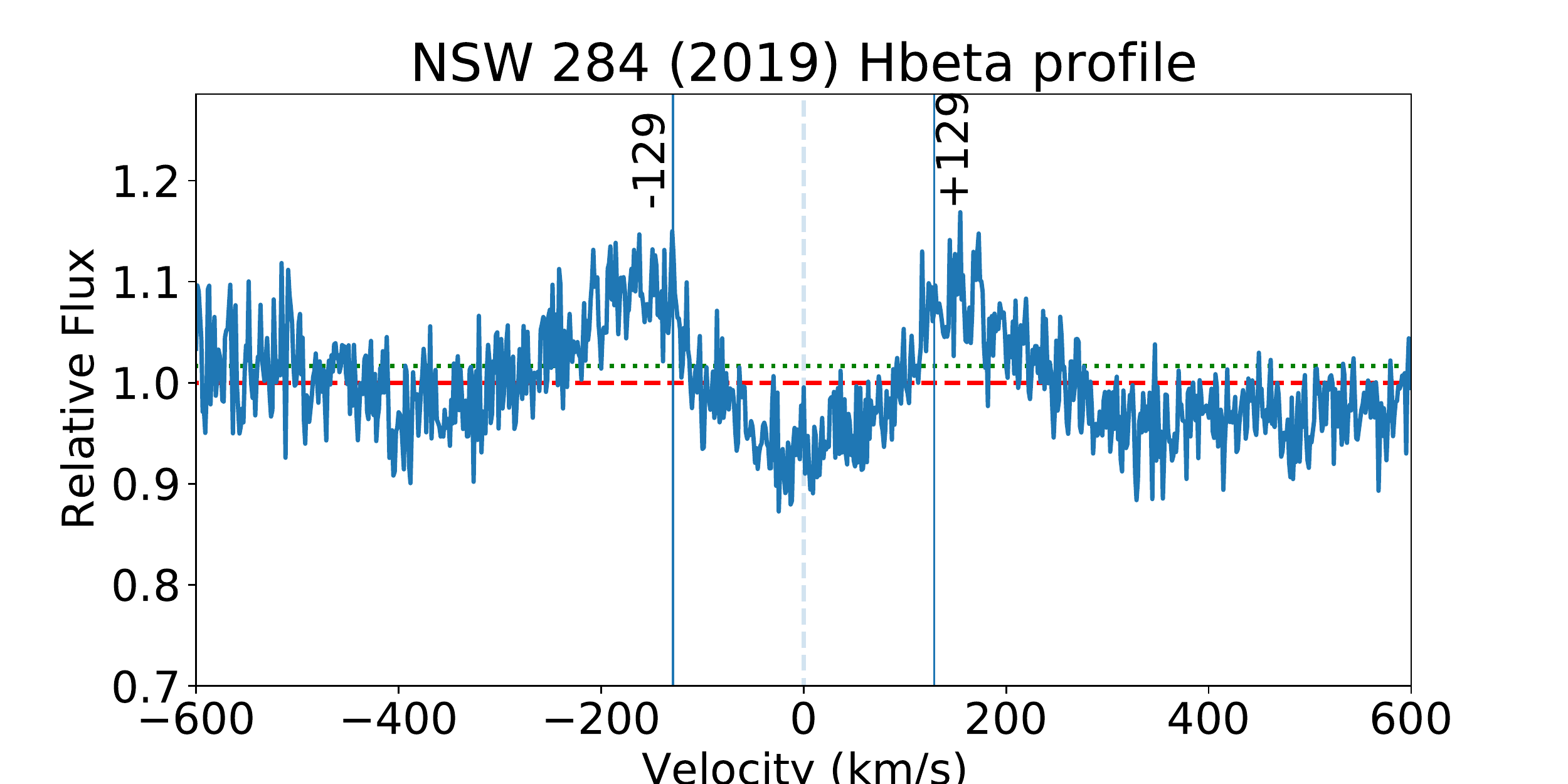} \\
\includegraphics[angle=0,width=\columnwidth]{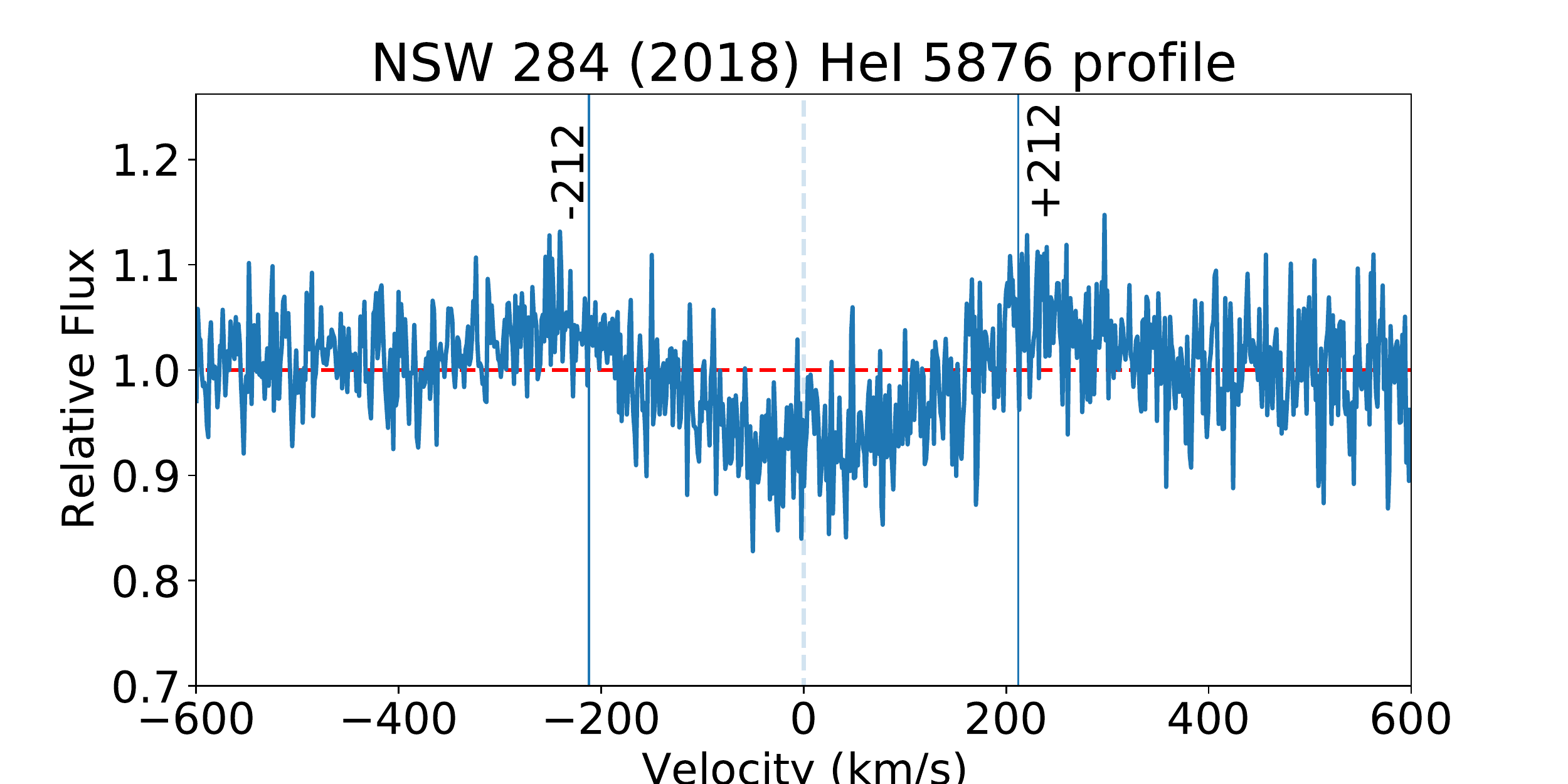} \hfill
\includegraphics[angle=0,width=\columnwidth]{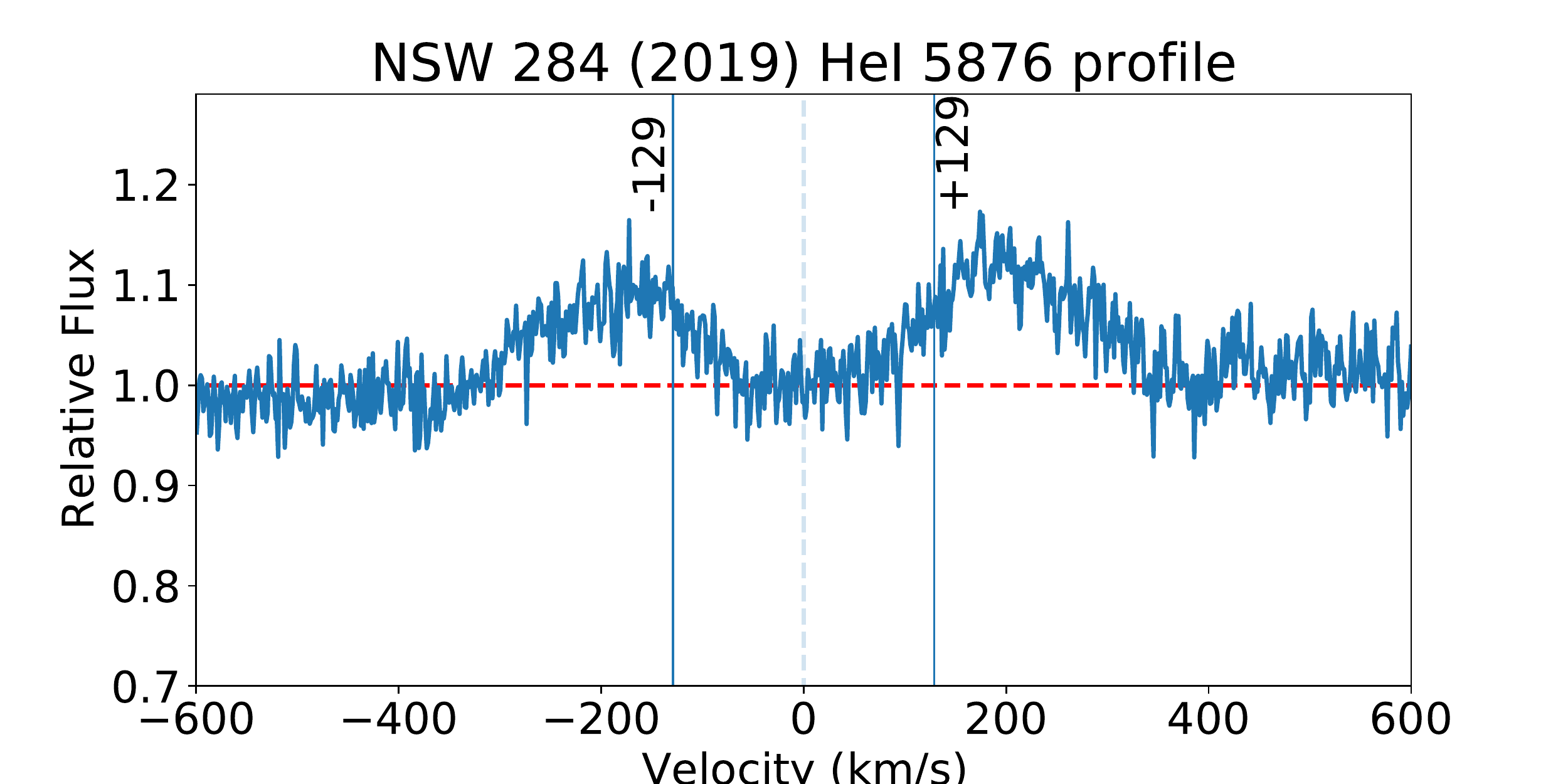} \\
\caption{H$\alpha$, H$\beta$, and \ion{He}{I} 5876\,\AA\ profiles for NSW\,284 from 2018 (left, quiescence just after weak burst B4) and 2019 (right, during brightness decline of strong burst B5). The red dashed line shows the profile normalization level while the green dotted line indicates the 10\,\% intensity. The vertical dashed line indicates zero stellocentric velocity, adopting a heliocentric radial velocity of -57.8\ \kms. Vertical lines at $\pm$212\ \kms\ in 2018 and $\pm129$\ \kms\ in 2019 correspond to the peaks of the double-horned profile in H$\alpha$. Note that the brighter photometric state corresponds to stronger lines but lower velocities.  Note also that the peaks in H$\beta$ seem to be at larger velocities than the peaks in H$\alpha$, and those in \ion{He}{I}\,5876\,\AA\ larger still. \label{fig:nsw284balmer}}
\end{figure*}

\subsubsection{Keck/HIRES}\label{nsw_temp}

Two high dispersion optical spectra were obtained using the W.M. Keck Observatory and HIRES \citep{Vogt1994} with wavelength coverage $\sim$4800\,--\,9200\,\AA\ at resolution R\,=\,25,000. The observations were obtained on 2018-11-03, and 2019-11-29 (UT) and were processed using the MAKEE reduction pipeline\footnote{\url{https://astro.caltech.edu/~tb/makee/}}  written by T. Barlow. Unfortunately, neither of the high-dispersion observations occurred during the peak of a photometric burst of NSW\,284. The first spectrum was obtained in a clearly quiescent period of the light curve, just after the weak burst B4. The second spectrum has been taken during the tail end of the 2019 burst B5, where the brightness in all filters is still clearly increased compared to the base level.

The spectra are fairly featureless, with hydrogen and helium the only stellar lines. The few absorption lines that are present are broad, and indicate rapid rotation. Between the two epochs, there is a change in the presentation of these features. Velocity profiles of the H$\alpha$ and H$\beta$ Balmer lines are illustrated in Figure~\ref{fig:nsw284balmer}.

In 2018 there was weak, doubled emission apparent at H$\alpha$ and H$\beta$, and similarly weak and broad Paschen lines apparent starting at 8865\,\AA. In absorption, there are notable strong DIB features at 5488, 5491, 5508, 5780, 5797, and 6614\,\AA. The only stellar absorption lines are from \ion{He}{I} (4921\,\AA\ with $W_\lambda$\,=\,0.93\,\AA; 5015\,\AA\ with $W_\lambda$\,=\,0.32\,\AA; 5876\,\AA\ with $W_\lambda$\,=\,0.57\,\AA; and 6678\,\AA\ with $W_\lambda$\,=\,0.55\,\AA). There is no \ion{He}{II} apparent. Using the equivalent width correlations to spectral type in \cite{Leone1998}, a B3\,--\,B4 spectral type is inferred. Relevant to our analysis below, this implies a temperature for the star of $T_{\rm eff} \approx 17,500$\,K. We note that the disk emission potentially could 'fill in' the lines, and that the \ion{He}{I} to H$\beta$ line ratio is indicative of temperatures in excess of 20\,000K. This is in line with the hotter temperature of 22\,666~K reported in Gaia\,DR3 (see Table\,\ref{source_table}), which would correspond to a spectral type $\sim$B1. In the table we can also see that Gaia estimates a temperature of 15\,415\,K for Gaia\,19eyy, which however, has almost the same absolute magnitude, extinction and optical colours as NSW\,284. It is thus not clear either how accurate the Gaia effective temperature estimates for these type of objects are. We thus use the estimated 17\,500\,K for NSW\,284 (and in turn Gaia\,19eyy) in our modelling later on. But we note that all model results are expressed in units of the assumed stellar temperature and can be easily scaled to other effective temperatures. Section\,\ref{burst_temp} discusses in detail how the assumption of a different effective temperature influences the quantitative results.

In 2019 the hydrogen Balmer and Paschen lines are all stronger, with clear double-horn emission peaks. Several of the helium lines now also exhibit emission, with a similar double-peak structure. Only the 4921\,\AA\ line remains in absorption, though it is weaker now at just $W_\lambda$\,=\,0.67\,\AA. An estimate for the stellar rotation can be estimated from the full width half maximum of this line, with $v~sin{i} \approx 280$\ \kms.

From analysis of the hydrogen emission lines (Figure~\ref{fig:nsw284balmer}) a heliocentric radial velocity of -57.8\ \kms\ is derived with an estimated error of at least 5\ \kms. For the 2019 spectrum, the velocity separation of the H$\alpha$ peaks is $\pm129$\ \kms, based on Gaussian fitting, while the full-width at 10\,\% intensity of the line is 719\ \kms. For H$\beta$ the peaks are at $\pm170$\ \kms, 32\,\% higher velocity than the H$\alpha$ peaks. In the 2018 spectrum, the velocity separation of the H$\alpha$ peaks is $\pm212$\ \kms; the H$\beta$ is hard to measure, though the peak separation again appears larger than for H$\alpha$, closer to 231\ \kms.  Between the two spectral epochs, the separation of the two velocity peaks clearly changes.

\begin{figure}
\centering
\includegraphics[angle=-90,width=1.1\columnwidth]{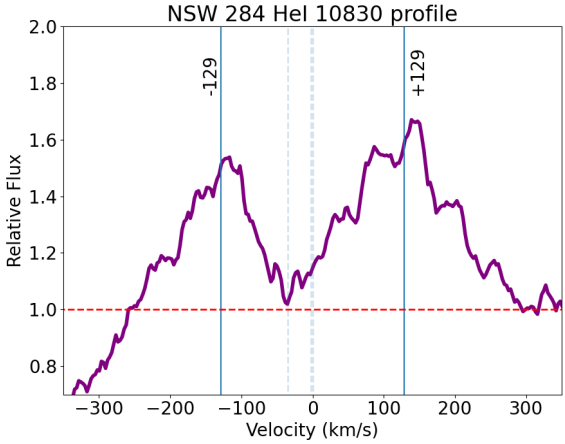} 
\caption{NIRSPEC \ion{He}{I} 10830\,\AA\ profile of NSW\,284, taken during the brightness decrease of burst B5. The three different components of the triplet line are marked at 0, -2.5, and -34.7\,km/s (dashed vertical lines). The double-horn morphology in the 10830\,\AA\ line is similar to that exhibited by the \ion{H}{I} Balmer $\alpha$ and $\beta$ lines, and the \ion{He}{I} 5876\,\AA\ line (Fig.~\ref{fig:nsw284balmer}); the same reference velocity as for the 2019 HIRES spectrum is marked at $\pm129$\,km/s (solid vertical lines). \label{fig:He10830}}
\end{figure}

\subsubsection{Keck / NIRSPEC}

Finally, a high dispersion near-infrared spectrum was obtained using the W.M. Keck Observatory and NIRSPEC \citep{McLean1998} spectrograph covering the $1\mu$m Y-band at resolution R\,=\,19,000. The observations were obtained on 2019-11-17 (during the brightness decrease of burst B5) in an ABBA nodding fashion, and were processed along with a telluric standard using the REDSPEC\footnote{written by L. Prato, S.S. Kim, \& I.S. McLean} package for executing the trace, extraction, wavelength calibration, and spectral combining.

Our main interest with these observations was in the region containing the \ion{He}{I} 10830\,\AA\ triplet line. Similar to the \ion{H}{I} and \ion{He}{I} lines in the optical HIRES spectrum, the \ion{He}{I} 10830\,\AA\ is seen in double-peaked emission, with a separation between the peaks that is consistent with that in the HIRES spectrum taken at about the same time. The order also covers Pa$\gamma$. Figure~\ref{fig:He10830} shows the \ion{He}{I} line.

\begin{figure}
\centering
\includegraphics[angle=0,width=\columnwidth]{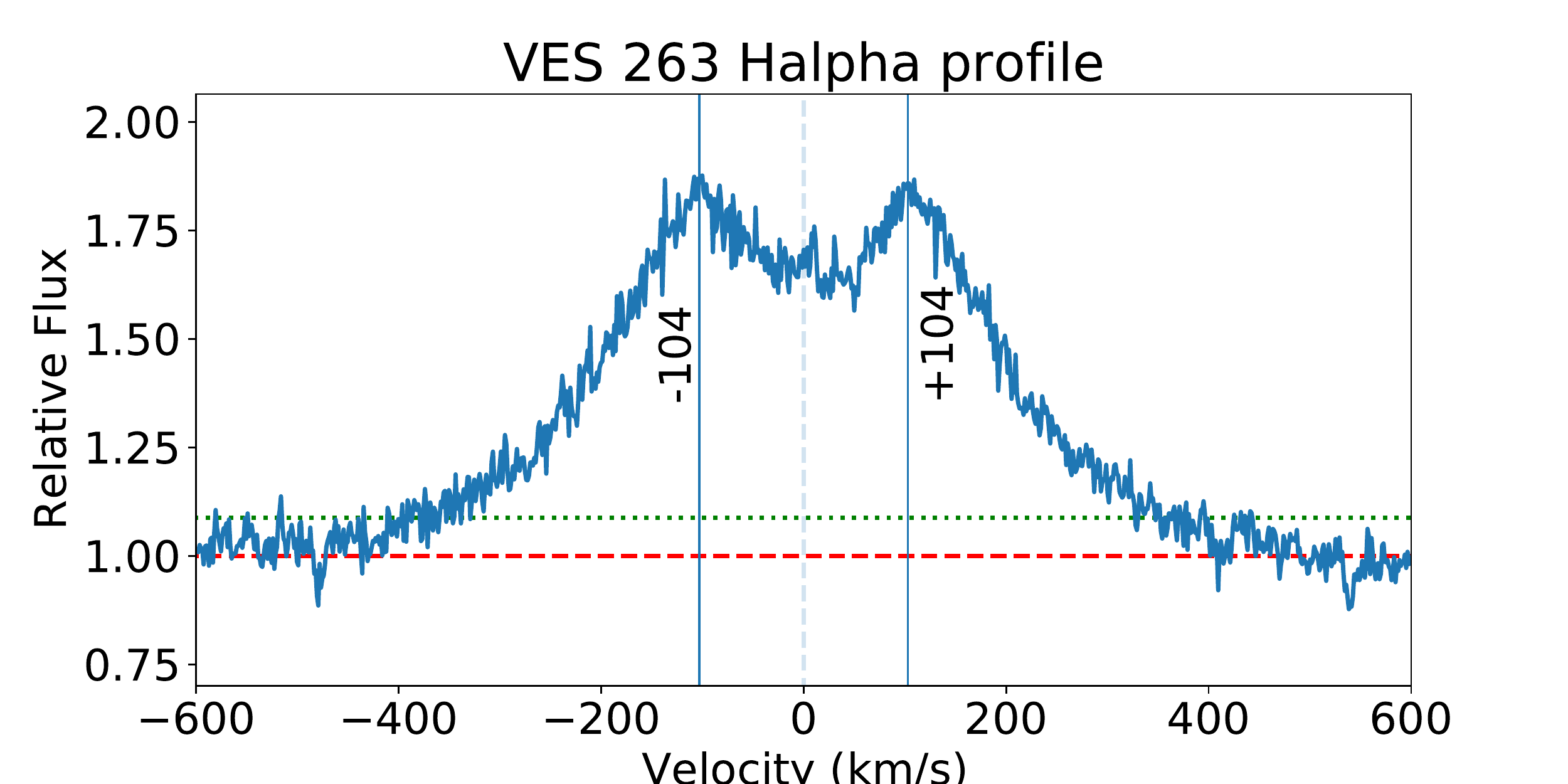} \\
\includegraphics[angle=0,width=\columnwidth]{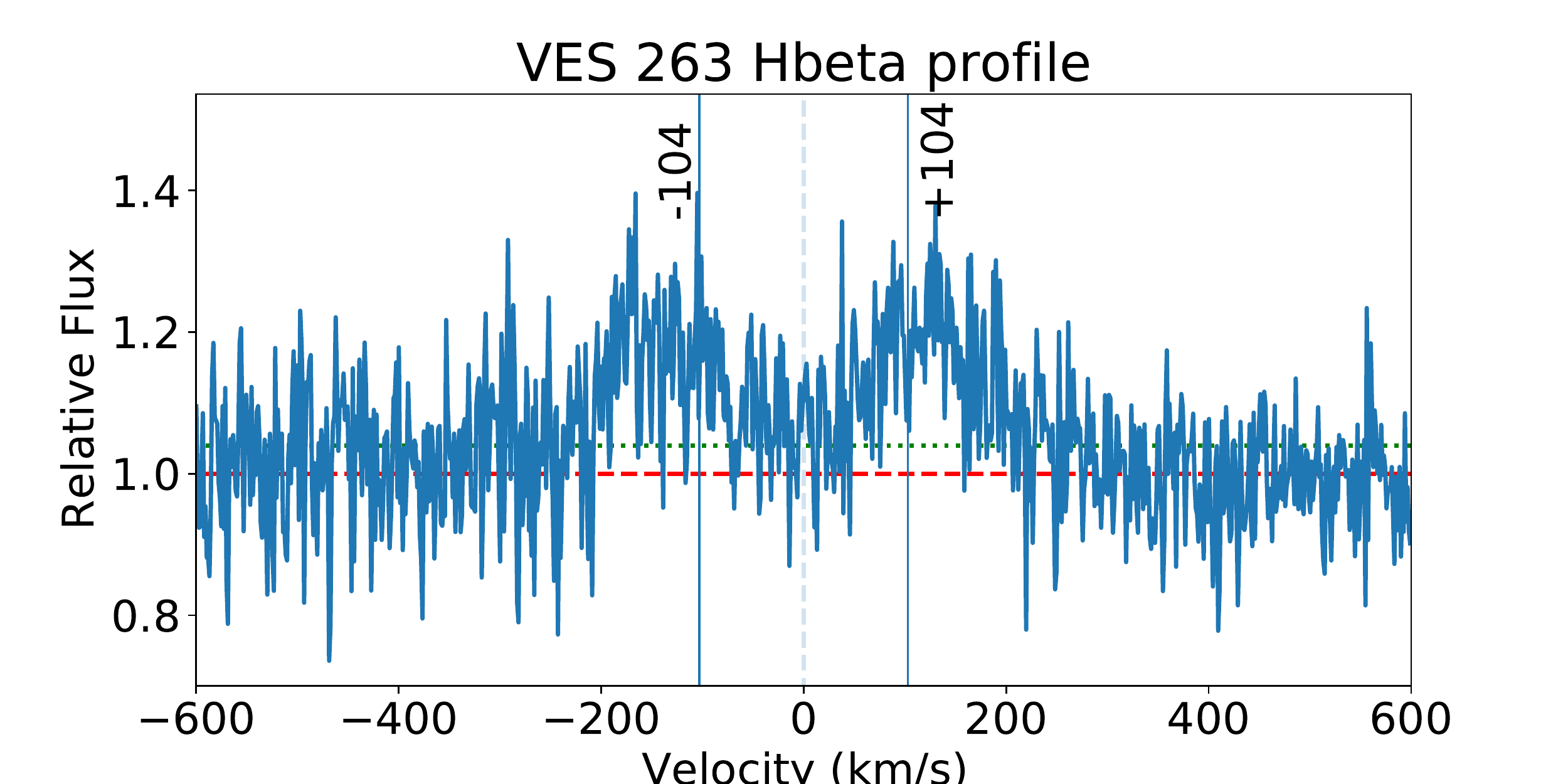} \\
\includegraphics[angle=0,width=\columnwidth]{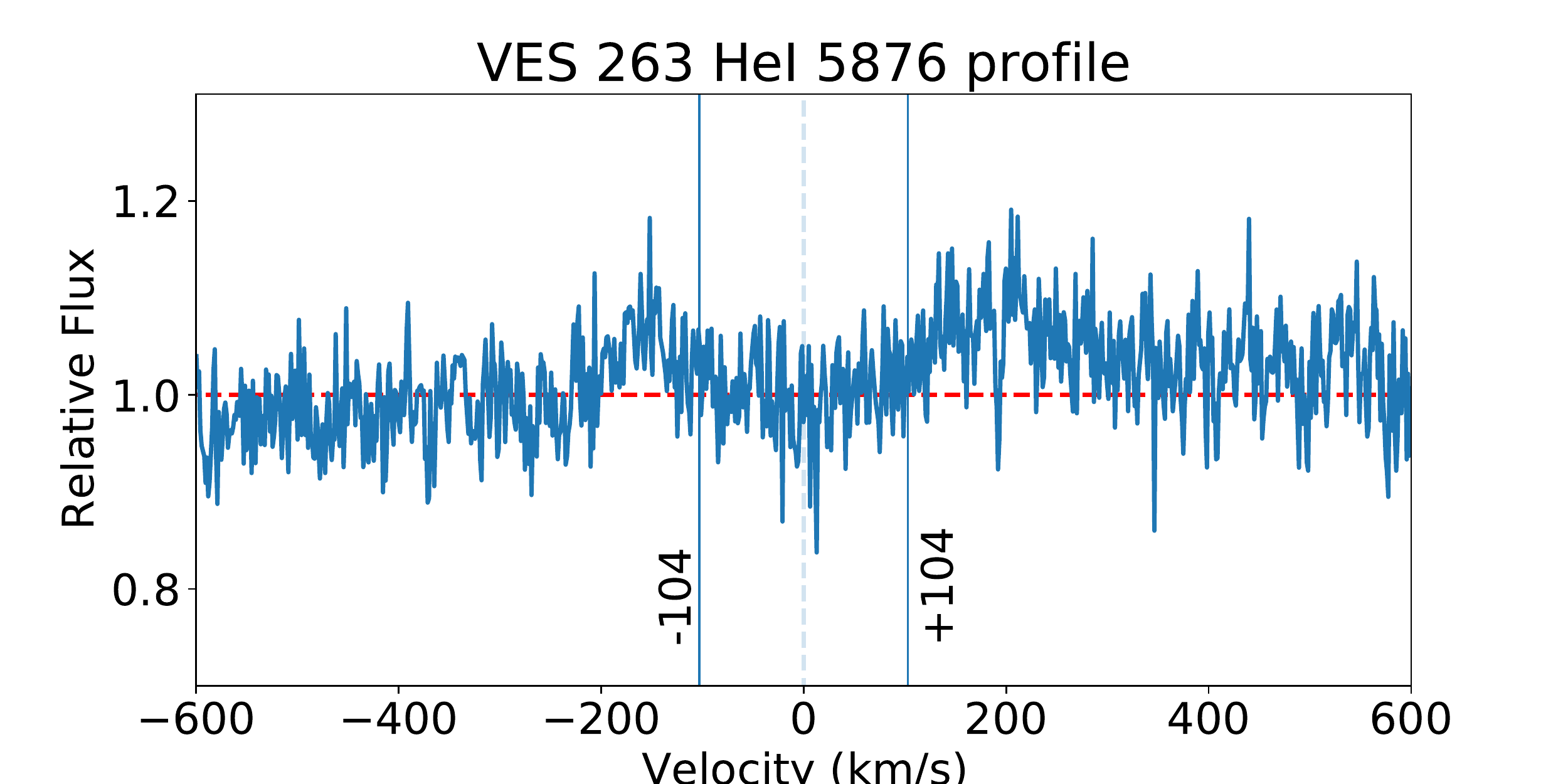} 
\caption{H$\alpha$, H$\beta$ and \ion{He}{I} 5876\,\AA\ profiles for VES\,263 taken during the transition from burst B10 to B11. The horizontal red dashed line shows the profile normalization level while the green dotted line indicates the 10\,\% intensity. The vertical dashed line indicates zero stellocentric velocity, adopting a heliocentric radial velocity of -4.1\,km/s. Vertical solid lines at $\pm104$\,km/s correspond to the peaks of the double-horned profile in H$\alpha$; the H$\beta$ peaks have larger separation, $\pm129$\,km/s and the \ion{He}{I} 5876\,\AA\ larger still. \label{fig:ves263balmer}}
\end{figure}

\subsection{VES~263}

On 2019-11-29 (UT) a Keck/HIRES spectrum of VES\,263 was obtained. This epoch in the light curve is a transition between bursts B10 and B11 of the source, and the first occasion where the source brightness did not fully return to the earlier baseline magnitudes. Our spectra show the Balmer lines of H$\alpha$ and H$\beta$ with a double-horned emission profile, with the Paschen series lines exhibiting similar morphology. \ion{He}{I} 5876\,\AA\ has a similar profile. \ion{He}{I} 4922\,\AA, a hotter line, is seen in absorption with $W_\lambda = 0.21$\,\AA. Otherwise, the spectrum shows only a number of strong DIBS lines, specifically those at 5488, 5491 (weak), 5508 (weak), 5780, 5797, 5849, 6269, 6379, 6284, 6614\,\AA, plus narrow interstellar absorption in \ion{Na}{I}\,D and \ion{K}{I}. 

From fitting of the hydrogen profiles (see Fig.~\ref{fig:ves263balmer}) a heliocentric radial velocity of -15.7\ \kms\ is derived with an estimated error of at least 5\ \kms. However, the value of -4.1\ \kms\ reported by \cite{Munari2019} provides better centering of the double-peaked profiles, and we thus adopt that. The velocity separation of the H$\alpha$ peaks is $\pm104$\ \kms, based on Gaussian fitting, which can be compared to the $\pm130-140$\ \kms\ found by \cite{Munari2019}. The full-width at 10\,\% intensity of the H$\alpha$ line is 713\ \kms. For H$\beta$ the peaks are at $\pm129$\ \kms, a 24\,\% higher velocity than the H$\alpha$ peaks. \cite{Munari2019} have also demonstrated that the {He}{I} 5876\,\AA\ double-peaked profiles are more separated than the H$\alpha$ peaks at their observing epochs.  They also find, similar to what we found for NSW 284, that the peak velocities change over time such that the higher velocities correspond to minimum brightness (although our interpretation of these findings differs; see our discussion below).

We do not have an infrared spectrum of VES\,263, but refer the reader to Fig.\,4 of \cite{Munari2019}. Similar to NSW\,284, there is \ion{H}{I} Paschen and Brackett line emission indicating high optical depth, as well as \ion{He}{I} emission. The spectrum was taken during a low state, and the emission appears stronger than that in our NSW\,284 high-state infrared spectrum. Additionally, there are a few metal lines that are present in emission in the infrared in VES\,263. 

\section{Investigating the Burst Properties}\label{burstsection}

\begin{figure*}
\centering
\includegraphics[angle=0,width=0.68\columnwidth]{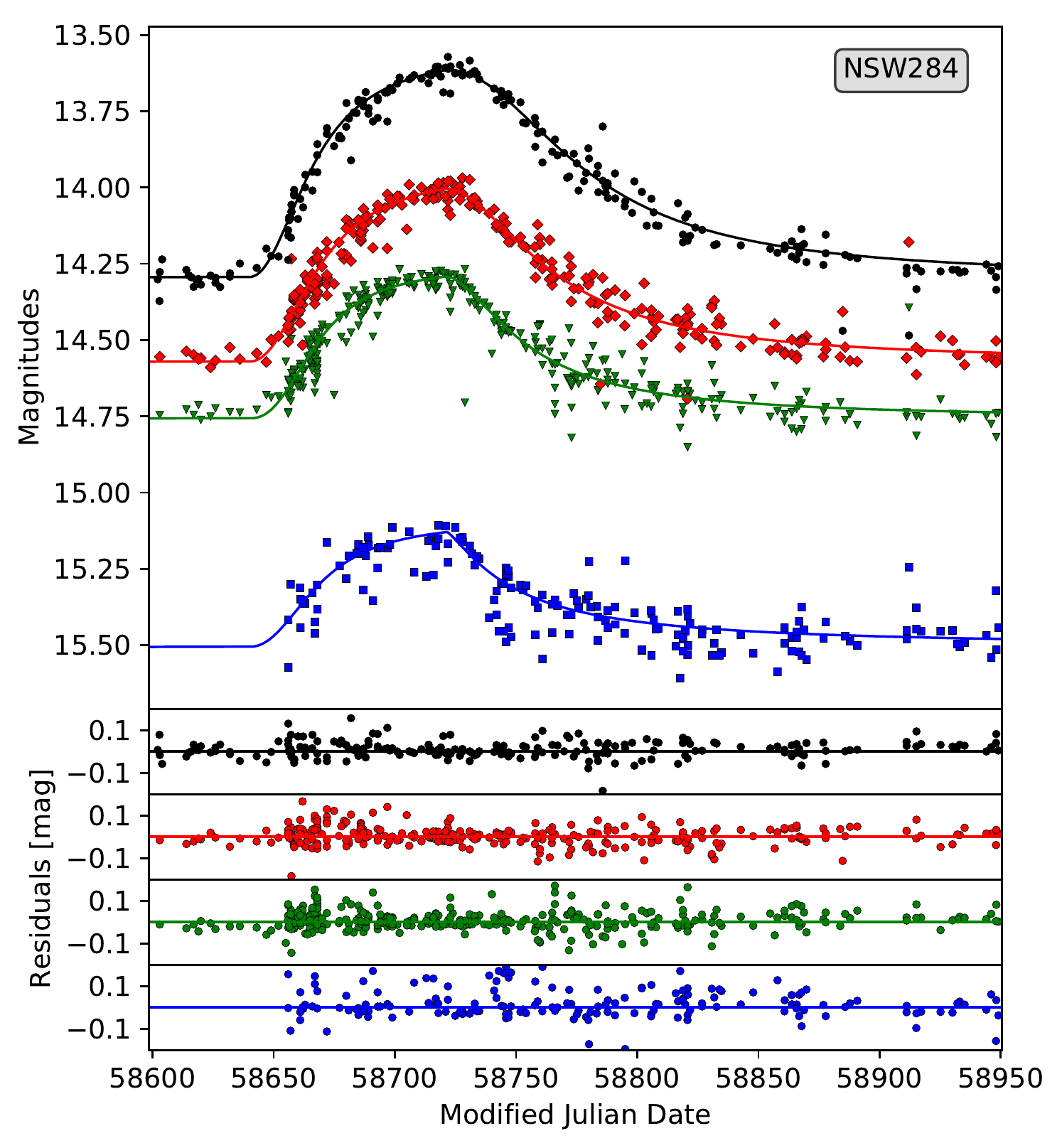} \hfill
\includegraphics[angle=0,width=0.66\columnwidth]{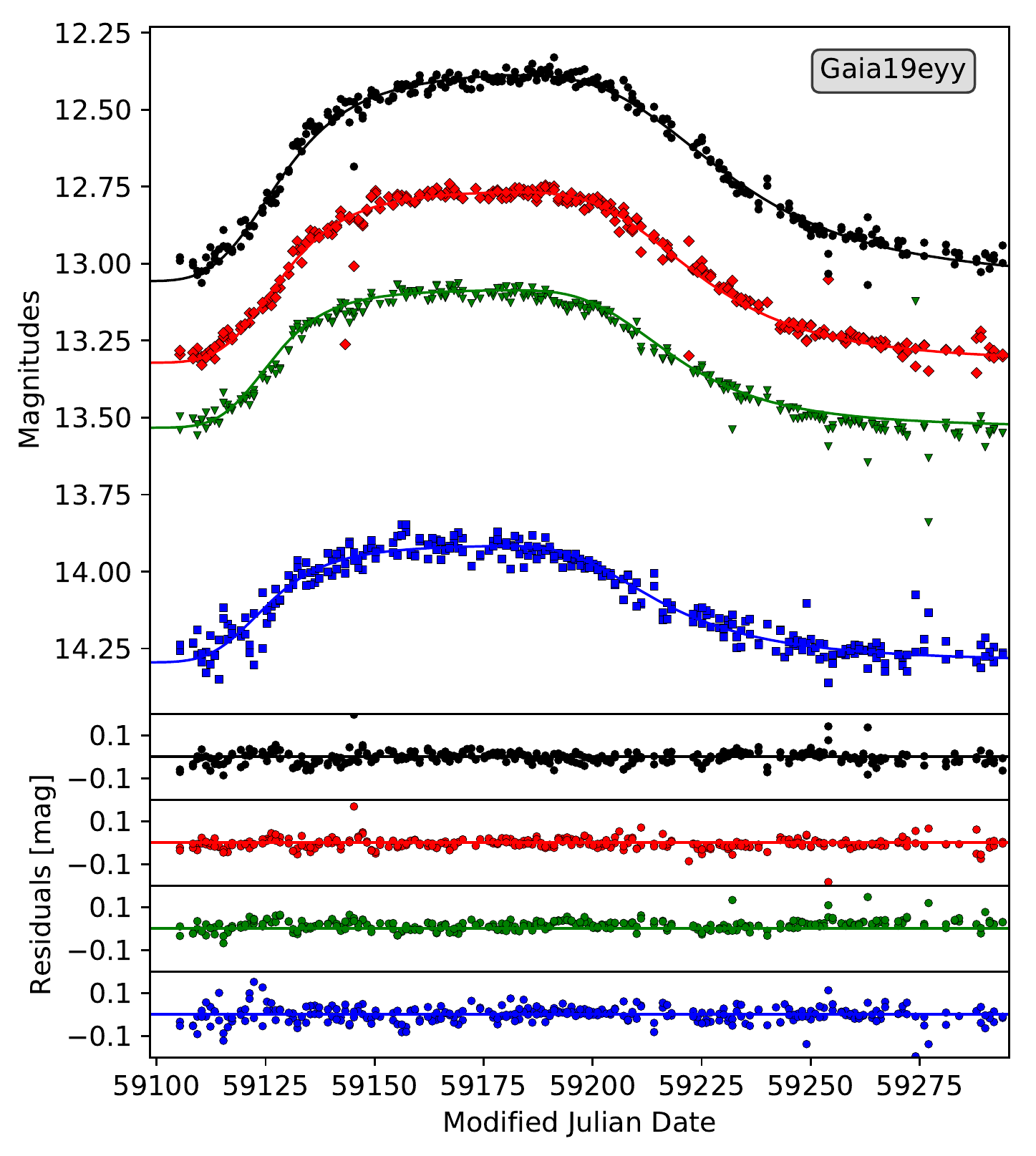} \hfill
\includegraphics[angle=0,width=0.65\columnwidth]{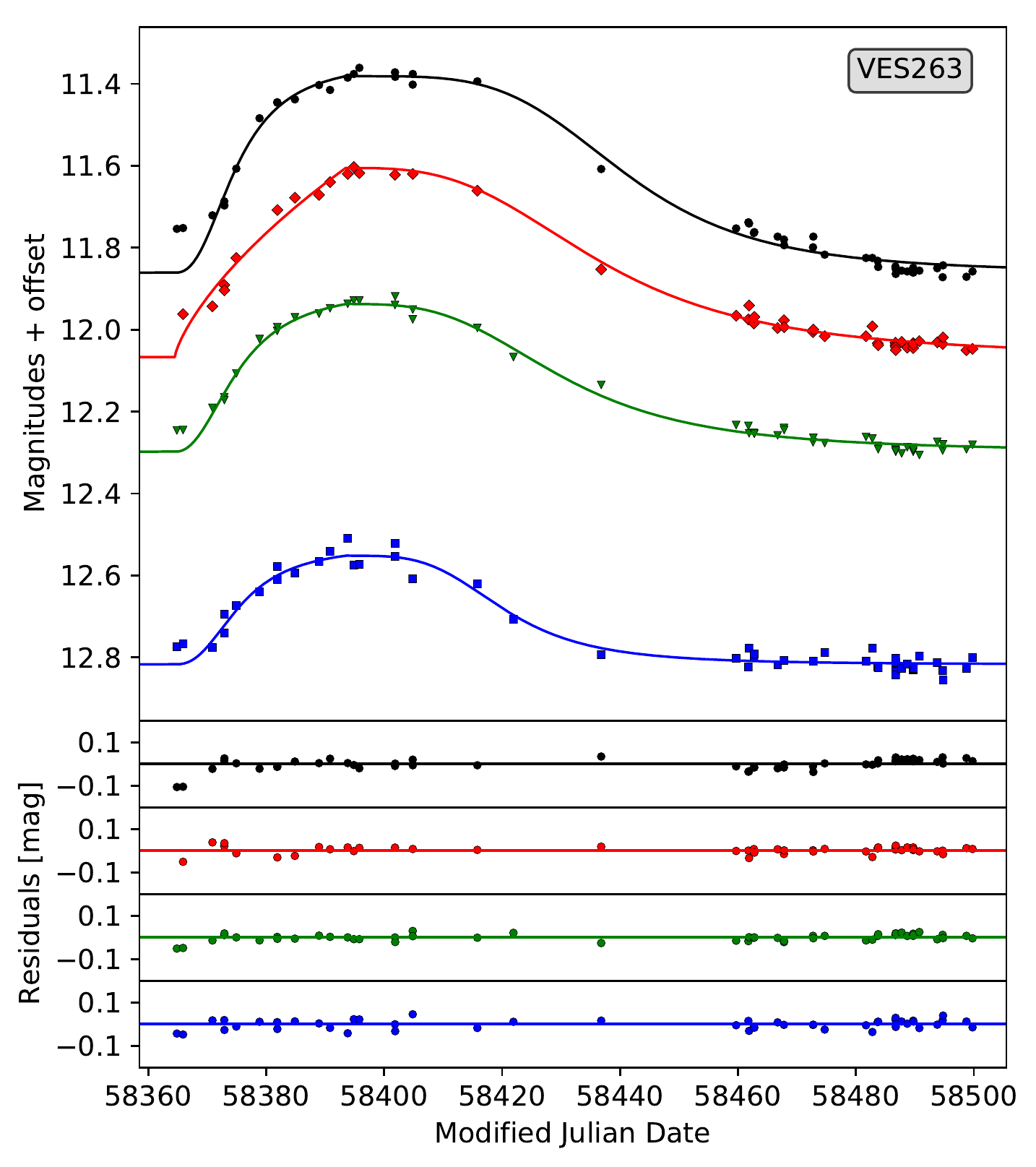} \\
\caption{Detailed optical photometry and model fit for one of the bursts in each object (left: NSW\,284, Burst B5; middle: Gaia\,19eyy, Burst B5; right: VES\,263, Burst B8). The symbols in the top panel show the HOYS photometry used in the light curve fit for the different filters (from bottom to top: $B$ - blue; $V$ - green; $R$ - red; $I$ - black). The solid lines are the best fits according to Eqs.\,\ref{dm_lesst2} and \ref{dm_moret2}. In the bottom panels we show the residuals of the data and fit with the same colour coding as in the top panel. Note that we have shifted the magnitudes of VES\,263 for better visibility as follows: $I$: $+0.8$\,mag, $V$: $-0.8$\,mag, $B$: $-2.1$\,mag. \label{example_burst}}
\end{figure*}

The long term light curves in Fig.\,\ref{hoys_lc} and our discussion in Sect.\,\ref{photometry} show that the sources have semi-periodic outbursts in brightness, with average properties reported in Table\,\ref{source_table}. The bursts repeat roughly between once or twice per year. The amplitudes and detailed shapes of each burst differ. Some bursts are very weak or absent, i.e. especially NSW\,284 seems to have a quiescent period between mid 2017 and mid 2019. One feature in common among all three of our sources is that -- in all bursts -- the amplitudes at longer wavelengths are higher than at shorter wavelengths, and the peak brightness occurs later at longer wavelengths.

All bursts show a relatively fast increase in brightness which levels off towards the maximum brightness. The decline back towards the quiescent state is always slower than the increase. During the bursts, all colours do increase towards the peak and then return to their normal values. This clearly indicates that the additional emission from the sources during the bursts originates from material that is cooler than the surface temperature of the stars. 

\subsection{Analysis of the burst shape}

All evidence is consistent with the interpretation that the objects investigated here are repeatedly outbursting Be stars. Numerical simulations and theoretical analysis of such objects by \citet[][R18 hereafter]{Rimulo2018} have provided an analytical description of the expected light curve shape. The model is based on the assumption that disk loading starts at a time $t_1$ and continues until time $t_2$, when the disk material is dispersed again. According to R18 the brightness increase $\Delta m$ in the light curve as a function of time $t$ should follow the equations:

\begin{equation}\label{dm_lesst2}
\displaystyle \Delta m = \Delta m^\infty \left( 1 - \frac{1}{1 + \left[ C_1 (t - t_1) \right]^{\eta_1}} \right)
\end{equation}

\begin{equation}\label{dm_moret2}
\displaystyle \Delta m = \Delta m^\infty \left( 1 - \frac{1}{1 + \left[ C_1 (t_2 - t_1) \right]^{\eta_1}} \right) \left( \frac{1}{1 + \left[ C_2 (t - t_2) \right]^{\eta_2}} \right)
\end{equation}

where Eq.\,\ref{dm_lesst2} is valid for for times $t_1 \leq t \leq t_2$, and Eq.\,\ref{dm_moret2} for times $t > t_2$. The value $\Delta m^\infty$ denotes the asymptotic magnitude increase (in the filter used) that would be achieved if $t_2 -t_1$ would be very large. The parameters $C$ and $\eta$ are free parameters which depend on the filter, the source properties and assumptions made about the disk (R18), e.g. its temperature relative to the star. The parameters differ for the brightness increase (index 1) and decrease (index 2). In particular, the $C$ parameters depend on the viscosity $\alpha$ of the disk as:

\begin{equation}
C_1 = \alpha_1 \frac{\zeta_1}{\alpha \tau} \hspace{0.5cm} {\rm and } \hspace{0.5cm} C_2 = \alpha_2 \frac{\zeta_2}{\alpha \tau}.
\end{equation}

Following R18, the $\zeta$ and $\eta$ values are determinable from numerical simulations and $\alpha \tau$ can be inferred from the intrinsic source properties as well as the orbital velocity and sound speed of the disk material according to Eq.\,\ref{alphatau}.

\begin{equation}\label{alphatau}
\alpha \tau = \sqrt{\frac{R^3_{eq}}{G M}} \frac{v^2_{orb}}{c^2_s}
\end{equation}

Here $R_{eq}$ is the equatorial radius of the star and $M$ the stellar mass. The orbital velocity of the constant temperature disk is $v_{orb} = (GM/R_{eq})^{1/2}$, and the sound speed is determined as $c_s = \sqrt{ \frac{\gamma k_B T_{disk}}{\mu m_H} }$. The adiabatic index $\gamma$ would be equal to 1.67 for atomic gas and the mean mass per particle $\mu = 1.3 $ for typical abundances. 

We have followed the description in R18 and fit the shape of all bursts for all objects using a least squares optimisation for Eqs.\,\ref{dm_lesst2} and \ref{dm_moret2} in all filters leaving all parameters to vary freely. We also included a part of the light curve prior to the burst, and allowed the fitting to also determine the baseline brightness $m_0$ of the star, so that the observed magnitudes $m$ in the light curve are $m = \Delta m + m_0$. 

When this completely unrestricted fit was applied to the data of the same burst in different filters, it returned slightly different values for the start $t_1$ of the burst and the end of the disk loading $t_2$. It is obvious that these times should be the same for the data of the same burst in all filters. Thus, we manually chose the values for $t_1$ and $t_2$ that corresponded best to the shape of the light curve. These are typically very close to the values returned for fitting the $I$-band data, as this has the highest amplitude in the optical filters. These manually selected values are then fixed for all filters and other parameters are evaluated again. 

The resulting fits resemble the shape of the bursts very well. We show the fit and residuals for one example burst for each source in Fig.\,\ref{example_burst}. Typically the residual root mean square (RMS) of the data and fit are of the order of 0.3\,--\,0.4 times the photometric uncertainty for NSW\,284 and Gaia\,19eyy, and 0.5\,--\,0.6\,$\sigma$ for VES\,263. There are some cases with small systematic deviations of the data from the fit, as e.g. during the first 30 days of the brightness increase in the burst B5 of Gaia\,19eyy (see middle panel of Fig.\,\ref{example_burst}). We note that the exact choice of the $t_1$ and $t_2$ values can change the best fitting parameters ($\eta_1, \eta_2, C_1, C_2$). Similarly, adding or removing individual data points can cause changes in the best fit parameters. However, in all cases the shape of the fit and the RMS are only marginally changed. Furthermore, one can fix some of the parameters in a wide range from the best values. This causes other parameters to change without any sizeable increase in the fit RMS. For example, forcing an increase in $\eta$ leads to a decrease in the corresponding $C$ value. Thus, for the objects and bursts investigated here, fitting all free parameters in Eqs.\,\ref{dm_lesst2} and \ref{dm_moret2} without any constraints on $\eta$ from numerical simulations, as in R18, does not allow us to investigate disk viscosity. It is beyond the scope of this paper to perform these simulations. We list our notional best fitting parameters for all objects, bursts and filters investigated in detail in Table\,\ref{burst_table} in the Appendix. 

\subsection{Analysis of the burst temperature}\label{burst_temp}

The fits of the observed light curves in all available filters discussed in the previous section provide a smooth description of the shape of each burst. We utilise those to investigate the properties of the additional emission from the disk. This can in principle also be done with the original light curve data, but the results would be more noisy. One obtains very similar qualitative and quantitative results for both cases. The fits hence allow us to determine the amplitudes $\Delta m$ in each filter at all times after $t_1$.  We use the PHOENIX \citep{2013A&A...553A...6H} and ATLAS9 \citep{2003IAUS..210P.A20C} stellar atmosphere models and blackbody radiation to simulate these amplitudes in our optical filters. We use the solar metallicity and $\log(g)$\,=\,4 atmosphere models for these calculations. Note that these choices are reasonable and changing the surface gravity or metallicity only has a minimal effect on the results.

We assume that the central star has a temperature of $T_S$ and a visible projected surface area $A_S$ of unity. The emission responsible for the burst comes from material with a temperature $T_b$ and a projected surface area $A_b$, in units of $A_S$. The model spectra are convolved with the filter transmission curves accessed through the astropy {\tt PySynphot} distribution \citep{2013ascl.soft03023S}. For each time $t$ along the burst (in one day intervals) we find the burst temperature and area that result in the best fitting (lowest RMS) amplitudes. We use a Monte Carlo approach that varies the amplitudes to fit by a standard deviation of 0.01mag - in accordance with the light curve fit accuracy - to determine the statistical uncertainty of the best fitting parameters. We note that our model only considers optically thick emission. The continuum emission from Be-disks is usually modelled as a combination of a pseudo-photosphere and a tenuous disk \citep[e.g.][]{2015MNRAS.454.2107V, 2017MNRAS.464.3071V}. The tenuous disk contributes typically 30\,\% of the flux via optically thin emission. Thus, we expect our model to work less well at longer wavelengths, as the optical emission will be dominated by optically thick emission based on the temperatures involved.

From the best fitting temperature and area of the emission causing the burst, we further determine the additional luminosity of the burst based on $L_b = A_b \cdot T^4_b$ in units of the luminosity of the central star, determined the same way. Note that the choice of stellar temperature, atmospheric model spectra and number and wavelengths of filters used only changes the resulting temperature and size evolution in a small systematic way. The qualitative behaviour does not change. Here we briefly discuss the extent to which these choices influence the quantitative results, by using burst B5 of NSW\,284 as an example. The same applies to all other bursts.

We first test the influence of the choice of the stellar atmosphere model. The detailed results are shown in Fig.\,\ref{fit_test_models} in the Appendix. All models show the same goodness of fit (RMS). The blackbody models typically result in a luminosity that is up to 50 percent higher compared to the PHOENIX or ATLAS models. The latter two are typically within 10 percent of each other, with the ATLAS models predicting slightly higher luminosities. Similarly, the temperature and effective surface area of the additional emission is highest when using the blackbody models, while the two sets of model atmospheres are in agreement within the uncertainties.

\begin{figure*}
\centering
\includegraphics[angle=0,width=0.68\columnwidth]{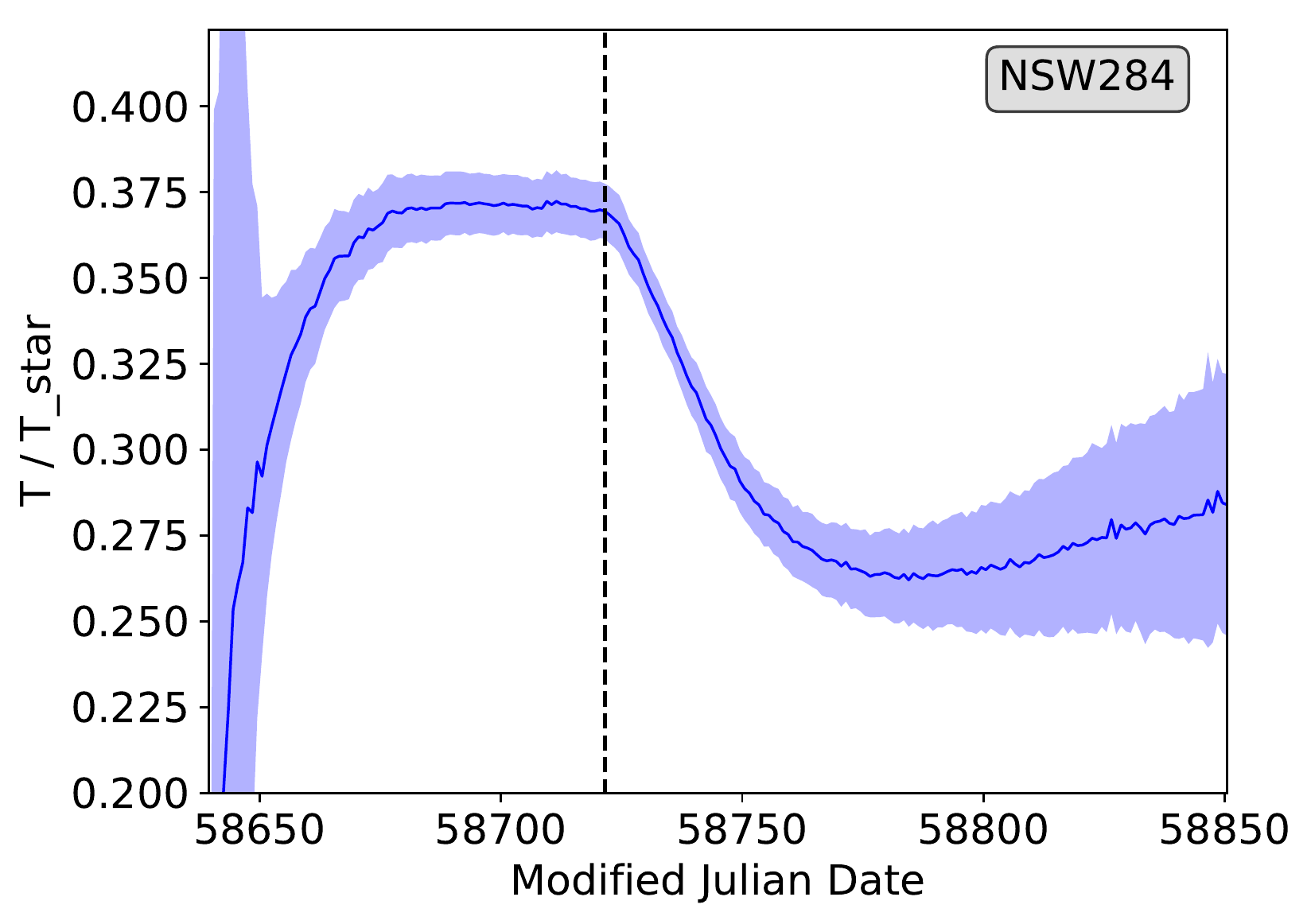} \hfill
\includegraphics[angle=0,width=0.66\columnwidth]{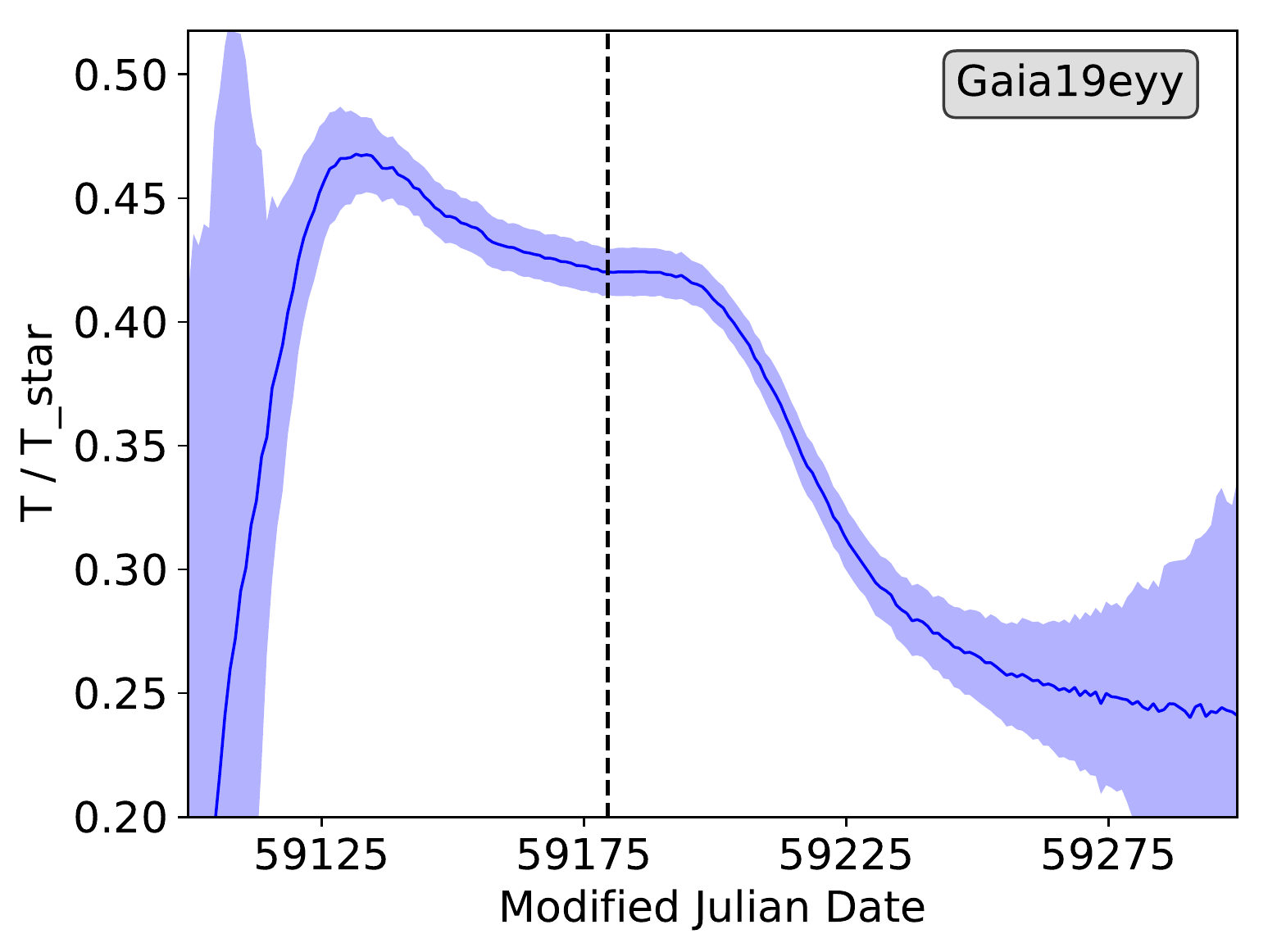} \hfill
\includegraphics[angle=0,width=0.66\columnwidth]{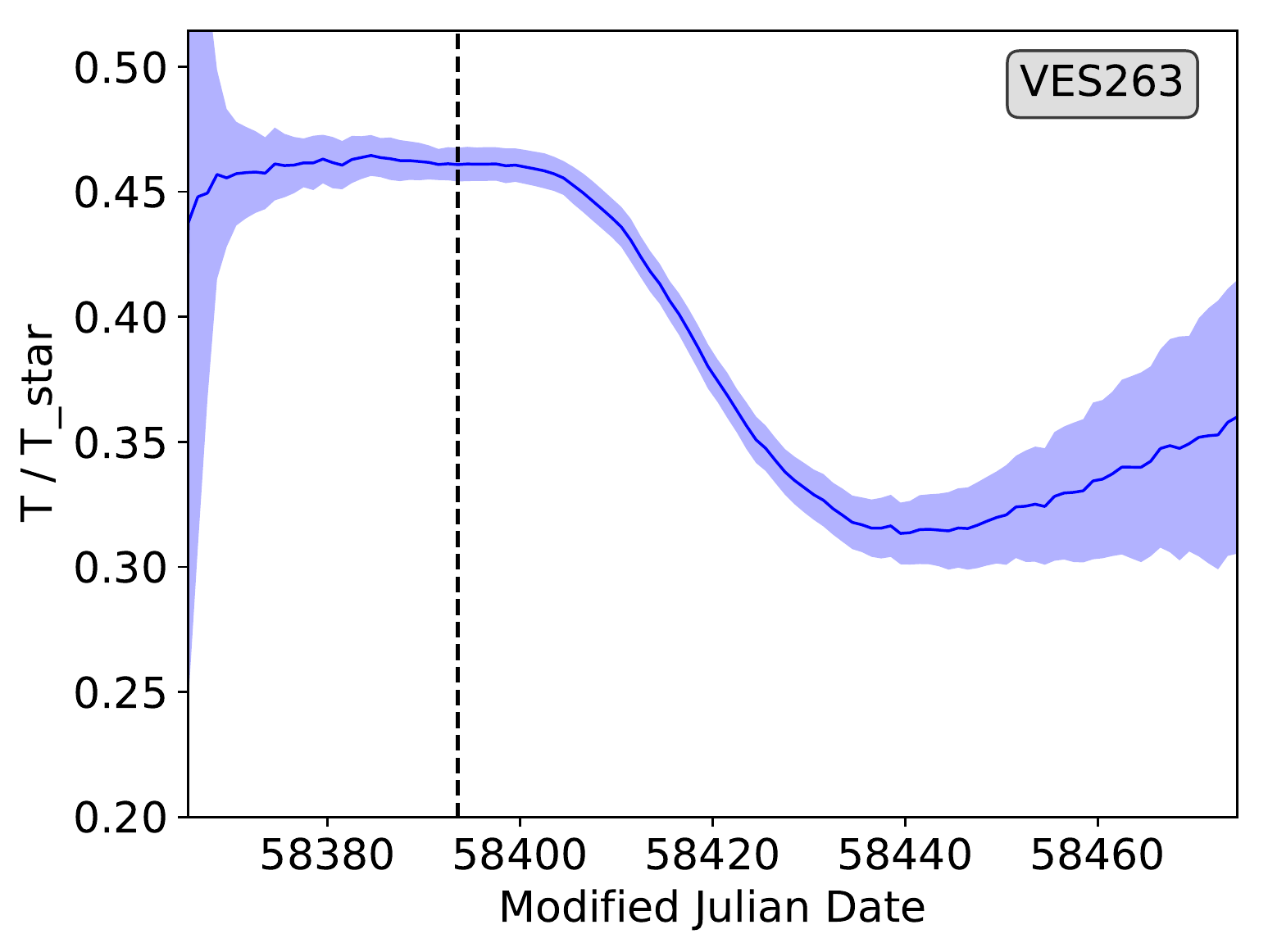} \\
\includegraphics[angle=0,width=0.68\columnwidth]{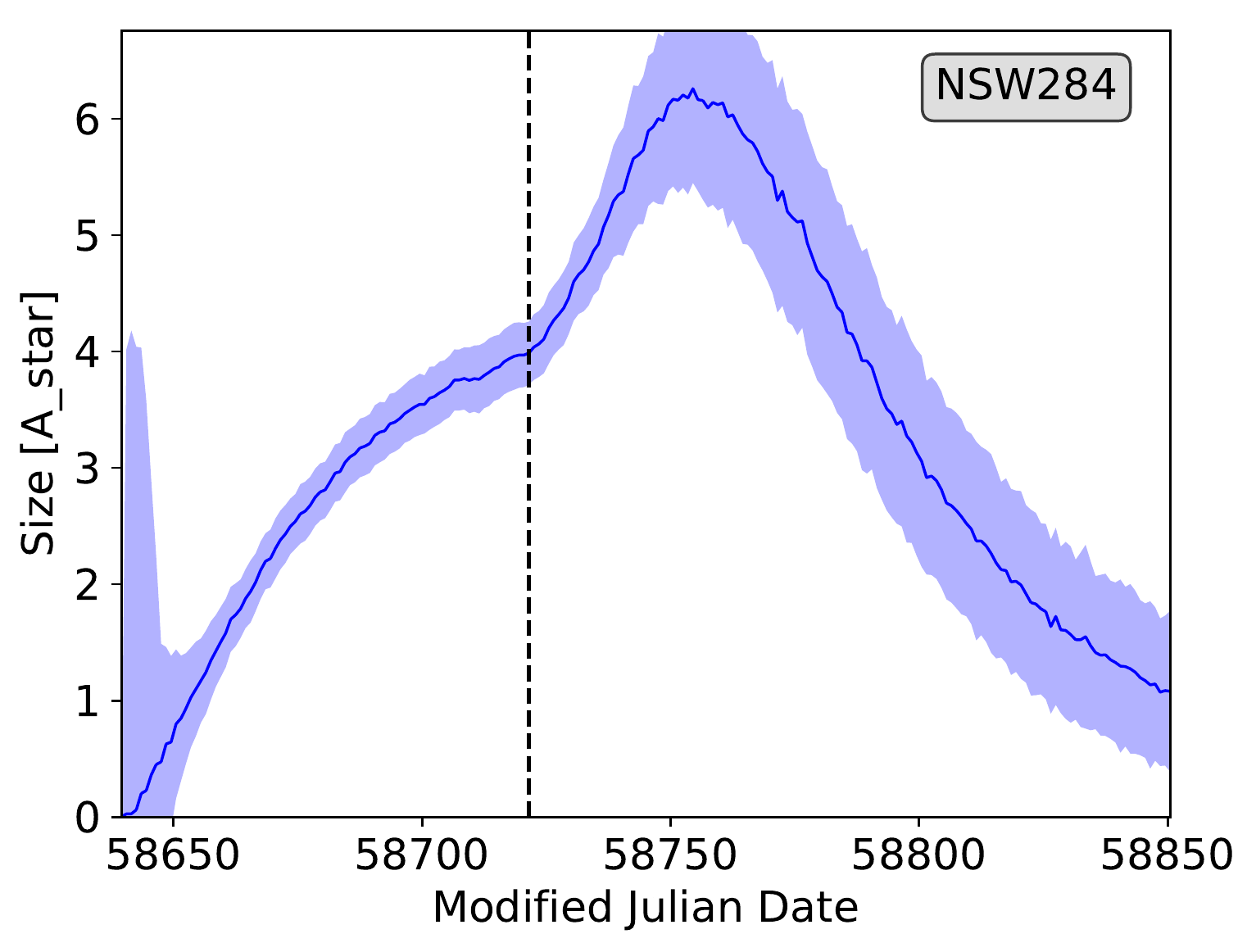} \hfill
\includegraphics[angle=0,width=0.66\columnwidth]{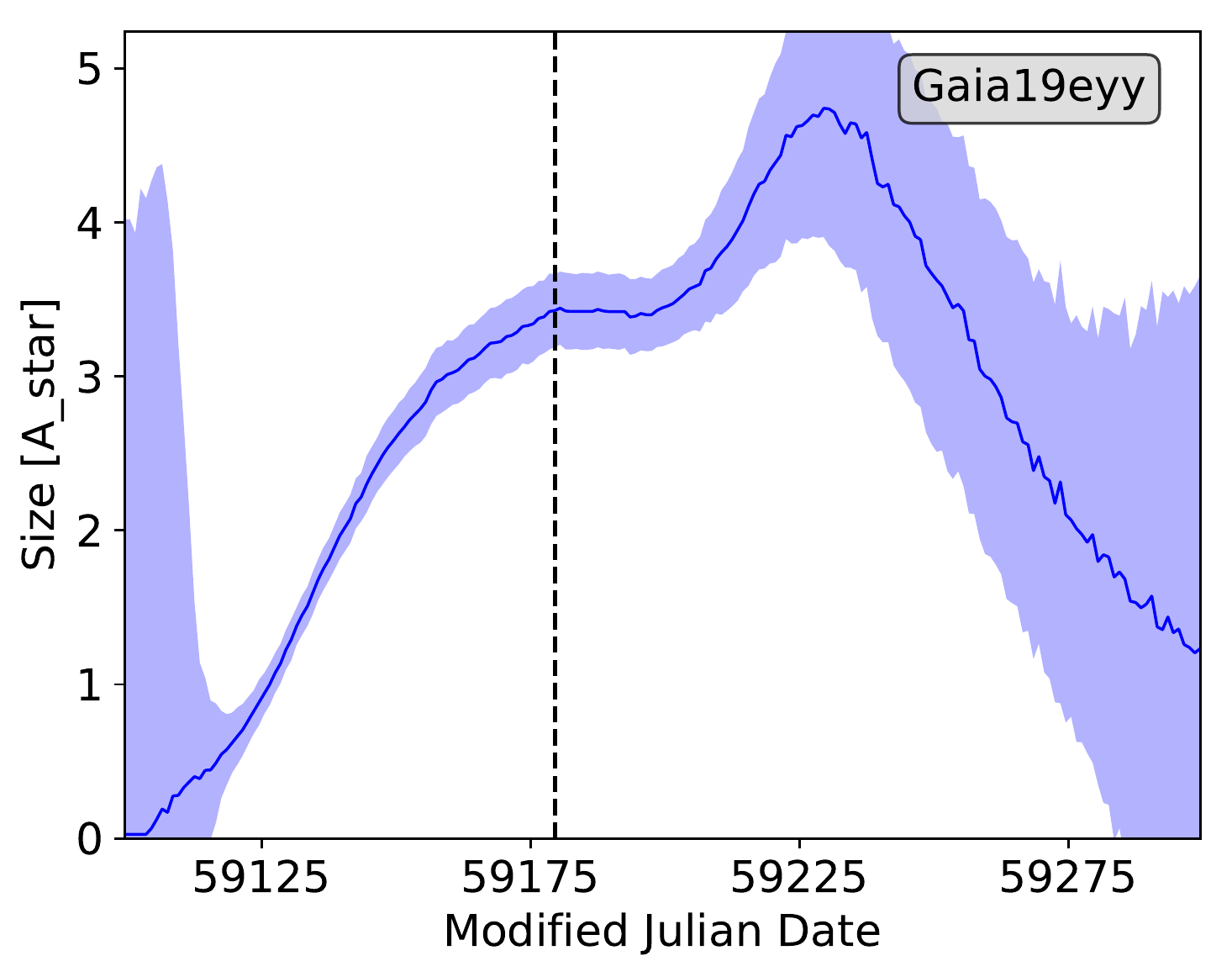} \hfill
\includegraphics[angle=0,width=0.66\columnwidth]{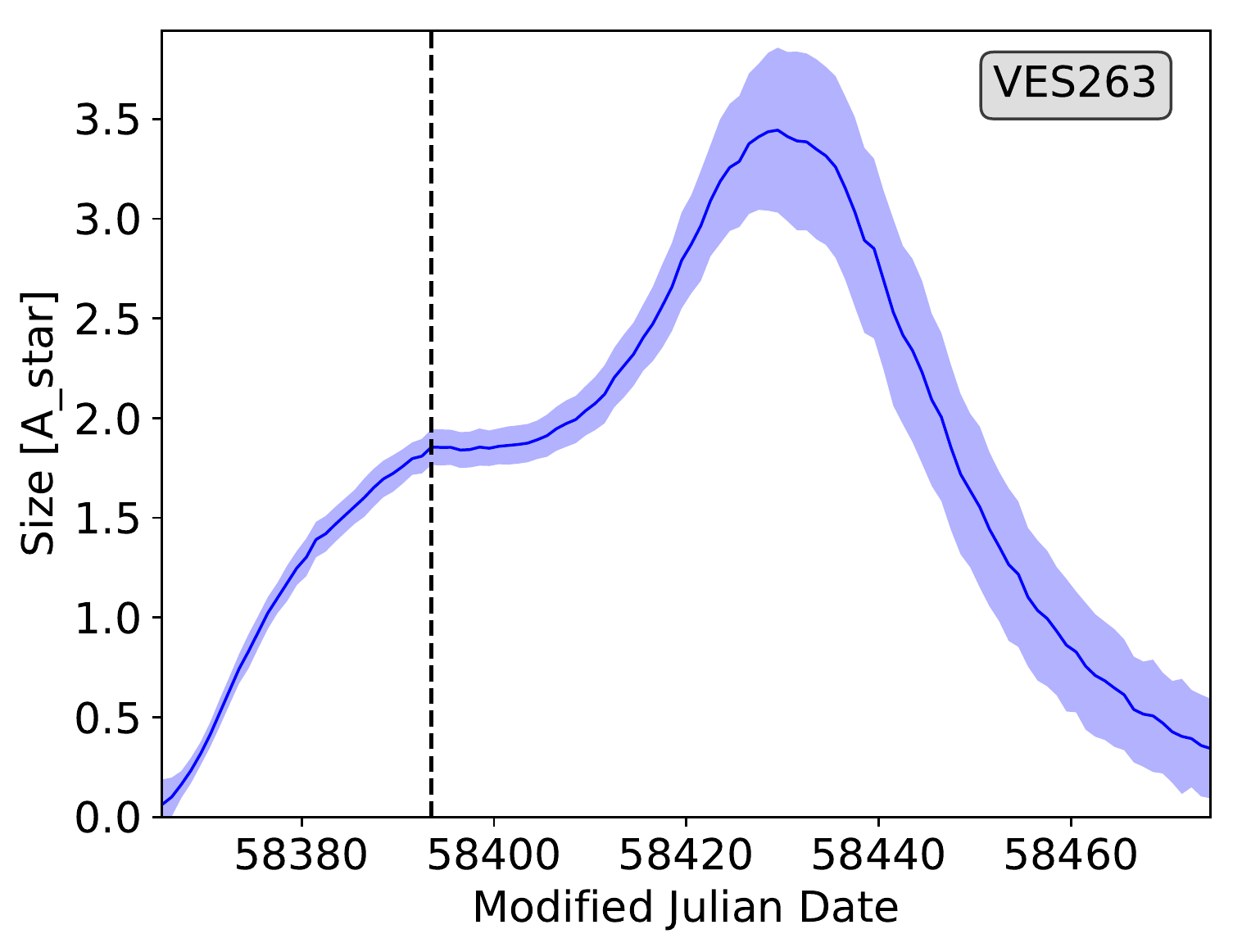} \\
\includegraphics[angle=0,width=0.68\columnwidth]{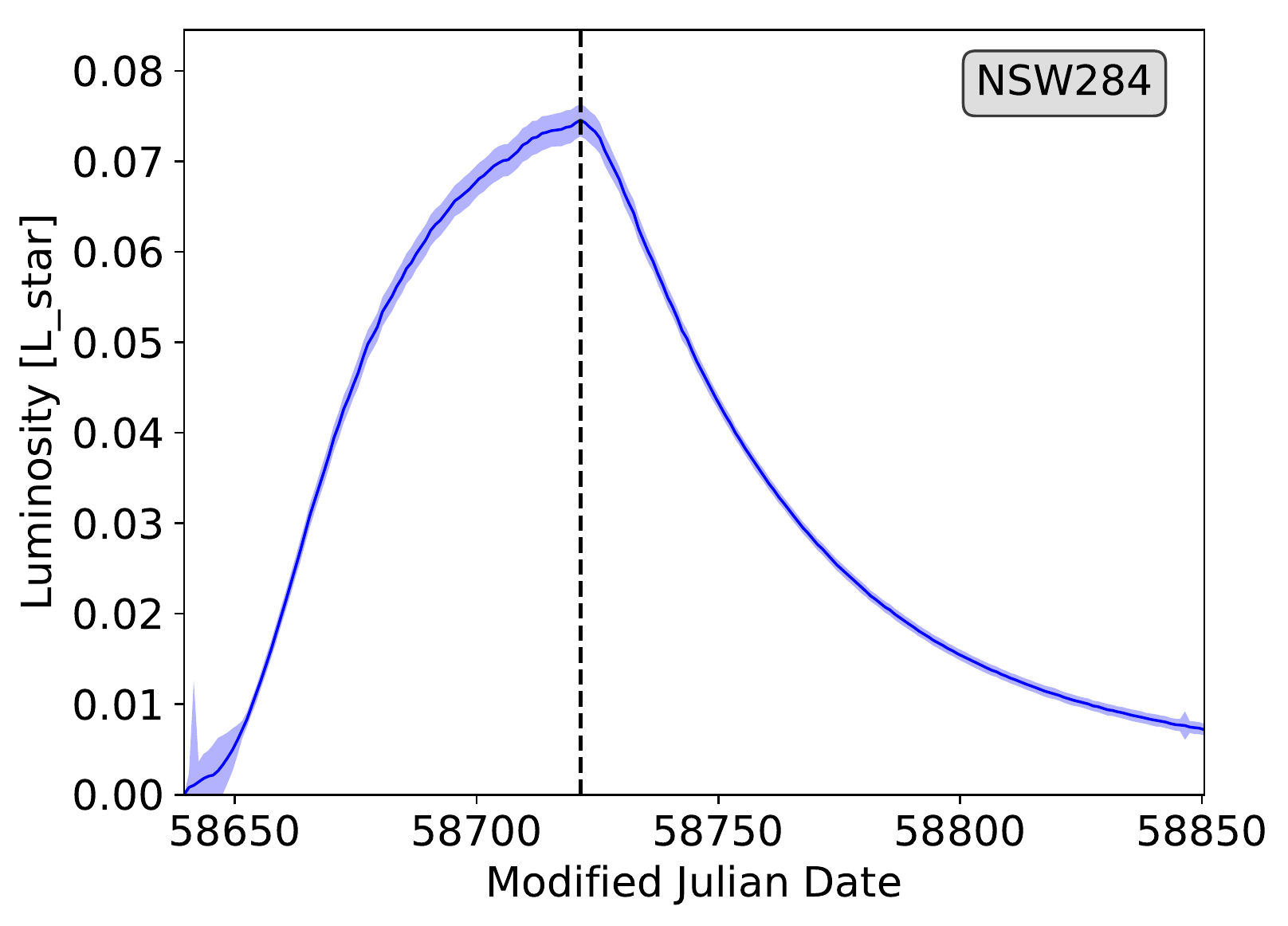} \hfill
\includegraphics[angle=0,width=0.66\columnwidth]{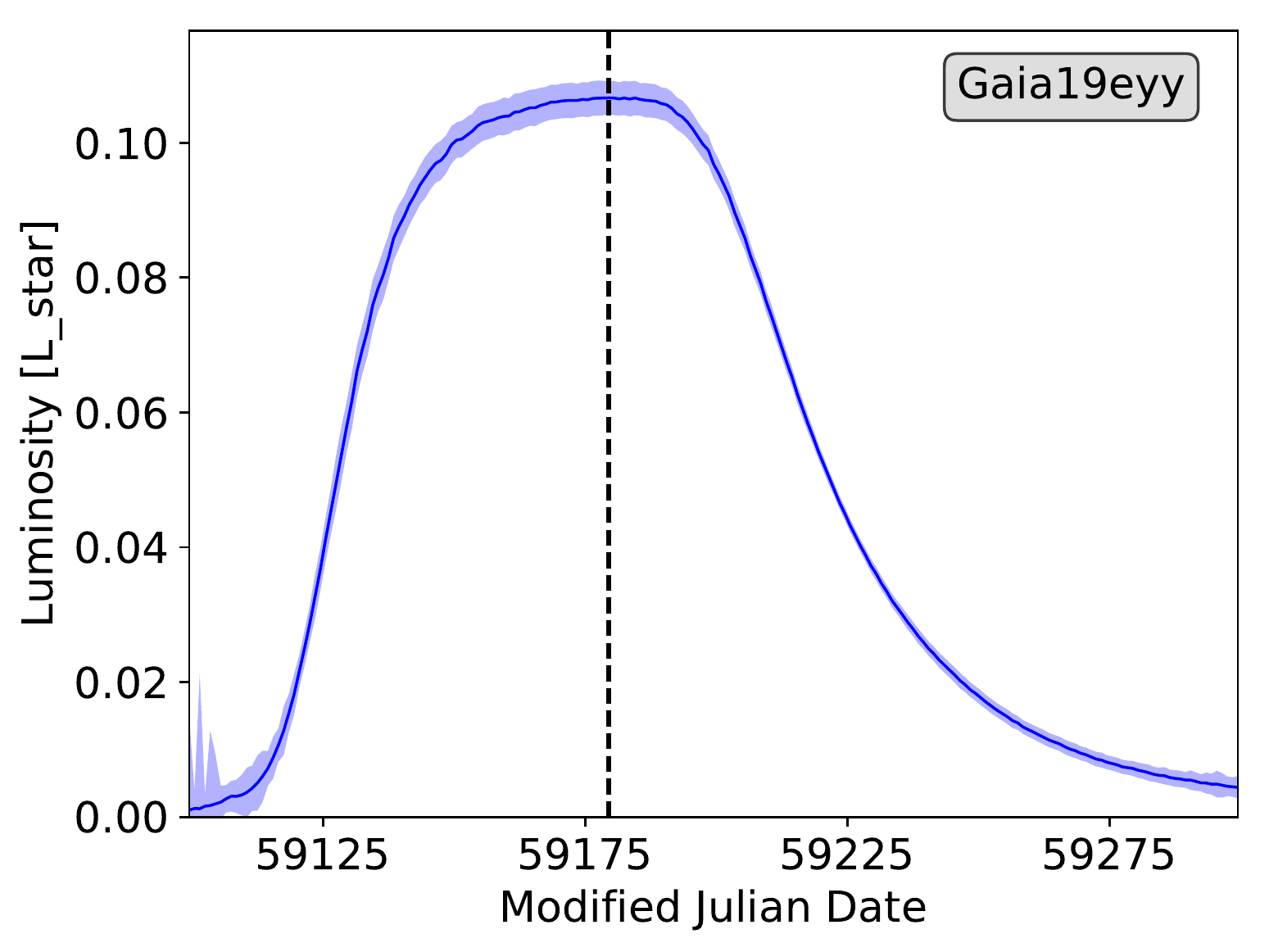} \hfill
\includegraphics[angle=0,width=0.66\columnwidth]{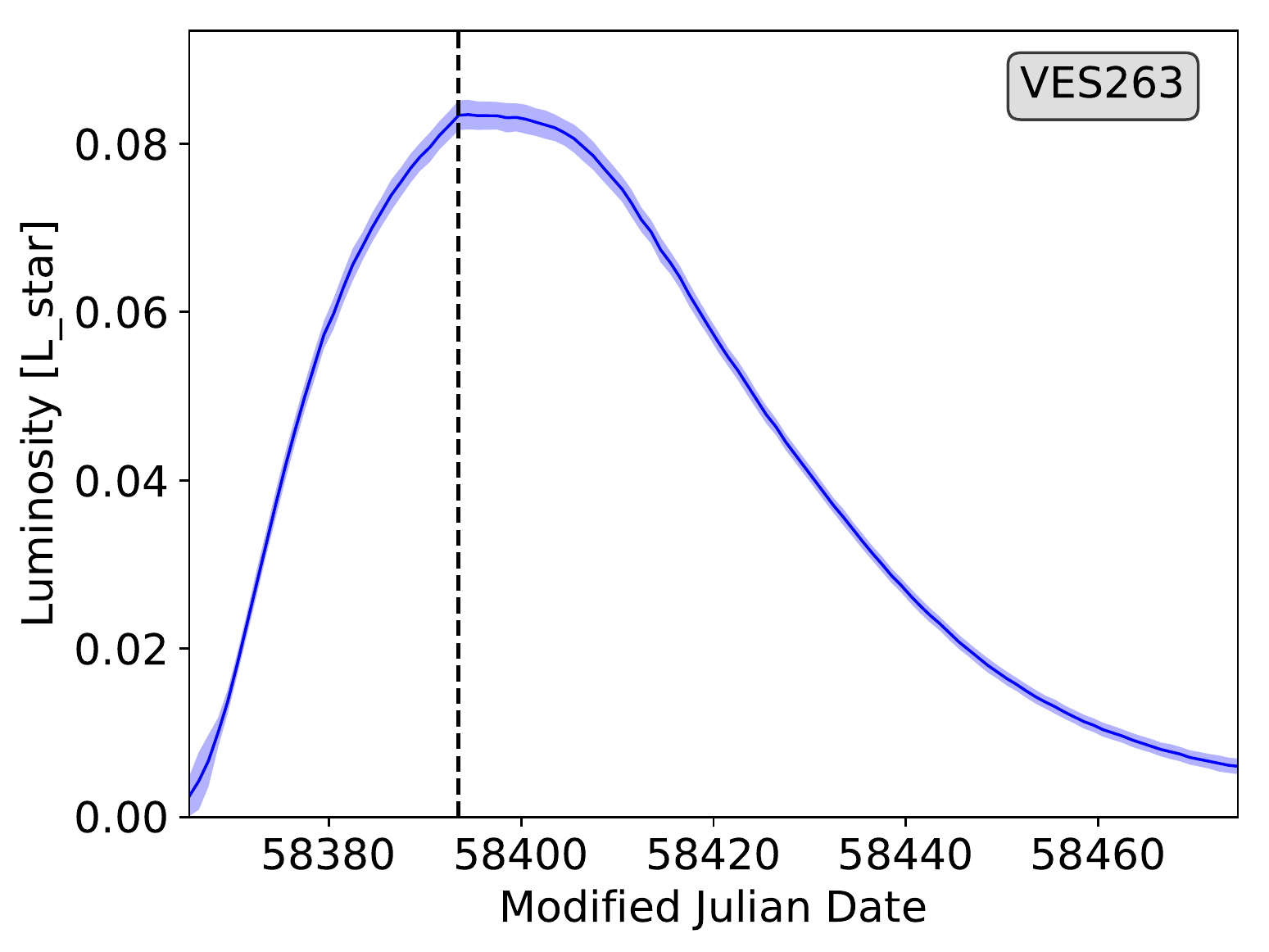} \\
\includegraphics[angle=0,width=0.68\columnwidth]{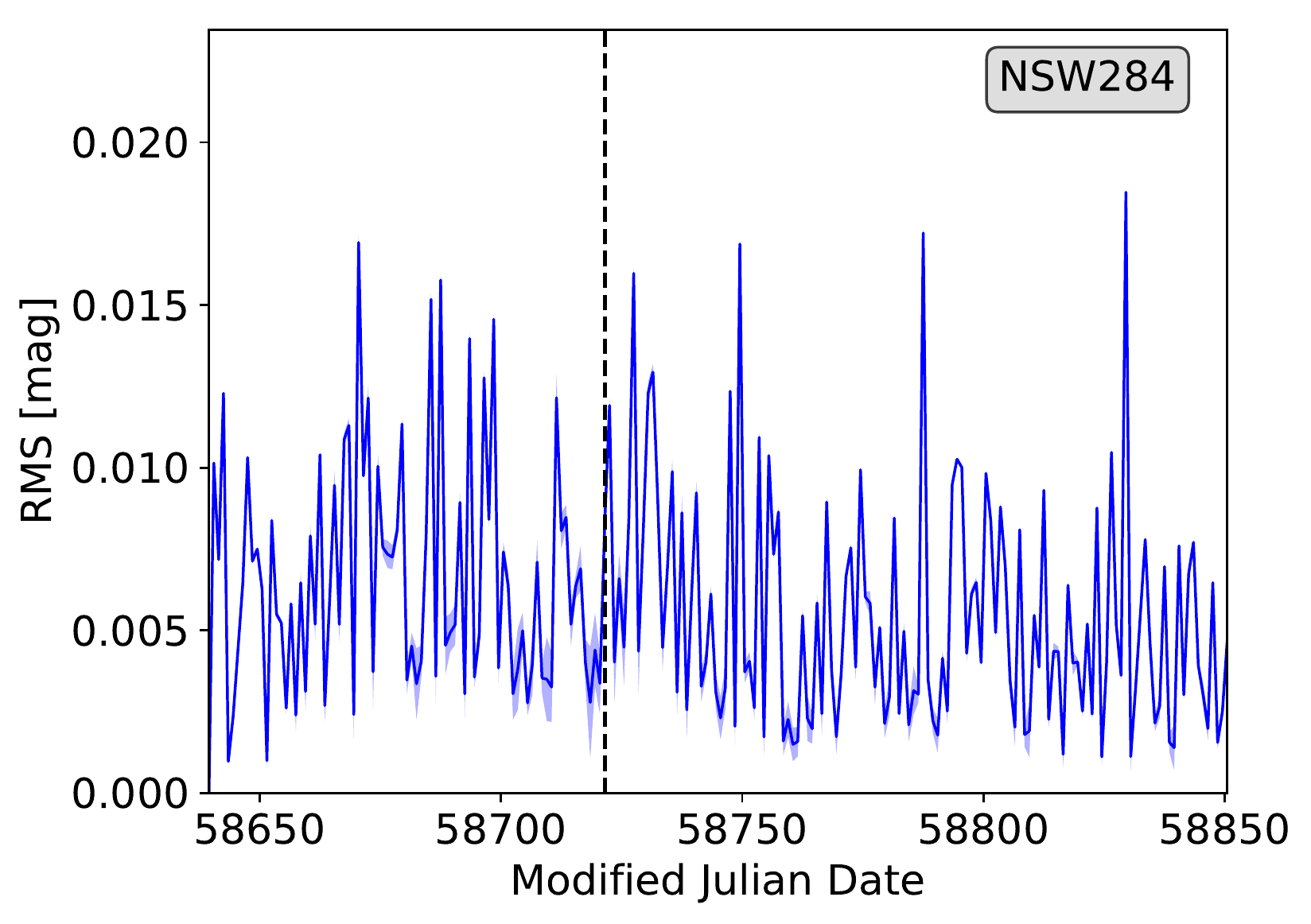} \hfill
\includegraphics[angle=0,width=0.66\columnwidth]{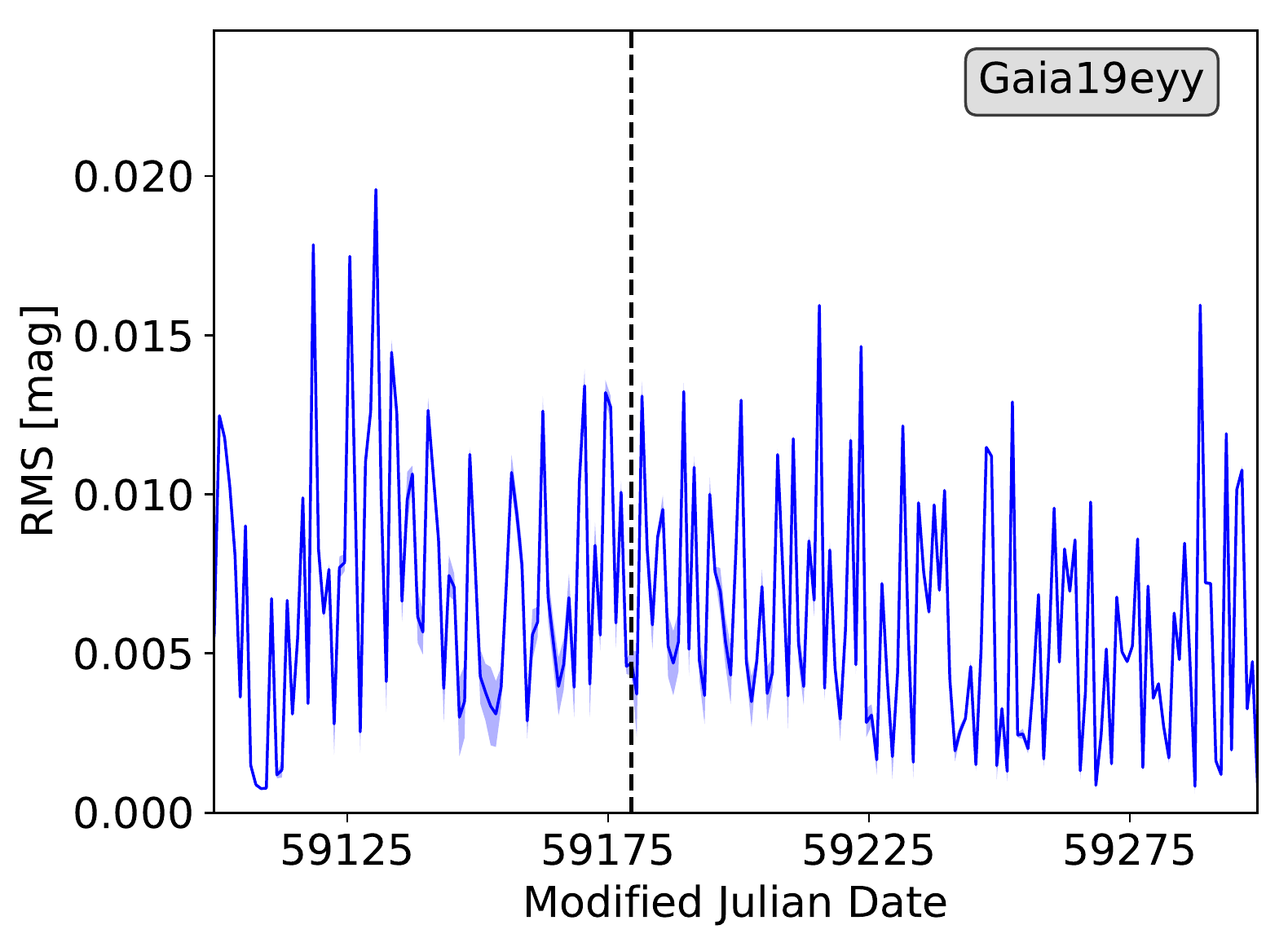} \hfill
\includegraphics[angle=0,width=0.66\columnwidth]{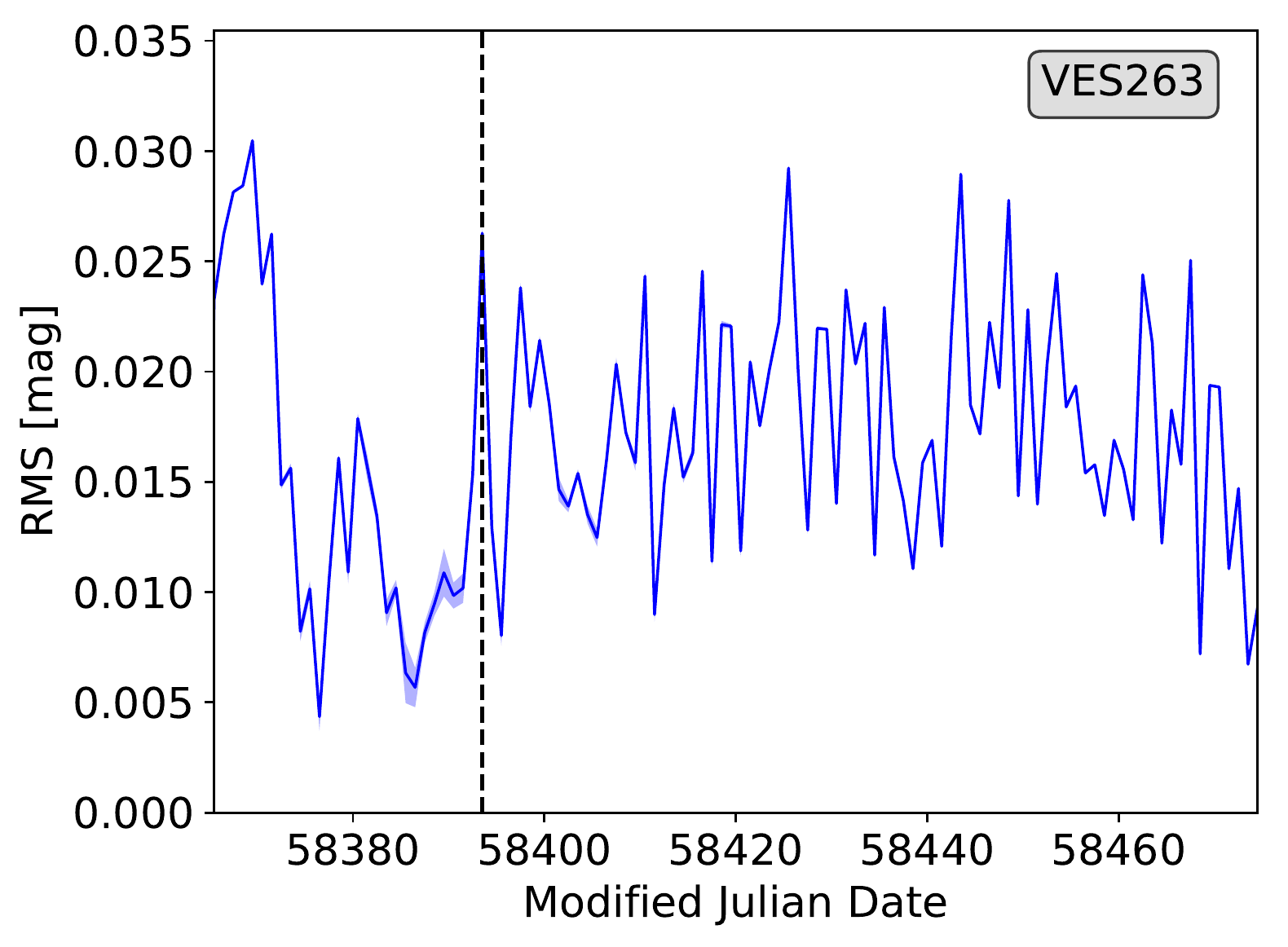} \\
\caption{Example plots for the burst modelling from fitted data with template spectra (left: NSW\,284, Burst B5; middle: Gaia\,19eyy, Burst B5; right: VES\,263, Burst B8). We show the temperature (1st row) , size (2nd row), resulting luminosity ($L = A \cdot T^4$; 3rd row) of the additional emitting area in units of the central star and the RMS (4th row) of the fit to the light curve. The central star has been assumed to have an effective temperature of 17500\,K for NSW\,284 and Gaia\,19eyy and 20500\,K for VES\,263. We show the fit results using the $VRI$ filters for NSW\,284 and Gaia\,19eyy and $BVRI$ for the hotter object VES\,263, and using the PHOENIX models. The dashed line indicates $t_2$. The values for all fit parameters for each burst in all sources are listed in Table\,\ref{burst_table} in the Appendix. \label{fit_burst}} 
\end{figure*}

When one includes the $B$-band amplitudes (which are not generally available for all bursts), there are some small systematic differences in the determined burst properties. This is shown in Fig.\,\ref{fit_test_filters} in the Appendix. The burst temperature is slightly higher when including the $B$-amplitudes. This is of the same order as the increase seen when using the blackbody models instead of the other two, i.e. about 10 percent. The size and luminosity also vary by this amount. The most important difference when including the $B$-amplitudes into the fit is that the RMS typically increases. In our example burst it is by almost a factor of three, from 0.007\,mag to about 0.02\,mag - which is still of the order of the photometry uncertainty of the data. This is most likely caused by the simplicity of our model, i.e. the assumption of a single temperature disk and all emission being thermal. The disk temperature will certainly change with distance from the star. And at the typical disk temperatures we derive, the $B$-filter covers roughly the peak of the SED. This is also notable in the individual fits shown in Fig.\,\ref{fit_burst}. There, one can see that the RMS can be higher at the start of the burst, during the mass loading phase compared to the decline in brightness at the end.

As a further test, we investigate how the choice of stellar temperature in the fits influences the results. This is summarised in Fig.\,\ref{fit_test_temps} in the Appendix. We varied the stellar temperature in steps of 1000\,K between 15500\,K and 20500\,K. The first thing to note is that the quality of the fit (RMS) is completely independent of the choice of stellar temperature. Furthermore, the general qualitative behaviour of all fit parameters is not changed at all, as for all the other tests. The inferred luminosity increases by approximately ten percent with each 1000\,K stellar temperature decrease. Similarly, the temperature of the emission decreases by the same amount. The effective size of the emitting region changes by even smaller amounts and is largest for the higher stellar temperatures. This shows that having the spectral type wrong by one sub-type (about 1000\,K, more for very early B0/B1 stars), does not change the qualitative behaviour and only marginally influences the quantitative behaviour. This is especially important for Gaia\,19eyy, where we have no spectra available. We note that our tested temperature range does not extend all the way to the potential value for NSW\,284 from Gaia\,DR3 (see the discussion in Sect.\,\ref{nsw_temp}). However, the qualitative results for the disk properties are unaffected and the quantitative values can be scaled to any stellar temperature chosen as discussed.

\begin{figure}
\centering
\includegraphics[angle=0,width=\columnwidth]{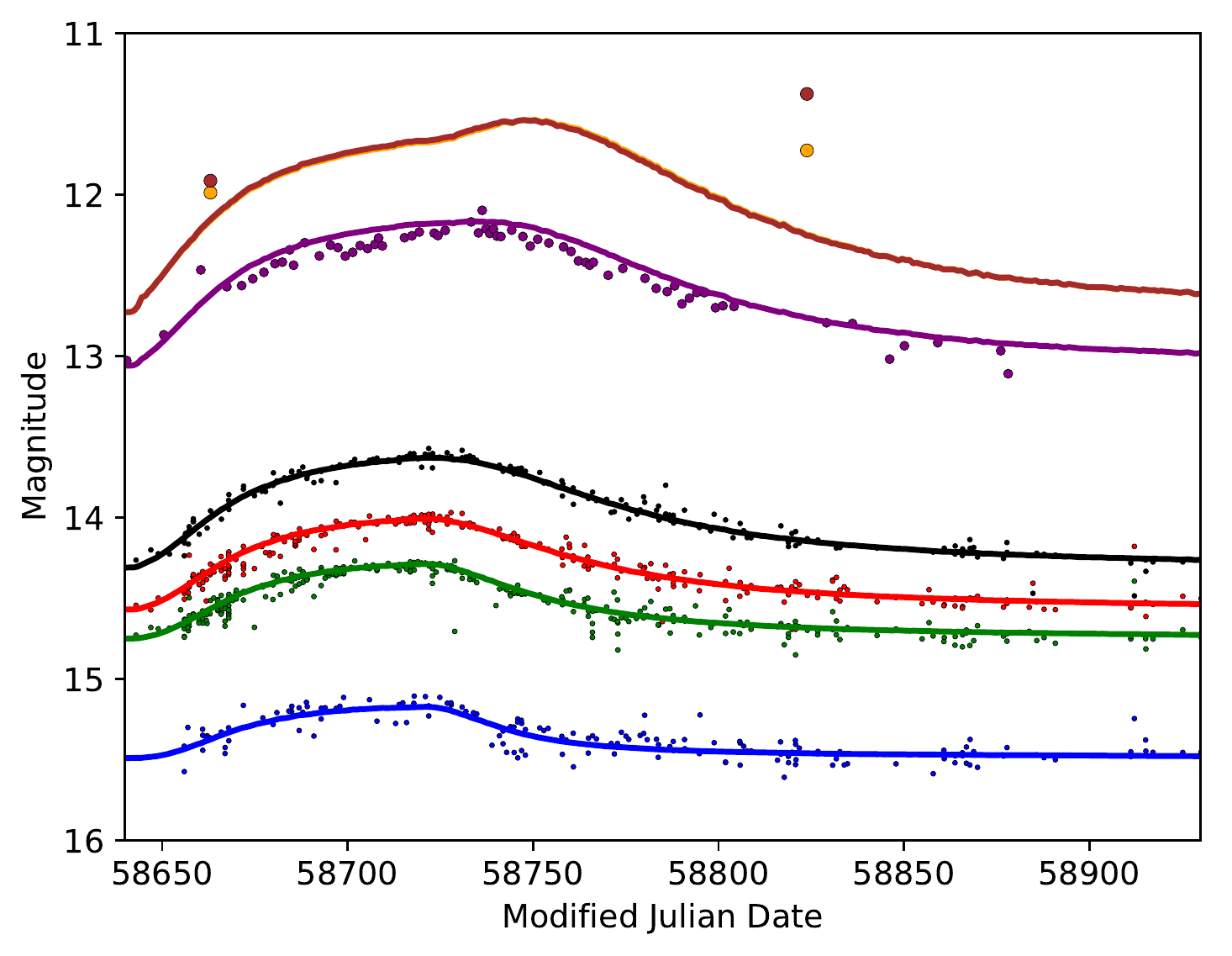} \\
\caption{Data for the burst B5 of NSW\,284 for the optical $B, V, R, I, J$ and $W1/W2$ filters (bottom to top) using the same colour codes as in Fig.\,\ref{hoys_lc}. We over plot as solid lines the predicted amplitudes in all filters based on the fit discussed in Sect.\,\ref{burst_temp} which was derived using only the observed $V, R,$ and $I$-band data (green, red, black). Nevertheless, the $B$-band and $J$-band photometry are well-matched.  The 3.6 and 4.5\,$\mu$m WISE observations are under-predicted, however, suggesting additional emission perhaps from a free-free component of the warm gas.
\label{fit_burst_ir}}
\end{figure}

As can be seen in Fig.\,\ref{hoys_lc}, for most bursts we have optical $V, R,$ and $I$-band data. The additional $B$ and $J$-band data is available only for the later bursts, and also not for all sources. Naturally the mid-IR WISE data are available for very few dates. We have hence tested how well the models that are fit to the optical data, predict the infrared data. We show one example of this in Fig.\,\ref{fit_burst_ir}. The figure shows the observational data for burst B5 in NSW\,284 in the same colour code as in Fig.\,\ref{hoys_lc}. We use the $V, R,$ and $I$-band data to model the temperature and area of the additional emission. These values are then used to predict the amplitudes and light curve in all bands from $B$ to the WISE filters. These predicted light curves are over plotted in Fig.\,\ref{fit_burst_ir} as solid lines. 

As one would expect, all optical magnitudes, incl. the $B$-band which was not used in the fit, are in very good agreement with the data. The shape of the $J$-band light curve is qualitatively in good agreement with the data. However, the measured brightness values are systematically 10\,--\,20\,\% lower than the model predictions. We used the 2MASS $J$-band filter curve for the model, which is very close to the Gattini-IR filter\footnote{Roger Smith; private communication}. Thus, at most only a very small fraction of the difference can be attributed to a difference in filter transmission curve. Furthermore, the WISE data points are much brighter than the predictions from the model, by up to half a magnitude. These differences, in particular the under-prediction of the WISE data points, indicate that our simple assumption of thermal, optically thick emission from the disk to explain the bursts is not entirely correct. At the longer wavelength, additional flux from other mechanisms is significantly contributing to the bursts. This is most likely caused by free-free or synchrotron emission from the warm disk material. A set of spectra covering the optical to mid-IR wavelength range over the entire burst duration will be able to verify the nature of the emission, in combination with the above discussed models including optically thin emission from a tenuous disk \citep{2015MNRAS.454.2107V, 2016MNRAS.455.2607V, 2012ApJ...756..156H}. Our sources are ideal for such an investigation, as they show bursts at predictable intervals.

\section{Discussion}\label{discussion}

\subsection{Summary of Bursting Behavior}

The three repeating Be star bursters investigated here share a number of commonalities. They all show semi-regular bursts that recur once or twice per year and last for months. The typical time gaps between consecutive bursts vary for the same source by about one quarter to one third of the average gap but vary from source to source. There are indications that the gaps between bursts slightly decrease over time, but only in VES\,263 do we find a significant change in behaviour. In this object also the amplitudes of the bursts change. This suggests a change from a (photometrically) disk-less inter-burst phase to a long-term brightness increase, indicating a slowly growing disk in size and density. The absence of any clear periodicity excludes any burst trigger mechanism caused by binary companions. Furthermore, with the exception of VES\,263, there is no discernible pattern to the amplitudes and duration of the bursts, but in all cases the rise time $t_2-t_1$ is faster than the brightness decline back to the quiescence level after $t_2$. Most consistently the bursts have amplitudes that increase with wavelength. The spectra of our sources show double-peaked symmetric emission that increases in strength during the photometric bursts.

\subsection{Evolution of Individual Bursts}

When fitting the burst photometry with our simplified model, all individual bursts follow a similar pattern (if the photometric signal-to-noise is large enough). We refer the reader to Fig.~\ref{fit_burst} in what follows.  The size of the emitting region in the disk increases from $t_1$ to $t_2$ while the temperatures remain roughly constant. There are some cases where the temperatures start high, but typically the SNR is too low for this to be significant. Temperatures of the disk are of the order of 0.4 times the stellar temperature, when using the adopted values for $T_S$ in Table\,\ref{TheStars}. This of course changes slightly if one uses a different stellar temperature, atmosphere model, or includes amplitudes measured at different wavelengths. However, for the most likely correct stellar temperatures, our value for the disk temperature during the burst is consistently below the 0.6 times the stellar temperature used in R18.

After the end of the mass loading at $t_2$, the disk temperature starts dropping in all cases. Similarly, the luminosity also peaks at this point. The disk temperature drop can be followed until the photometry SNR gets too low. The indications are that towards the end of the brightness decline, the disk temperature levels off at about a quarter to one third of the stellar temperature. In all cases the emitting area increases from zero at $t_1$ to larger values at $t_2$. The values reached depend on the burst amplitude. After the time $t_2$ the best model solutions in all cases show a further increase of the emitting area. The peak is then reached at roughly the point where the decreasing disk temperature starts to level off. Following that, the emitting area shrinks until the end of the burst to effectively zero in all cases. If our model is correct, this would indicate that after mass loading stops, the main emitting area of the disk gets pushed away from the star. This expands the emitting area and cools the material, before the radial thickness of the emitting area in the disk shrinks. There are some simulations of the radial disk density profile during bursts roughly similar to our scenario \citep[e.g. top panel of Fig.\,7 in][]{2012ApJ...756..156H}. However, they do not show the evolution after the mass loading stops and we are thus not able to compare them directly to our observations.

The rate of increase of the burst luminosity at the point $t_2$ shows that the end of the mass loading occurs at different parts of the burst. In the majority of the cases the luminosity increase just prior to $t_2$ is still very steep or has just started to level off. In some cases, most notable for burst B5 in Gaia\,19eyy, the luminosity has plateaued before $t_2$ is reached. This shows, that whatever mechanism stops the mass loading process can occur at any time after $t_1$. Since only very few of the bursts reach a stable, steady-state luminosity and the gap between bursts is much longer than the rise time, the trigger mechanism to stop the mass loading must occur much more frequently than the trigger to start the mass loading process. This is further emphasised by the occasional very weak burst, e.g. in NSW\,284, where the mass loading most likely stops immediately after it has been triggered.

The fit to the amplitudes using Eqs.\,\ref{dm_lesst2} and \ref{dm_moret2} results in all cases in an RMS that is smaller than the typical photometric uncertainties of the individual brightness measurements\footnote{There are some minor exceptions such as the start of the burst B5 in Gaia\,19eyy.}. This is strong evidence that the theoretical framework in R18 correctly predicts the shape of the bursts. The accuracy by which the model fits the optical and even the unconstrained near infrared amplitudes (see Fig.\,\ref{fit_burst_ir}), further justifies this approach. Furthermore, the high optical depth inferred from our NIR spectrum of NSW\,284 (Sect.\,\ref{triplespec}) also supports our attempt at modelling the burst continuum emission with stellar atmosphere or black body models. 

The comparison of the spectra taken in quiescence  and burst show that emission does not just increase in brightness. The peak emission also moves to lower velocities, i.e. happens further out in the disk. This fits well with the decretion disk model. Our photometry modelling further supports this picture, indicating cooling, expansion and probably radial outward drift of the main emitting part of the disk after $t_2$. Using the measured velocities and modeled emitting surface area, one finds that the main emission is confined to a narrow ring with a radial thickness of the order of 0.5 to one solar radius.

\subsection{Uncertain Physical Parameters}

Despite the apparent successes in matching a heuristic model to optical lightcurve data, it is clear that there are limitations to our simplified model. 

First, it is not possible to fit accurately all the parameters in these equations at the same time. Thus, it is not possible to investigate the viscosity of the disks without additional assumptions or simulations as in R18, \citet{Haubois2012} or \citet{2012ApJ...744L..15C}.

Second, the model considers only a single temperature disk rather than a small Keplerian disk. We have seen from the high resolution spectra (Sect.\,\ref{spectroscopy}) that the peaks of the double-horned line profiles are at higher velocities for higher excitation transitions. Thus, there are clear indications of a radial temperature gradient in the disk during the burst, and presumably as well in quiescence. This could explain the 10\,--\,20\,\% over-prediction of the $J$-band magnitudes.

Our model furthermore does not use any real radiative transfer. It thus will not be able to handle any outbursting Be stars where we observe the disk edge-on, which usually results in brightness dips at the shorter wavelengths. For the MIR data the model under-predicts the observed amplitudes in the WISE filters by up to half a magnitude. This strong MIR excess emission compared to our optically thick emission model indicates significant contribution from free-free emission in these Be disks. This is in line with previous works showing that typical Be stars have infrared access governed by free-free emission \citep{Finkenzeller1984}.

\subsection{VES~263 in Context}

VES\,263 is the  only one of our sources that has been studied in any detail previously. \citet[][in Sect.\,8]{Munari2019} interpret their data for this object in terms of Herbig~Ae/Be star accretion, but we take the photometric and spectroscopic evidence as indicative of a decretion disk, with gas flowing outward rather than inward.

\cite{Munari2019} also model the SED of VES\,263 during quiescence and burst. They fit quiescence to a 20,000~K template, and add blackbody emission at 4500~K and 7500~K for the two different brighter phases. This can be compared to the approximately 9000\,K disk temperature inferred from our light curve fit here. As one can see from Fig.\,\ref{fit_burst}, we find that during the burst the luminosity increases by up to eight percent. This equates to roughly 1000\,L$_\odot$, which is in line with the sum of the two additional luminosities ($120+860$\,L$_\odot$) inferred by \cite{Munari2019}. Note that the \cite{Munari2019} model also includes, in the quiescent state, a dust continuum excess fit to two AKARI data points, that is a blackbody disk with maximum temperature of 400~K and luminosity of 12~L$_\odot$. This emission is most likely always present, but becomes insignificant during the bursts.

The presence of cold dust does not necessarily rule out our interpretation of the nature of this source. Dust has been shown to form in many other hot environments, such as e.g. novae \citep{2017MNRAS.469.1314D} or Wolf-Rayet binaries \citep{1991MNRAS.252...49U}. In those cases dust formation episodes can occur in dense shocks formed due to the interaction of outflowing material or winds with circumstellar material. While outflow velocities and densities need to be high for this, this mechanism remains a possible source for dust formation in VES\,263.

\subsection{The Broader Context}

The objects investigated here provide text book examples of regularly outbursting Be stars. Indeed there are very few really regularly outbursting Be stars known. A short list is presented in \citet{2017ASPC..508...93B} and a handful are listed in \citet{2022AJ....163..226L}. But these objects vary on shorter timescales (up to tens of days) between bursts and at most a few percent amplitudes. Semi-regular variations are much more common \citep[e.g.][]{2017AJ....153..252L}. However, typically these objects show a much larger variation of the times between bursts, large burst to burst amplitude variations, slightly smaller amplitudes than the objects discussed here, and the bursts do not follow the general shape from Eqs.\,\ref{dm_lesst2} and \ref{dm_moret2}. Thus, the sources discussed here, represent rare cases of truly regularly outbursting Be stars.

While we have been able to piece together light curves from several different long-term photometric surveys, the spectroscopic coverage is less than desirable. For example, we do not have time series high resolution spectra available at multiple stages of any of the bursts. This would enable us to study in detail the temperature and radial evolution of the emitting disk material during the burst. Additionally valuable observations would be $JHK$ spectroscopic monitoring throughout a burst. This would enable the modelling of the thermal and non-thermal contributions to the spectrum as the burst evolves. However, because the timing of the bursts can be predicted in advance -- in most cases within a few weeks -- future observations can be planned.   

We also note that there are other likely outbursting Be objects that are being revealed through ongoing time domain surveys, but remain poorly studied. We have identified the following potentially similar objects to those we have studied, based on their public light curves: Gaia22bre\footnote{\url{http://gsaweb.ast.cam.ac.uk/alerts/alert/Gaia22bre/}}, Gaia22axt\footnote{\url{http://gsaweb.ast.cam.ac.uk/alerts/alert/Gaia22axt/}}, ASASSN-V\,J210822.32+584613.4\footnote{\url{https://asas-sn.osu.edu/variables/587872}}, and potentially AP565575\footnote{\url{https://asas-sn.osu.edu/photometry/f5962fee-2400-55ba-b692-4946613cf59a}}. Again, spectroscopic follow-up is warranted. 

Considering the larger population of early-type stars, we note that the vast majority of OB-type stars are binaries \citep{Moe2017}. There are no clear indications in any of our spectra that the sources we have investigated are binaries. There are no general radial velocity shifts between spectra taken at different times. The disk emission profiles furthermore, rule out any close, low mass companion, which would disturb the disk. As discussed above, the non-periodic nature of the bursts excludes the presence of highly elliptical companions with small perihelion distances. We cannot rule out any wide, lower mass companions with the data available. The spectral energy distributions give no indication of a second component. The above suggested high resolution time series spectroscopy could provide more stringent constraints on the multiplicity of the sources. 

Returning to the viscous decretion disk model described in the introduction, our observations and modelling support a general picture in which Be star outbursts correspond to a rapid temperature rise in the disks over an initially small area, and a gradual increase in the total area participating in the higher temperatures.

\section{Conclusions}

We collected long term optical photometric data for the three variable sources NSW\,284, Gaia19eyy, and VES\,263. These are supplemented by NIR and NEO-WISE photometry, and high resolution spectra. We characterise these sources as rare cases of truly regularly outbursting Be stars.

The bursts in these sources occur semi-regularly with a cadence between half to one year and a duration from about 120 to 200 days. The burst amplitudes are variable and reach a maximum of about half a magnitude in the $R$-band. In all cases the amplitudes increase towards longer wavelengths, and the peak of the bursts shifts to later times at longer wavelengths.

We fit the individual burst light curves with the theoretical burst shape model from \citet{Rimulo2018}. This provides an excellent fit to the data in all cases. The typical RMS of the fit is of the order or better than half the photometric uncertainty in the data. However, despite this excellent fit, the fit parameters cannot be constrained accurately as they depend on each other. Thus, in order to study e.g. the disk viscosity during and after mass loading, the photometry needs to be supplemented by numerical simulations, as performed by \citet{Rimulo2018}.

The burst shapes in all cases show a steep increase in brightness and a longer lasting fall back. The duration of the brightness increase represents the typical duration between a (mass loading) burst trigger event and the trigger that stops the mass loading. This duration (50\,--\,80\,d) is always much shorter than separation of consecutive bursts (120\,--\,200\,d). Thus, the trigger that stops the mass loading occurs on average 3\,--\,5 times more frequent than the trigger that starts the mass loading. Thus, any physical explanation for the occurrence of these bursts and their end will have to account for this. In some cases, the bursts stop almost immediately after their start, leading to very low amplitudes or even undetected bursts, e.g. the '?' and B4 bursts in NSW\,284.

We model the smoothed burst light curves with a simple model that attempts to reproduce the amplitude increase in the different filters over time with a simple single temperature emitter surrounding the host star. The host star and additional emitter are modeled with different template atmospheric spectra or black body emission. The models predict the size (relative to the host star) and temperature of the additional emitter during the burst. The resulting values for those are only marginally (quantitatively) influenced by the exact value of the stellar surface temperature chosen - which is not known exactly. The qualitative behavior of size and temperature evolution of the additional emitter during the burst is always well constrained. In particular we find that the surface temperature of the additional emitter is of the order of 40\,\% of the stellar temperature. This is lower than the 60\,\% assumed in the work by \citet{Rimulo2018}. Our simple model under-predicts the observed amplitudes in the WISE filters by up to half a magnitude, indicating a significant contribution from free-free emission in these Be disks.

All three sources investigated in this work show very regular and predictable burst behaviour, albeit with unpredictable variations in amplitude, over the better part of a decade. They thus present excellent candidates for follow up high cadence spectroscopy and photometry of their future bursts. This would allow more detailed models of the outbursting Be phenomenon to be developed to identify in particular the triggering mechanism for the start and the stop of the disk mass loading process in these sources.

A notable aspect of our sample is that two of the three objects studied here have been previously discussed in the literature as young Herbig~Ae/Be stars, rather than as evolved Be stars. It is thus worth emphasizing the value of long term light curves, such as those produced by the HOYS survey and other long-term photometric surveys such as ASAS-SN and PTF/ZTF.

\section*{Acknowledgements}

Our referee is acknowledged for a helpful report.
We would like to thank all contributors of optical photometric data for their efforts towards the success of the HOYS project.
We also thank the entire Gattini-IR team for access to infrared photometry of our sources.
We acknowledge Tony Rodriguez for assistance in acquiring and reducing the Palomar/DoubleSpec data for NSW\,284 that is reported here, and Adolfo Carvalho for assistance with DIBs feature identification in the HIRES spectra. JCW was funded by the European Union (ERC, WANDA, 101039452). This work has made use of data from the European Space Agency (ESA) mission {\it Gaia} (\url{https://www.cosmos.esa.int/gaia}), processed by the {\it Gaia} Data Processing and Analysis Consortium (DPAC, \url{https://www.cosmos.esa.int/web/gaia/dpac/consortium}). Funding for the DPAC has been provided by national institutions, in particular the institutions participating in the {\it Gaia} Multilateral Agreement.
All authors from Silesian University of Technology were responsible for data processing and automation of observations at SUTO observatories and were financed by grant BK-246/RAu-11/2022. Piotr Jóźwik-Wabik also acknowledges support from grant BKM-574/RAu-11/2022 and 32/014/SDU/10-22-20.

\section*{Data Availability Statement}

Some of the photometry data underlying this article are available in the HOYS database at http://astro.kent.ac.uk/HOYS-CAPS/.
Some of the spectroscopic data is available in the Keck Observatory Archive (KOA).


\bibliographystyle{mnras}
\bibliography{bibliography} 

\begin{thebibliography}{}
\makeatletter
\relax
\def\mn@urlcharsother{\let\do\@makeother \do\$\do\&\do\#\do\^\do\_\do\%\do\~}
\def\mn@doi{\begingroup\mn@urlcharsother \@ifnextchar [ {\mn@doi@}
  {\mn@doi@[]}}
\def\mn@doi@[#1]#2{\def\@tempa{#1}\ifx\@tempa\@empty \href
  {http://dx.doi.org/#2} {doi:#2}\else \href {http://dx.doi.org/#2} {#1}\fi
  \endgroup}
\def\mn@eprint#1#2{\mn@eprint@#1:#2::\@nil}
\def\mn@eprint@arXiv#1{\href {http://arxiv.org/abs/#1} {{\tt arXiv:#1}}}
\def\mn@eprint@dblp#1{\href {http://dblp.uni-trier.de/rec/bibtex/#1.xml}
  {dblp:#1}}
\def\mn@eprint@#1:#2:#3:#4\@nil{\def\@tempa {#1}\def\@tempb {#2}\def\@tempc
  {#3}\ifx \@tempc \@empty \let \@tempc \@tempb \let \@tempb \@tempa \fi \ifx
  \@tempb \@empty \def\@tempb {arXiv}\fi \@ifundefined
  {mn@eprint@\@tempb}{\@tempb:\@tempc}{\expandafter \expandafter \csname
  mn@eprint@\@tempb\endcsname \expandafter{\@tempc}}}

\bibitem[\protect\citeauthoryear{{Baade}, {Rivinius}, {Pigulski}, {Carciofi}
  \& {BEST Collaboration}}{{Baade} et~al.}{2017}]{2017ASPC..508...93B}
{Baade} D.,  {Rivinius} T.,  {Pigulski} A.,  {Carciofi} A.,   {BEST
  Collaboration} 2017, in {Miroshnichenko} A.,  {Zharikov} S.,
  {Kor{\v{c}}{\'a}kov{\'a}} D.,   {Wolf} M.,  eds,  Astronomical Society of the
  Pacific Conference Series Vol. 508, The B[e] Phenomenon: Forty Years of
  Studies. p.~93 (\mn@eprint {arXiv} {1610.02200})

\bibitem[\protect\citeauthoryear{{Bellm} et~al.,}{{Bellm}
  et~al.}{2019}]{ztf_overview}
{Bellm} E.~C.,  et~al., 2019, \mn@doi [\pasp] {10.1088/1538-3873/aaecbe}, \href
  {https://ui.adsabs.harvard.edu/abs/2019PASP..131a8002B} {131, 018002}

\bibitem[\protect\citeauthoryear{{Berlanas}, {Wright}, {Herrero}, {Drew}  \&
  {Lennon}}{{Berlanas} et~al.}{2019}]{Berlanas2019}
{Berlanas} S.~R.,  {Wright} N.~J.,  {Herrero} A.,  {Drew} J.~E.,   {Lennon}
  D.~J.,  2019, \mn@doi [\mnras] {10.1093/mnras/stz117}, \href
  {https://ui.adsabs.harvard.edu/abs/2019MNRAS.484.1838B} {484, 1838}

\bibitem[\protect\citeauthoryear{{Bernhard}, {Otero}, {H{\"u}mmerich},
  {Kaltcheva}, {Paunzen}  \& {Bohlsen}}{{Bernhard} et~al.}{2018}]{Bernhard2018}
{Bernhard} K.,  {Otero} S.,  {H{\"u}mmerich} S.,  {Kaltcheva} N.,  {Paunzen}
  E.,   {Bohlsen} T.,  2018, \mn@doi [\mnras] {10.1093/mnras/sty1320}, \href
  {https://ui.adsabs.harvard.edu/abs/2018MNRAS.479.2909B} {479, 2909}

\bibitem[\protect\citeauthoryear{{Carciofi}, {Bjorkman}, {Otero}, {Okazaki},
  {{\v{S}}tefl}, {Rivinius}, {Baade}  \& {Haubois}}{{Carciofi}
  et~al.}{2012}]{2012ApJ...744L..15C}
{Carciofi} A.~C.,  {Bjorkman} J.~E.,  {Otero} S.~A.,  {Okazaki} A.~T.,
  {{\v{S}}tefl} S.,  {Rivinius} T.,  {Baade} D.,   {Haubois} X.,  2012, \mn@doi
  [\apjl] {10.1088/2041-8205/744/1/L15}, \href
  {https://ui.adsabs.harvard.edu/abs/2012ApJ...744L..15C} {744, L15}

\bibitem[\protect\citeauthoryear{{Carvalho} \& {Hillenbrand}}{{Carvalho} \&
  {Hillenbrand}}{2022}]{Carvalho2022}
{Carvalho} A.~S.,  {Hillenbrand} L.~A.,  2022, arXiv e-prints, \href
  {https://ui.adsabs.harvard.edu/abs/2022arXiv220406061C} {p. arXiv:2204.06061}

\bibitem[\protect\citeauthoryear{{Castelli} \& {Kurucz}}{{Castelli} \&
  {Kurucz}}{2003}]{2003IAUS..210P.A20C}
{Castelli} F.,  {Kurucz} R.~L.,  2003, in {Piskunov} N.,  {Weiss} W.~W.,
  {Gray} D.~F.,  eds,  IAU Symposium 210 Vol. 210, Modelling of Stellar
  Atmospheres. p.~A20 (\mn@eprint {arXiv} {astro-ph/0405087})

\bibitem[\protect\citeauthoryear{{Chojnowski} et~al.,}{{Chojnowski}
  et~al.}{2015}]{Chojnowski2015}
{Chojnowski} S.~D.,  et~al., 2015, \mn@doi [\aj] {10.1088/0004-6256/149/1/7},
  \href {https://ui.adsabs.harvard.edu/abs/2015AJ....149....7C} {149, 7}

\bibitem[\protect\citeauthoryear{{Cochetti}, {Arias}, {Cidale}, {Granada}  \&
  {Torres}}{{Cochetti} et~al.}{2022}]{Cochetti2022}
{Cochetti} Y.~R.,  {Arias} M.~L.,  {Cidale} L.~S.,  {Granada} A.,   {Torres}
  A.~F.,  2022, arXiv e-prints, \href
  {https://ui.adsabs.harvard.edu/abs/2022arXiv220603819C} {p. arXiv:2206.03819}

\bibitem[\protect\citeauthoryear{{Comer{\'o}n} \& {Pasquali}}{{Comer{\'o}n} \&
  {Pasquali}}{2012}]{Comeron2012}
{Comer{\'o}n} F.,  {Pasquali} A.,  2012, \mn@doi [\aap]
  {10.1051/0004-6361/201219022}, \href
  {https://ui.adsabs.harvard.edu/abs/2012A&A...543A.101C} {543, A101}

\bibitem[\protect\citeauthoryear{{Cushing}, {Vacca}  \& {Rayner}}{{Cushing}
  et~al.}{2004}]{Cushing2004}
{Cushing} M.~C.,  {Vacca} W.~D.,   {Rayner} J.~T.,  2004, \mn@doi [\pasp]
  {10.1086/382907}, \href
  {https://ui.adsabs.harvard.edu/abs/2004PASP..116..362C} {116, 362}

\bibitem[\protect\citeauthoryear{{Cutri} \& {et al.}}{{Cutri} \& {et
  al.}}{2012}]{Cutri2012}
{Cutri} R.~M.,  {et al.} 2012, VizieR Online Data Catalog, \href
  {https://ui.adsabs.harvard.edu/abs/2012yCat.2311....0C} {p. II/311}

\bibitem[\protect\citeauthoryear{{Cutri} et~al.,}{{Cutri}
  et~al.}{2021}]{2014yCat.2328....0C}
{Cutri} R.~M.,  et~al., 2021, VizieR Online Data Catalog, \href
  {https://ui.adsabs.harvard.edu/abs/2014yCat.2328....0C} {p. II/328}

\bibitem[\protect\citeauthoryear{{De} et~al.,}{{De} et~al.}{2020}]{de2020}
{De} K.,  et~al., 2020, \mn@doi [\pasp] {10.1088/1538-3873/ab6069}, \href
  {https://ui.adsabs.harvard.edu/abs/2020PASP..132b5001D} {132, 025001}

\bibitem[\protect\citeauthoryear{{Derdzinski}, {Metzger}  \&
  {Lazzati}}{{Derdzinski} et~al.}{2017}]{2017MNRAS.469.1314D}
{Derdzinski} A.~M.,  {Metzger} B.~D.,   {Lazzati} D.,  2017, \mn@doi [\mnras]
  {10.1093/mnras/stx829}, \href
  {https://ui.adsabs.harvard.edu/abs/2017MNRAS.469.1314D} {469, 1314}

\bibitem[\protect\citeauthoryear{{Drew} et~al.,}{{Drew}
  et~al.}{2005}]{2005MNRAS.362..753D}
{Drew} J.~E.,  et~al., 2005, \mn@doi [\mnras]
  {10.1111/j.1365-2966.2005.09330.x}, \href
  {https://ui.adsabs.harvard.edu/abs/2005MNRAS.362..753D} {362, 753}

\bibitem[\protect\citeauthoryear{{Evitts} et~al.,}{{Evitts}
  et~al.}{2020}]{2020MNRAS.493..184E}
{Evitts} J.~J.,  et~al., 2020, \mn@doi [\mnras] {10.1093/mnras/staa158}, \href
  {https://ui.adsabs.harvard.edu/abs/2020MNRAS.493..184E} {493, 184}

\bibitem[\protect\citeauthoryear{{Finkenzeller} \& {Mundt}}{{Finkenzeller} \&
  {Mundt}}{1984}]{Finkenzeller1984}
{Finkenzeller} U.,  {Mundt} R.,  1984, \aaps, \href
  {https://ui.adsabs.harvard.edu/abs/1984A&AS...55..109F} {55, 109}

\bibitem[\protect\citeauthoryear{{Froebrich} et~al.,}{{Froebrich}
  et~al.}{2018}]{2018MNRAS.478.5091F}
{Froebrich} D.,  et~al., 2018, \mn@doi [\mnras] {10.1093/mnras/sty1350}, \href
  {https://ui.adsabs.harvard.edu/abs/2018MNRAS.478.5091F} {478, 5091}

\bibitem[\protect\citeauthoryear{{Gaia Collaboration}}{{Gaia
  Collaboration}}{2022}]{2022yCat.1355....0G}
{Gaia Collaboration} 2022, VizieR Online Data Catalog, \href
  {https://ui.adsabs.harvard.edu/abs/2022yCat.1355....0G} {p. I/355}

\bibitem[\protect\citeauthoryear{{Gaia Collaboration} et~al.,}{{Gaia
  Collaboration} et~al.}{2016}]{2016A&A...595A...1G}
{Gaia Collaboration} et~al., 2016, \mn@doi [\aap]
  {10.1051/0004-6361/201629272}, \href
  {https://ui.adsabs.harvard.edu/abs/2016A&A...595A...1G} {595, A1}

\bibitem[\protect\citeauthoryear{{Ghoreyshi} et~al.,}{{Ghoreyshi}
  et~al.}{2018}]{Ghoreyshi2018}
{Ghoreyshi} M.~R.,  et~al., 2018, \mn@doi [\mnras] {10.1093/mnras/sty1577},
  \href {https://ui.adsabs.harvard.edu/abs/2018MNRAS.479.2214G} {479, 2214}

\bibitem[\protect\citeauthoryear{{Hambsch}}{{Hambsch}}{2012}]{2012JAVSO..40.1003H}
{Hambsch} F.~J.,  2012, Journal of the American Association of Variable Star
  Observers (JAAVSO), \href
  {https://ui.adsabs.harvard.edu/abs/2012JAVSO..40.1003H} {40, 1003}

\bibitem[\protect\citeauthoryear{{Haubois}, {Carciofi}, {Rivinius}, {Okazaki}
  \& {Bjorkman}}{{Haubois} et~al.}{2012a}]{2012ApJ...756..156H}
{Haubois} X.,  {Carciofi} A.~C.,  {Rivinius} T.,  {Okazaki} A.~T.,   {Bjorkman}
  J.~E.,  2012a, \mn@doi [\apj] {10.1088/0004-637X/756/2/156}, \href
  {https://ui.adsabs.harvard.edu/abs/2012ApJ...756..156H} {756, 156}

\bibitem[\protect\citeauthoryear{{Haubois}, {Carciofi}, {Rivinius}, {Okazaki}
  \& {Bjorkman}}{{Haubois} et~al.}{2012b}]{Haubois2012}
{Haubois} X.,  {Carciofi} A.~C.,  {Rivinius} T.,  {Okazaki} A.~T.,   {Bjorkman}
  J.~E.,  2012b, \mn@doi [\apj] {10.1088/0004-637X/756/2/156}, \href
  {https://ui.adsabs.harvard.edu/abs/2012ApJ...756..156H} {756, 156}

\bibitem[\protect\citeauthoryear{{Herter} et~al.,}{{Herter}
  et~al.}{2008}]{Herter2008}
{Herter} T.~L.,  et~al., 2008, in {McLean} I.~S.,  {Casali} M.~M.,  eds,
  Society of Photo-Optical Instrumentation Engineers (SPIE) Conference Series
  Vol. 7014, Ground-based and Airborne Instrumentation for Astronomy II. p.
  70140X, \mn@doi{10.1117/12.789660}

\bibitem[\protect\citeauthoryear{{Hodgkin} et~al.,}{{Hodgkin}
  et~al.}{2021}]{Hodgkin2021}
{Hodgkin} S.~T.,  et~al., 2021, \mn@doi [\aap] {10.1051/0004-6361/202140735},
  \href {https://ui.adsabs.harvard.edu/abs/2021A&A...652A..76H} {652, A76}

\bibitem[\protect\citeauthoryear{{Husser}, {Wende-von Berg}, {Dreizler},
  {Homeier}, {Reiners}, {Barman}  \& {Hauschildt}}{{Husser}
  et~al.}{2013}]{2013A&A...553A...6H}
{Husser} T.~O.,  {Wende-von Berg} S.,  {Dreizler} S.,  {Homeier} D.,  {Reiners}
  A.,  {Barman} T.,   {Hauschildt} P.~H.,  2013, \mn@doi [\aap]
  {10.1051/0004-6361/201219058}, \href
  {https://ui.adsabs.harvard.edu/abs/2013A&A...553A...6H} {553, A6}

\bibitem[\protect\citeauthoryear{{Kochanek} et~al.,}{{Kochanek}
  et~al.}{2017}]{2017PASP..129j4502K}
{Kochanek} C.~S.,  et~al., 2017, \mn@doi [\pasp] {10.1088/1538-3873/aa80d9},
  \href {https://ui.adsabs.harvard.edu/abs/2017PASP..129j4502K} {129, 104502}

\bibitem[\protect\citeauthoryear{{Kohoutek} \& {Wehmeyer}}{{Kohoutek} \&
  {Wehmeyer}}{1997}]{KW1997}
{Kohoutek} L.,  {Wehmeyer} R.,  1997, Astronomische Abhandlungen der Hamburger
  Sternwarte, \href {https://ui.adsabs.harvard.edu/abs/1997AAHam..11.....K}
  {11, 1}

\bibitem[\protect\citeauthoryear{{Labadie-Bartz} et~al.,}{{Labadie-Bartz}
  et~al.}{2017a}]{Labadie-Bartz2017}
{Labadie-Bartz} J.,  et~al., 2017a, \mn@doi [\aj] {10.3847/1538-3881/aa6396},
  \href {https://ui.adsabs.harvard.edu/abs/2017AJ....153..252L} {153, 252}

\bibitem[\protect\citeauthoryear{{Labadie-Bartz} et~al.,}{{Labadie-Bartz}
  et~al.}{2017b}]{2017AJ....153..252L}
{Labadie-Bartz} J.,  et~al., 2017b, \mn@doi [\aj] {10.3847/1538-3881/aa6396},
  \href {https://ui.adsabs.harvard.edu/abs/2017AJ....153..252L} {153, 252}

\bibitem[\protect\citeauthoryear{{Labadie-Bartz} et~al.,}{{Labadie-Bartz}
  et~al.}{2018}]{Labadie-Bartz2018}
{Labadie-Bartz} J.,  et~al., 2018, \mn@doi [\aj] {10.3847/1538-3881/aa9c7e},
  \href {https://ui.adsabs.harvard.edu/abs/2018AJ....155...53L} {155, 53}

\bibitem[\protect\citeauthoryear{{Labadie-Bartz}, {Carciofi}, {Henrique de
  Amorim}, {Rubio}, {Luiz Figueiredo}, {Ticiani dos Santos}  \&
  {Thomson-Paressant}}{{Labadie-Bartz} et~al.}{2022a}]{Labadie-Bartz2022}
{Labadie-Bartz} J.,  {Carciofi} A.~C.,  {Henrique de Amorim} T.,  {Rubio} A.,
  {Luiz Figueiredo} A.,  {Ticiani dos Santos} P.,   {Thomson-Paressant} K.,
  2022a, \mn@doi [\aj] {10.3847/1538-3881/ac5abd}, \href
  {https://ui.adsabs.harvard.edu/abs/2022AJ....163..226L} {163, 226}

\bibitem[\protect\citeauthoryear{{Labadie-Bartz}, {Carciofi}, {Henrique de
  Amorim}, {Rubio}, {Luiz Figueiredo}, {Ticiani dos Santos}  \&
  {Thomson-Paressant}}{{Labadie-Bartz} et~al.}{2022b}]{2022AJ....163..226L}
{Labadie-Bartz} J.,  {Carciofi} A.~C.,  {Henrique de Amorim} T.,  {Rubio} A.,
  {Luiz Figueiredo} A.,  {Ticiani dos Santos} P.,   {Thomson-Paressant} K.,
  2022b, \mn@doi [\aj] {10.3847/1538-3881/ac5abd}, \href
  {https://ui.adsabs.harvard.edu/abs/2022AJ....163..226L} {163, 226}

\bibitem[\protect\citeauthoryear{{Law} et~al.,}{{Law} et~al.}{2009}]{Law2009}
{Law} N.~M.,  et~al., 2009, \mn@doi [\pasp] {10.1086/648598}, \href
  {https://ui.adsabs.harvard.edu/abs/2009PASP..121.1395L} {121, 1395}

\bibitem[\protect\citeauthoryear{{Lee}, {Osaki}  \& {Saio}}{{Lee}
  et~al.}{1991}]{Lee1991}
{Lee} U.,  {Osaki} Y.,   {Saio} H.,  1991, \mn@doi [\mnras]
  {10.1093/mnras/250.2.432}, \href
  {https://ui.adsabs.harvard.edu/abs/1991MNRAS.250..432L} {250, 432}

\bibitem[\protect\citeauthoryear{{Leone} \& {Lanzafame}}{{Leone} \&
  {Lanzafame}}{1998}]{Leone1998}
{Leone} F.,  {Lanzafame} A.~C.,  1998, \aap, \href
  {https://ui.adsabs.harvard.edu/abs/1998A&A...330..306L} {330, 306}

\bibitem[\protect\citeauthoryear{{Lim}, Diaz  \& Laidler}{{Lim}
  et~al.}{2015}]{2013ascl.soft03023S}
{Lim} P.~L.,  Diaz R.~I.,   Laidler V.,  2015, {PySynphot User's Guider}
  (\mn@eprint {ascl} {1303.023})

\bibitem[\protect\citeauthoryear{{Mainzer} et~al.,}{{Mainzer}
  et~al.}{2011}]{Mainzer2011}
{Mainzer} A.,  et~al., 2011, \mn@doi [\apj] {10.1088/0004-637X/731/1/53}, \href
  {https://ui.adsabs.harvard.edu/abs/2011ApJ...731...53M} {731, 53}

\bibitem[\protect\citeauthoryear{{Masci} et~al.,}{{Masci}
  et~al.}{2019}]{ztf_data}
{Masci} F.~J.,  et~al., 2019, \mn@doi [\pasp] {10.1088/1538-3873/aae8ac}, \href
  {https://ui.adsabs.harvard.edu/abs/2019PASP..131a8003M} {131, 018003}

\bibitem[\protect\citeauthoryear{{McLean} et~al.,}{{McLean}
  et~al.}{1998}]{McLean1998}
{McLean} I.~S.,  et~al., 1998, in {Fowler} A.~M.,  ed.,  Society of
  Photo-Optical Instrumentation Engineers (SPIE) Conference Series Vol. 3354,
  Infrared Astronomical Instrumentation. pp 566--578,
  \mn@doi{10.1117/12.317283}

\bibitem[\protect\citeauthoryear{{Mennickent}, {Pietrzy{\'n}ski}, {Gieren}  \&
  {Szewczyk}}{{Mennickent} et~al.}{2002}]{Mennickent2000}
{Mennickent} R.~E.,  {Pietrzy{\'n}ski} G.,  {Gieren} W.,   {Szewczyk} O.,
  2002, \mn@doi [\aap] {10.1051/0004-6361:20020916}, \href
  {https://ui.adsabs.harvard.edu/abs/2002A&A...393..887M} {393, 887}

\bibitem[\protect\citeauthoryear{{Moe} \& {Di Stefano}}{{Moe} \& {Di
  Stefano}}{2017}]{Moe2017}
{Moe} M.,  {Di Stefano} R.,  2017, \mn@doi [\apjs] {10.3847/1538-4365/aa6fb6},
  \href {https://ui.adsabs.harvard.edu/abs/2017ApJS..230...15M} {230, 15}

\bibitem[\protect\citeauthoryear{{Moore} \& {Kasliwal}}{{Moore} \&
  {Kasliwal}}{2019}]{2019NatAs...3..109M}
{Moore} A.~M.,  {Kasliwal} M.~M.,  2019, \mn@doi [Nature Astronomy]
  {10.1038/s41550-018-0675-x}, \href
  {https://ui.adsabs.harvard.edu/abs/2019NatAs...3..109M} {3, 109}

\bibitem[\protect\citeauthoryear{{Munari} et~al.,}{{Munari}
  et~al.}{2019}]{Munari2019}
{Munari} U.,  et~al., 2019, \mn@doi [\mnras] {10.1093/mnras/stz2078}, \href
  {https://ui.adsabs.harvard.edu/abs/2019MNRAS.488.5536M} {488, 5536}

\bibitem[\protect\citeauthoryear{{Nakano}, {Sugitani}, {Watanabe}, {Fukuda},
  {Ishihara}  \& {Ueno}}{{Nakano} et~al.}{2012}]{Nakano2012}
{Nakano} M.,  {Sugitani} K.,  {Watanabe} M.,  {Fukuda} N.,  {Ishihara} D.,
  {Ueno} M.,  2012, \mn@doi [\aj] {10.1088/0004-6256/143/3/61}, \href
  {https://ui.adsabs.harvard.edu/abs/2012AJ....143...61N} {143, 61}

\bibitem[\protect\citeauthoryear{{Neiner}, {de Batz}, {Cochard}, {Floquet},
  {Mekkas}  \& {Desnoux}}{{Neiner} et~al.}{2011}]{Neiner2011}
{Neiner} C.,  {de Batz} B.,  {Cochard} F.,  {Floquet} M.,  {Mekkas} A.,
  {Desnoux} V.,  2011, \mn@doi [\aj] {10.1088/0004-6256/142/5/149}, \href
  {https://ui.adsabs.harvard.edu/abs/2011AJ....142..149N} {142, 149}

\bibitem[\protect\citeauthoryear{{Oke} \& {Gunn}}{{Oke} \&
  {Gunn}}{1982}]{OG1982}
{Oke} J.~B.,  {Gunn} J.~E.,  1982, \mn@doi [\pasp] {10.1086/131027}, \href
  {https://ui.adsabs.harvard.edu/abs/1982PASP...94..586O} {94, 586}

\bibitem[\protect\citeauthoryear{{Porter}}{{Porter}}{1996}]{Porter1996}
{Porter} J.~M.,  1996, \mn@doi [\mnras] {10.1093/mnras/280.3.L31}, \href
  {https://ui.adsabs.harvard.edu/abs/1996MNRAS.280L..31P} {280, L31}

\bibitem[\protect\citeauthoryear{{Ressler}}{{Ressler}}{2021}]{Ressler2021}
{Ressler} S.~M.,  2021, \mn@doi [\mnras] {10.1093/mnras/stab2880}, \href
  {https://ui.adsabs.harvard.edu/abs/2021MNRAS.508.4887R} {508, 4887}

\bibitem[\protect\citeauthoryear{{Richardson} et~al.,}{{Richardson}
  et~al.}{2021}]{Richardson2021}
{Richardson} N.~D.,  et~al., 2021, \mn@doi [\mnras] {10.1093/mnras/stab2759},
  \href {https://ui.adsabs.harvard.edu/abs/2021MNRAS.508.2002R} {508, 2002}

\bibitem[\protect\citeauthoryear{{Ricker} et~al.,}{{Ricker}
  et~al.}{2015}]{2015JATIS...1a4003R}
{Ricker} G.~R.,  et~al., 2015, \mn@doi [Journal of Astronomical Telescopes,
  Instruments, and Systems] {10.1117/1.JATIS.1.1.014003}, \href
  {https://ui.adsabs.harvard.edu/abs/2015JATIS...1a4003R} {1, 014003}

\bibitem[\protect\citeauthoryear{{R{\'\i}mulo} et~al.,}{{R{\'\i}mulo}
  et~al.}{2018}]{Rimulo2018}
{R{\'\i}mulo} L.~R.,  et~al., 2018, \mn@doi [\mnras] {10.1093/mnras/sty431},
  \href {https://ui.adsabs.harvard.edu/abs/2018MNRAS.476.3555R} {476, 3555}

\bibitem[\protect\citeauthoryear{{Rivinius}, {Carciofi}  \&
  {Martayan}}{{Rivinius} et~al.}{2013}]{Rivinius2013}
{Rivinius} T.,  {Carciofi} A.~C.,   {Martayan} C.,  2013, \mn@doi [\aapr]
  {10.1007/s00159-013-0069-0}, \href
  {https://ui.adsabs.harvard.edu/abs/2013A&ARv..21...69R} {21, 69}

\bibitem[\protect\citeauthoryear{{Rivinius}, {Baade}  \& {Carciofi}}{{Rivinius}
  et~al.}{2016}]{Rivinius2016}
{Rivinius} T.,  {Baade} D.,   {Carciofi} A.~C.,  2016, \mn@doi [\aap]
  {10.1051/0004-6361/201628411}, \href
  {https://ui.adsabs.harvard.edu/abs/2016A&A...593A.106R} {593, A106}

\bibitem[\protect\citeauthoryear{{Semaan}, {Hubert}, {Zorec},
  {Guti{\'e}rrez-Soto}, {Fr{\'e}mat}, {Martayan}, {Fabregat}  \&
  {Eggenberger}}{{Semaan} et~al.}{2018}]{Seamann2018}
{Semaan} T.,  {Hubert} A.~M.,  {Zorec} J.,  {Guti{\'e}rrez-Soto} J.,
  {Fr{\'e}mat} Y.,  {Martayan} C.,  {Fabregat} J.,   {Eggenberger} P.,  2018,
  \mn@doi [\aap] {10.1051/0004-6361/201629243}, \href
  {https://ui.adsabs.harvard.edu/abs/2018A&A...613A..70S} {613, A70}

\bibitem[\protect\citeauthoryear{{Shappee} et~al.,}{{Shappee}
  et~al.}{2014}]{2014ApJ...788...48S}
{Shappee} B.~J.,  et~al., 2014, \mn@doi [\apj] {10.1088/0004-637X/788/1/48},
  \href {https://ui.adsabs.harvard.edu/abs/2014ApJ...788...48S} {788, 48}

\bibitem[\protect\citeauthoryear{{Slettebak}}{{Slettebak}}{1982}]{Slettebak1982}
{Slettebak} A.,  1982, \mn@doi [\apjs] {10.1086/190820}, \href
  {https://ui.adsabs.harvard.edu/abs/1982ApJS...50...55S} {50, 55}

\bibitem[\protect\citeauthoryear{{Slettebak}, {Collins}  \&
  {Truax}}{{Slettebak} et~al.}{1992}]{Slettebak1992}
{Slettebak} A.,  {Collins} George~W. I.,   {Truax} R.,  1992, \mn@doi [\apjs]
  {10.1086/191696}, \href
  {https://ui.adsabs.harvard.edu/abs/1992ApJS...81..335S} {81, 335}

\bibitem[\protect\citeauthoryear{Tonry et~al.,}{Tonry et~al.}{2018}]{Tonry2018}
Tonry J.~L.,  et~al., 2018, Publications of the Astronomical Society of the
  Pacific, 130, 064505

\bibitem[\protect\citeauthoryear{{Usov}}{{Usov}}{1991}]{1991MNRAS.252...49U}
{Usov} V.~V.,  1991, \mn@doi [\mnras] {10.1093/mnras/252.1.49}, \href
  {https://ui.adsabs.harvard.edu/abs/1991MNRAS.252...49U} {252, 49}

\bibitem[\protect\citeauthoryear{{Vacca}, {Cushing}  \& {Rayner}}{{Vacca}
  et~al.}{2003}]{Vacca2003}
{Vacca} W.~D.,  {Cushing} M.~C.,   {Rayner} J.~T.,  2003, \mn@doi [\pasp]
  {10.1086/346193}, \href
  {https://ui.adsabs.harvard.edu/abs/2003PASP..115..389V} {115, 389}

\bibitem[\protect\citeauthoryear{{Vieira}, {Carciofi}  \& {Bjorkman}}{{Vieira}
  et~al.}{2015}]{2015MNRAS.454.2107V}
{Vieira} R.~G.,  {Carciofi} A.~C.,   {Bjorkman} J.~E.,  2015, \mn@doi [\mnras]
  {10.1093/mnras/stv2074}, \href
  {https://ui.adsabs.harvard.edu/abs/2015MNRAS.454.2107V} {454, 2107}

\bibitem[\protect\citeauthoryear{{Vieira}, {Carciofi}  \& {Bjorkman}}{{Vieira}
  et~al.}{2016}]{2016MNRAS.455.2607V}
{Vieira} R.~G.,  {Carciofi} A.~C.,   {Bjorkman} J.~E.,  2016, \mn@doi [\mnras]
  {10.1093/mnras/stv2498}, \href
  {https://ui.adsabs.harvard.edu/abs/2016MNRAS.455.2607V} {455, 2607}

\bibitem[\protect\citeauthoryear{{Vieira}, {Carciofi}, {Bjorkman}, {Rivinius},
  {Baade}  \& {R{\'\i}mulo}}{{Vieira} et~al.}{2017}]{2017MNRAS.464.3071V}
{Vieira} R.~G.,  {Carciofi} A.~C.,  {Bjorkman} J.~E.,  {Rivinius} T.,  {Baade}
  D.,   {R{\'\i}mulo} L.~R.,  2017, \mn@doi [\mnras] {10.1093/mnras/stw2542},
  \href {https://ui.adsabs.harvard.edu/abs/2017MNRAS.464.3071V} {464, 3071}

\bibitem[\protect\citeauthoryear{{Vogt} et~al.,}{{Vogt}
  et~al.}{1994}]{Vogt1994}
{Vogt} S.~S.,  et~al., 1994, in {Crawford} D.~L.,  {Craine} E.~R.,  eds,
  Society of Photo-Optical Instrumentation Engineers (SPIE) Conference Series
  Vol. 2198, Instrumentation in Astronomy VIII. p.~362,
  \mn@doi{10.1117/12.176725}

\bibitem[\protect\citeauthoryear{{Wang} et~al.,}{{Wang}
  et~al.}{2022}]{Wang2022}
{Wang} L.,  et~al., 2022, \mn@doi [\apjs] {10.3847/1538-4365/ac617a}, \href
  {https://ui.adsabs.harvard.edu/abs/2022ApJS..260...35W} {260, 35}

\bibitem[\protect\citeauthoryear{{Wright} et~al.,}{{Wright}
  et~al.}{2010}]{Wright2010}
{Wright} E.~L.,  et~al., 2010, \mn@doi [\aj] {10.1088/0004-6256/140/6/1868},
  \href {https://ui.adsabs.harvard.edu/abs/2010AJ....140.1868W} {140, 1868}

\makeatother
\end{thebibliography}


\appendix

\begin{table*}
\section{Best burst fit parameters}
\caption{\label{burst_table} Table listing the fit results from Eqs.\,\ref{dm_lesst2} and \ref{dm_moret2} for the bursts with sufficient HOYS data available. For each burst we list the filter, the baseline magnitude, the asymptotic magnitude increase, the manually determined start $t_1$ and end $t_2$ time of mass loading, its duration, as well as the $\eta$ and $C$ parameters.}
\centering
\begin{tabular}{|cccccccccc|}
\hline
Filter & $m_0$ & $\Delta m^\infty$ & $t_1$ & $t_2$ & $t_2-t_1$ & $\eta_1$ & $\eta_2$ & $C_1$ & $C_2$ \\ 
 & [mag] & [mag] & [MJD] & [MJD] & [d] &  &  & [$d^{-1}$] & [$d^{-1}$] \\ 
\hline \multicolumn{10}{|l|}{NSW\,284, burst B3} \\ \hline
I & 14.305 & $-$0.639 & 57666 & 57750 & 84 &  2.361 &  1.604 &  0.038 &  0.022 \\
R & 14.572 & $-$0.803 & 57666 & 57750 & 84 &  1.621 &  1.325 &  0.021 &  0.043 \\
V & 14.772 & $-$0.431 & 57666 & 57750 & 84 &  2.280 &  1.292 &  0.037 &  0.037 \\
\hline \multicolumn{10}{|l|}{NSW\,284, burst B5} \\ \hline
I & 14.294 & $-$0.740 & 58640 & 58722 & 82 &  2.191 &  2.047 &  0.037 &  0.017 \\
R & 14.571 & $-$0.625 & 58640 & 58722 & 82 &  2.241 &  1.815 &  0.034 &  0.022 \\
V & 14.757 & $-$0.498 & 58640 & 58722 & 82 &  2.526 &  1.649 &  0.035 &  0.029 \\
B & 15.506 & $-$0.413 & 58640 & 58722 & 82 &  2.282 &  1.254 &  0.034 &  0.035 \\
\hline \multicolumn{10}{|l|}{NSW\,284, burst B6} \\ \hline
I & 14.274 & $-$0.723 & 58983 & 59035 & 52 &  1.378 &  2.061 &  0.034 &  0.029 \\
R & 14.561 & $-$0.466 & 58983 & 59035 & 52 &  2.317 &  1.712 &  0.044 &  0.035 \\
V & 14.755 & $-$0.505 & 58983 & 59035 & 52 &  1.400 &  1.745 &  0.029 &  0.040 \\
B & 15.481 & $-$0.376 & 58983 & 59035 & 52 &  1.154 &  1.307 &  0.031 &  0.045 \\
\hline \multicolumn{10}{|l|}{NSW\,284, burst B8} \\ \hline
I & 14.269 & $-$0.666 & 59550 & 59650 & 100 & 0.984 & 2.577 & 0.023 & 0.020 \\
R & 14.562 & $-$0.569 & 59550 & 59650 & 100 & 1.184 & 2.418 & 0.023 & 0.018 \\
V & 14.762 & $-$0.458 & 59550 & 59650 & 100 & 0.544 & 1.928 & 0.023 & 0.017 \\
\hline \multicolumn{10}{|l|}{Gaia\,19eyy, burst B5} \\ \hline
I & 13.057 & $-$0.686 & 59095 & 59180 & 85 &  4.046 &  3.202 &  0.029 &  0.019 \\
R & 13.322 & $-$0.563 & 59095 & 59180 & 85 &  4.590 &  3.398 &  0.029 &  0.022 \\
V & 13.533 & $-$0.450 & 59095 & 59180 & 85 &  5.035 &  3.414 &  0.031 &  0.025 \\
B & 14.295 & $-$0.383 & 59095 & 59180 & 85 &  4.424 &  2.848 &  0.032 &  0.027 \\
\hline \multicolumn{10}{|l|}{Gaia\,19eyy, burst B6} \\ \hline
I & 13.020 & -1.115 & 59507 & 59570 & 63 &  0.712 &  3.062 &  0.032 &  0.015 \\
R & 13.297 & -2.576 & 59507 & 59570 & 63 &  0.615 &  3.432 &  0.003 &  0.014 \\
V & 13.538 & -2.260 & 59507 & 59570 & 63 &  0.723 &  3.668 &  0.004 &  0.016 \\
B & 14.278 & -0.633 & 59507 & 59570 & 63 &  1.160 &  2.641 &  0.043 &  0.020 \\
\hline \multicolumn{10}{|l|}{VES\,263, burst B8} \\ \hline
I & 11.061 & -0.519 & 58365 & 58394 & 29 &  2.471 &  4.193 &  0.095 &  0.021 \\
R & 12.067 & -3.605 & 58365 & 58394 & 29 &  0.753 &  3.082 &  0.003 &  0.023 \\
V & 13.098 & -0.400 & 58365 & 58394 & 29 &  2.257 &  3.163 &  0.092 &  0.027 \\
B & 14.917 & -0.291 & 58365 & 58394 & 29 &  2.468 &  3.684 &  0.089 &  0.037 \\
\hline
\end{tabular}
\end{table*}

\begin{figure*}
\section{Burst Temperature analysis plots}
\centering
\includegraphics[angle=0,width=\columnwidth]{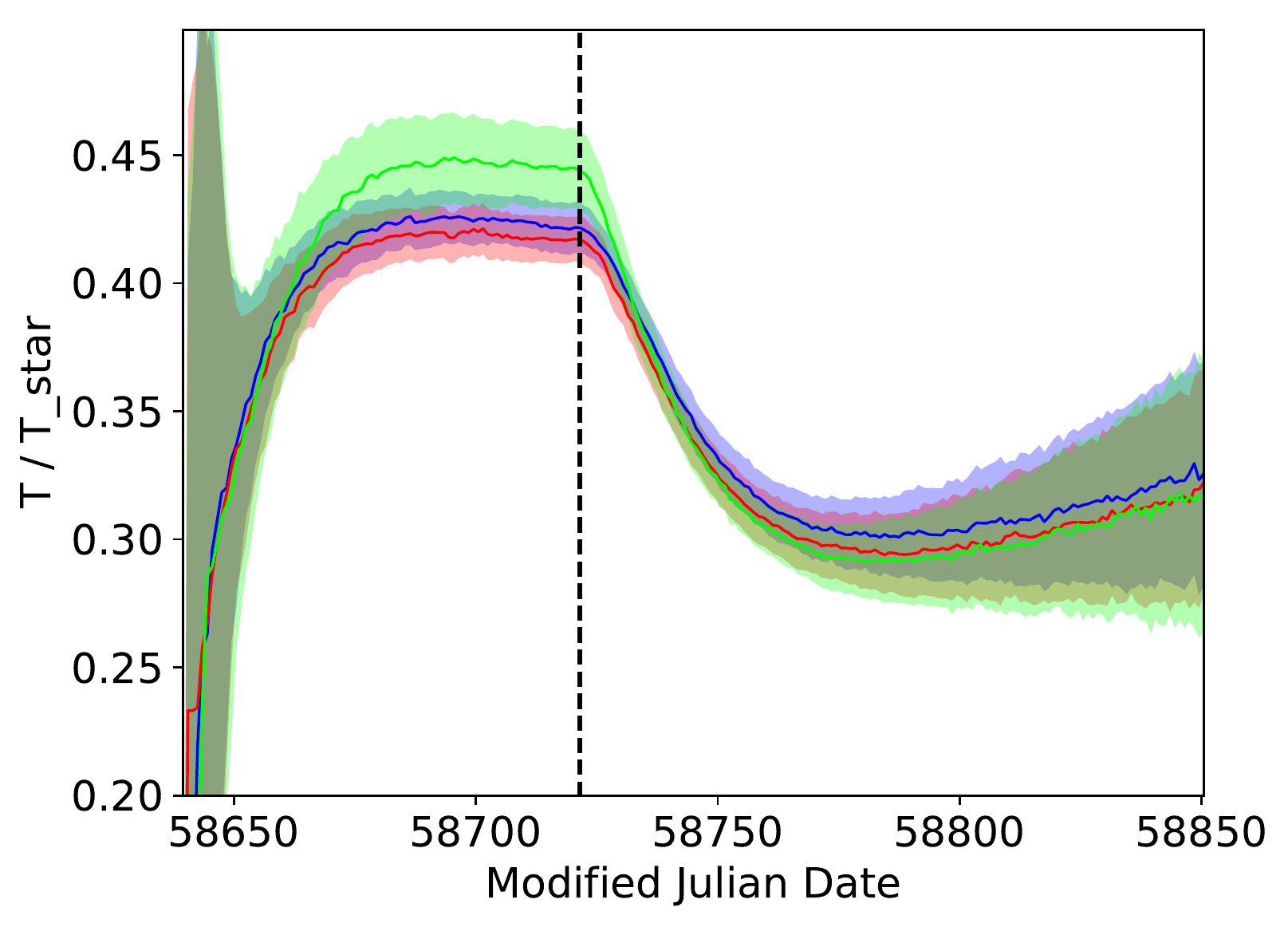} \hfill
\includegraphics[angle=0,width=\columnwidth]{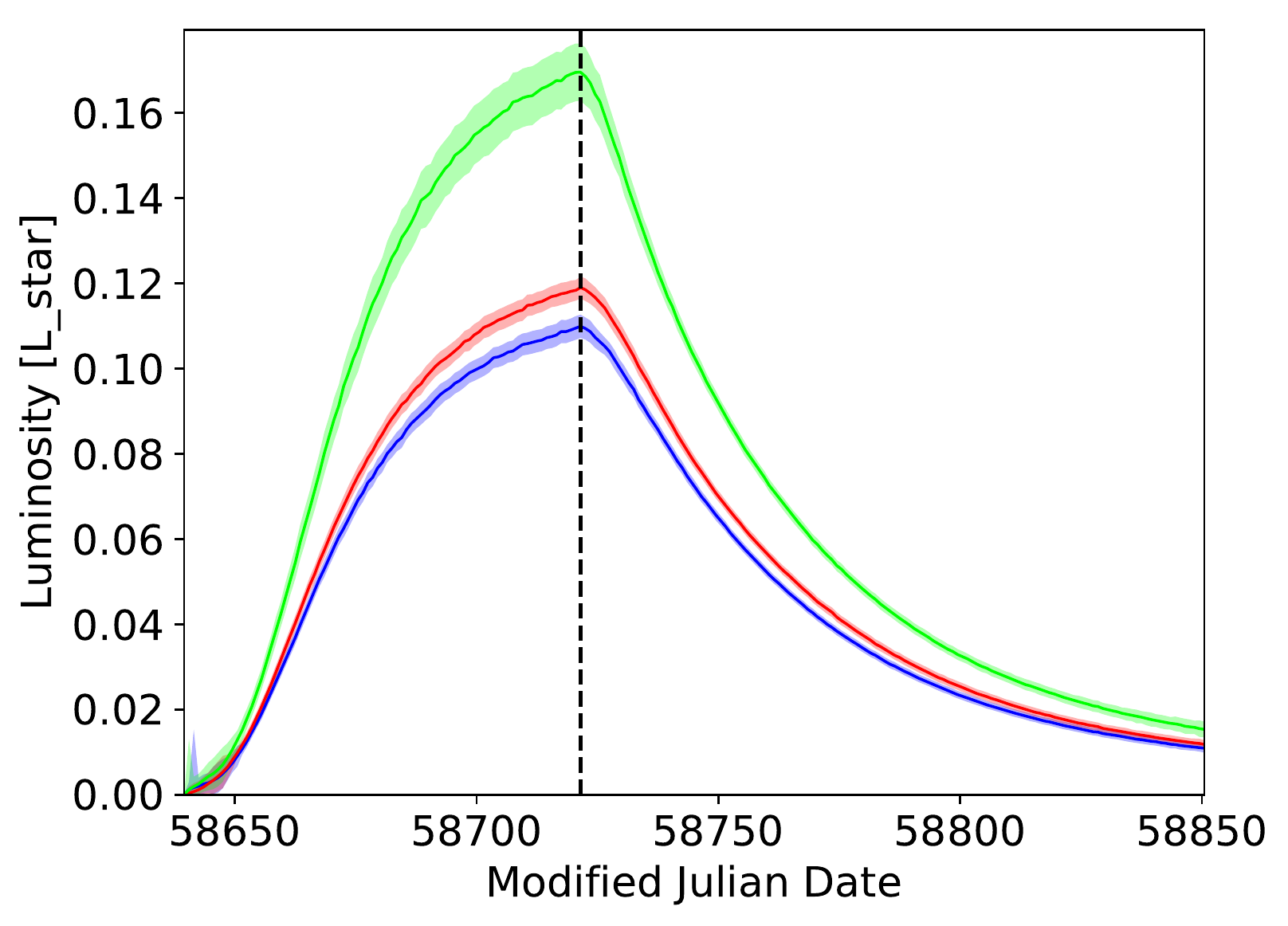} \\
\includegraphics[angle=0,width=\columnwidth]{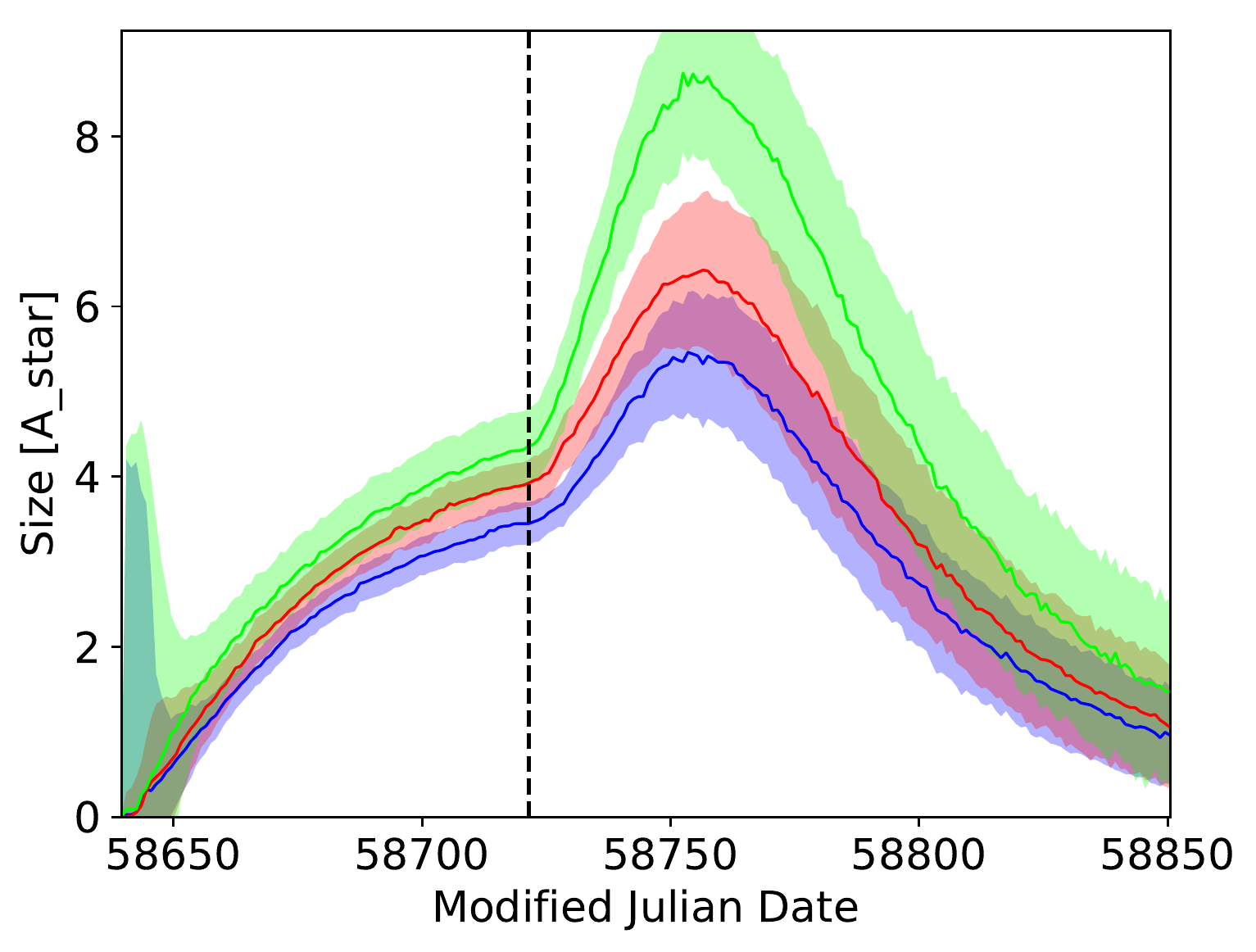} \hfill
\includegraphics[angle=0,width=\columnwidth]{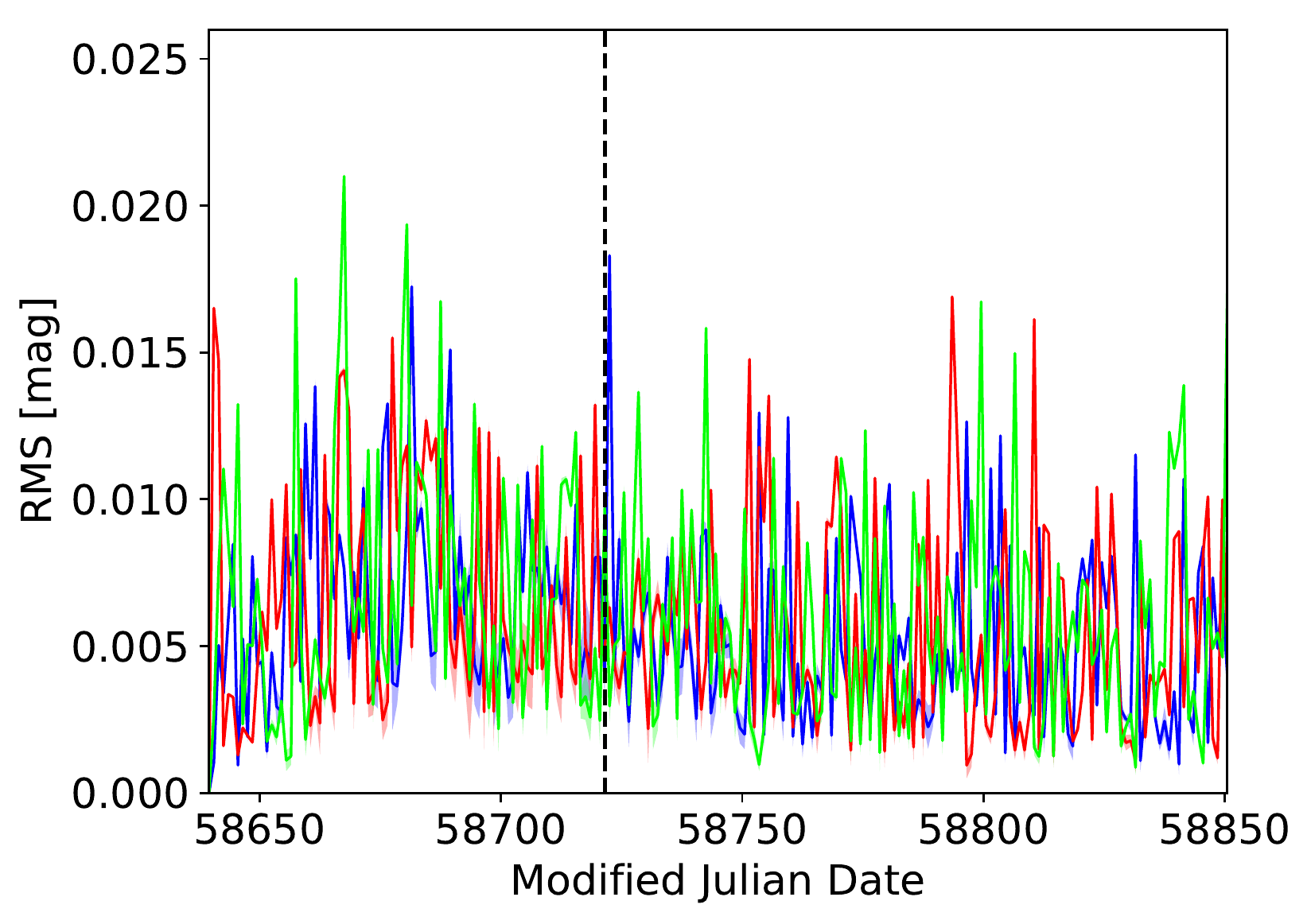} \\
\caption{Example of a burst fit using different model stellar atmospheres. We show the fit to burst B5 of NSW\,284, using a stellar temperature of 17500\,K and the $VRI$ photometry. If applicable we use $\log(g) = 4.0$ and [M/H]\,=\,0.0. The colours indicate the use of the PHOENIX (blue), ATLAS (red) and Blackbody (green) stellar atmosphere models and the lightly shaded areas the uncertainties. \label{fit_test_models}}
\end{figure*}

\begin{figure*}
\centering
\includegraphics[angle=0,width=\columnwidth]{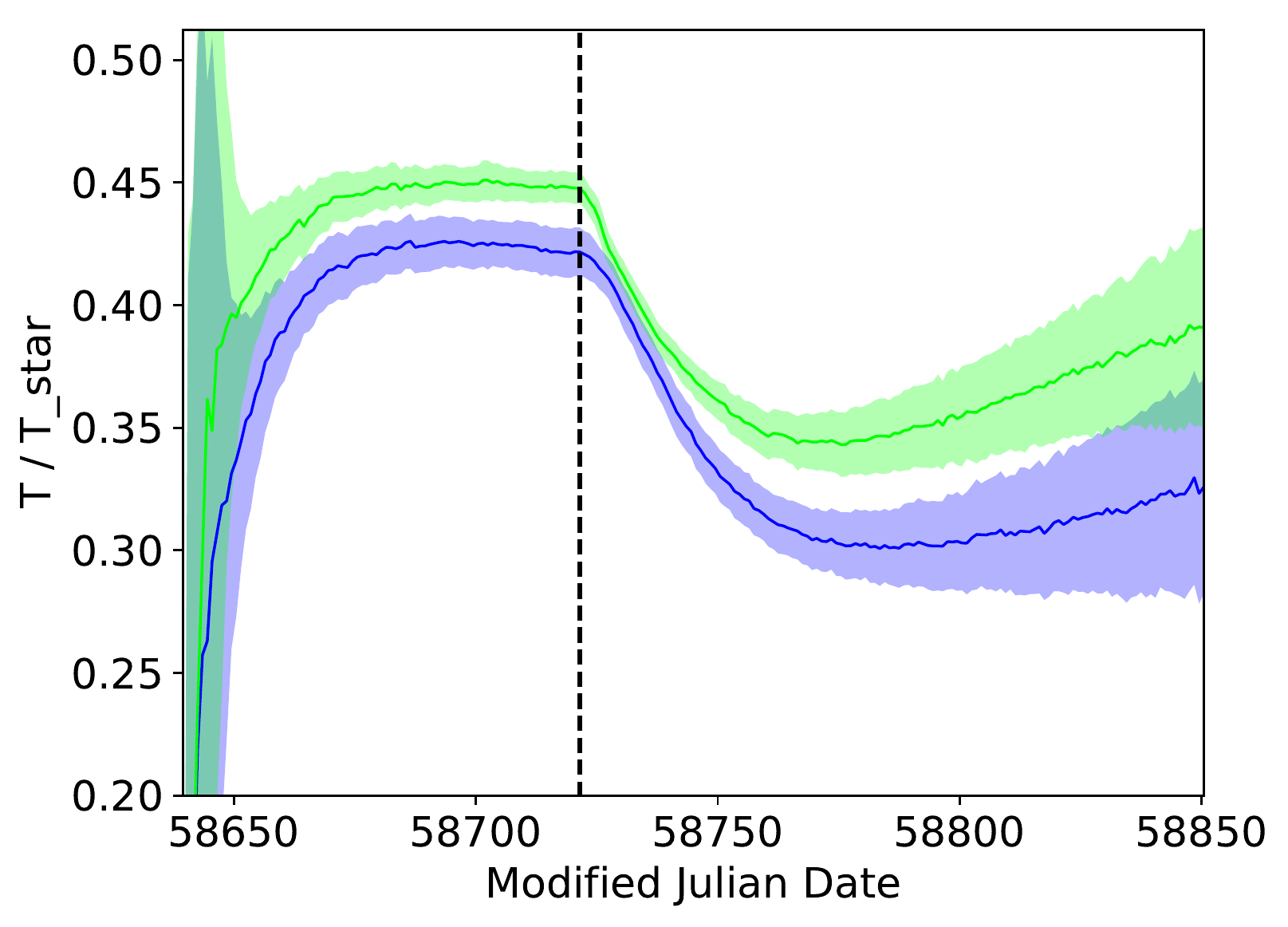} \hfill
\includegraphics[angle=0,width=\columnwidth]{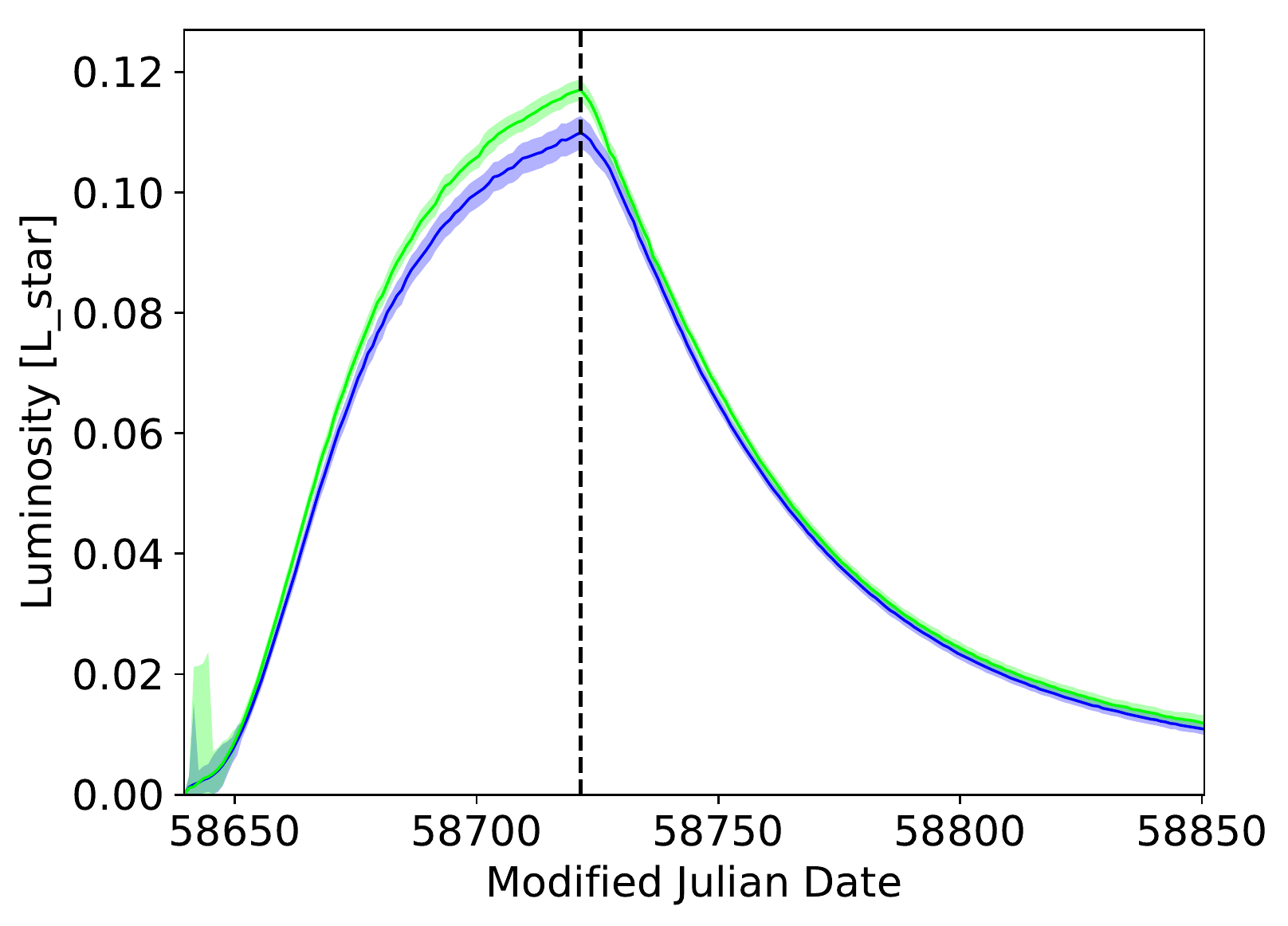} \\
\includegraphics[angle=0,width=\columnwidth]{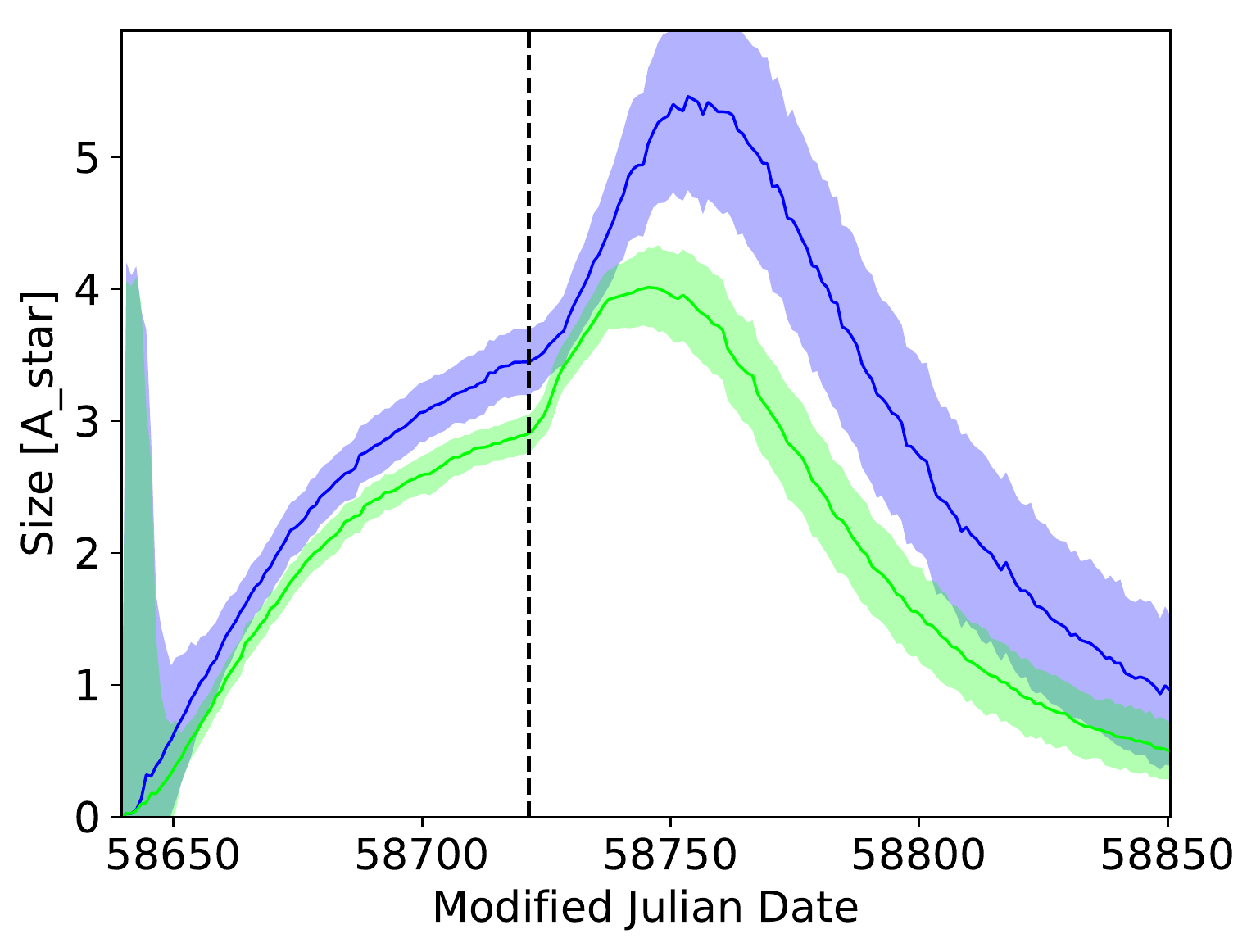} \hfill
\includegraphics[angle=0,width=\columnwidth]{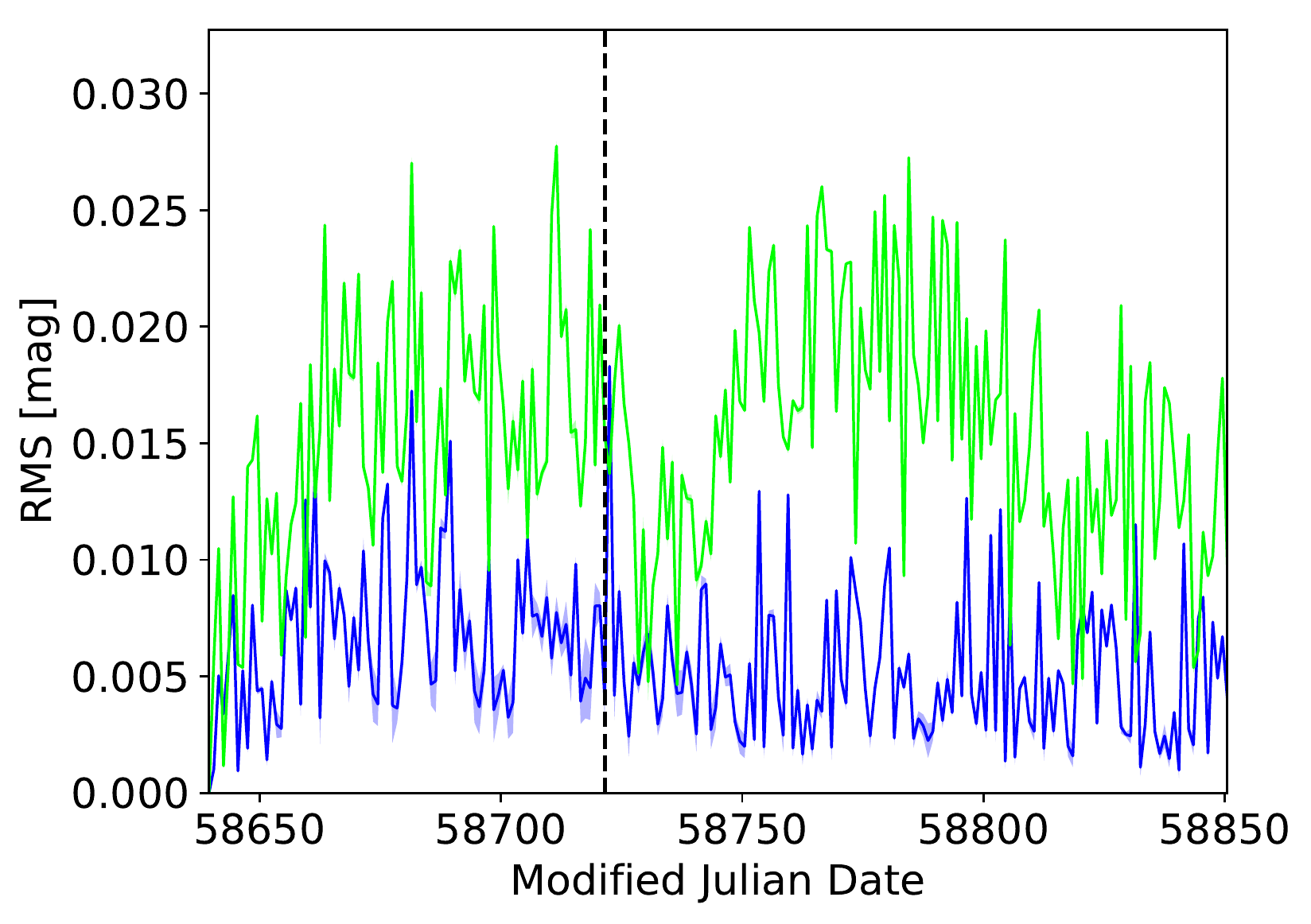} \\
\caption{Example of a burst fit using different sets of filters. We show the fit to burst B5 of NSW\,284, using a stellar temperature of 17500\,K and the PHOENIX stellar atmosphere models. We use $\log(g) = 4.0$ and [M/H]\,=\,0.0. The colours indicate the use of the $VRI$ amplitudes (blue) and the $BVRI$ amplitudes (green) and the lightly shaded areas the uncertainties. \label{fit_test_filters}}
\end{figure*}

\begin{figure*}
\centering
\includegraphics[angle=0,width=\columnwidth]{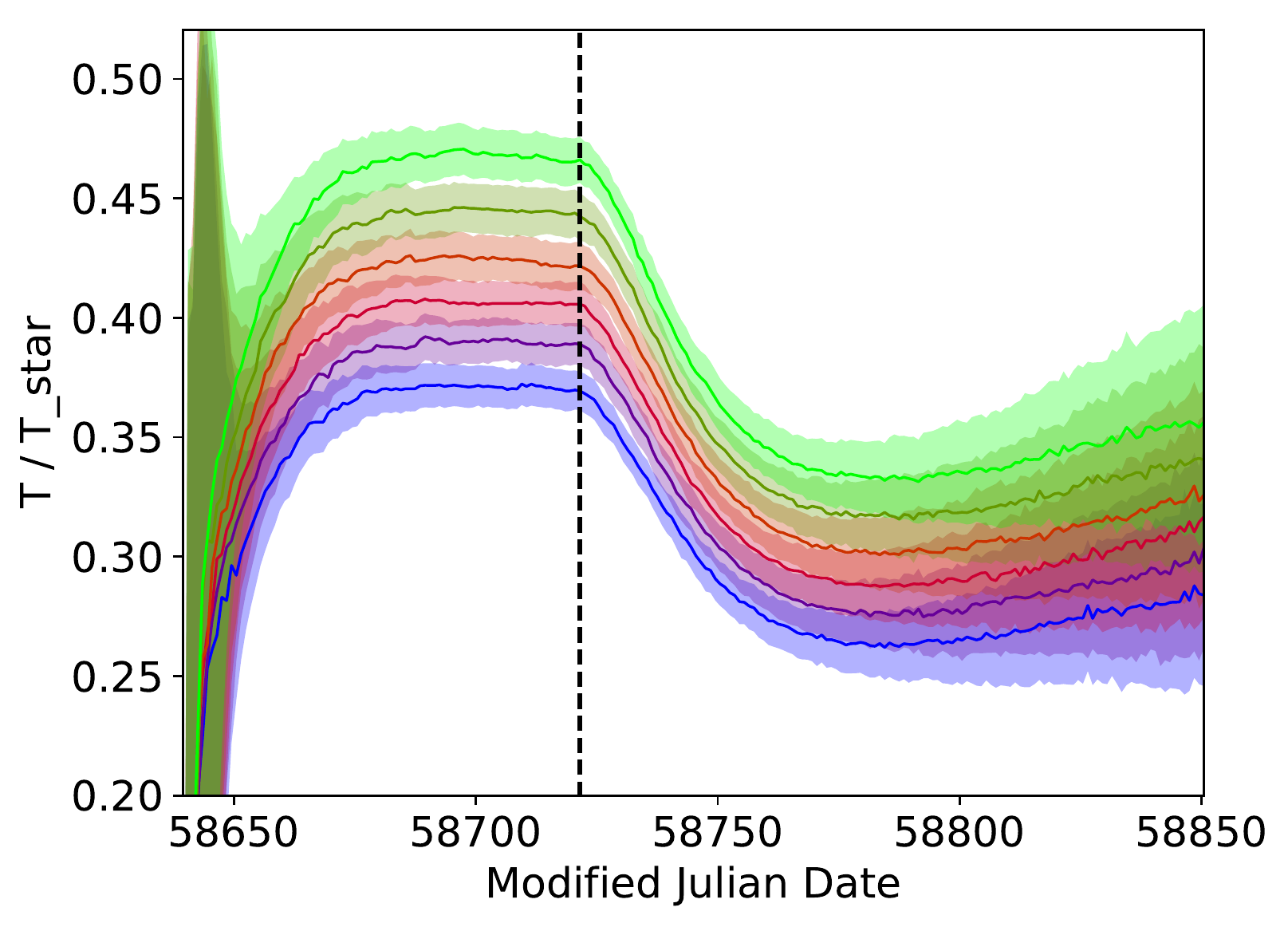} \hfill
\includegraphics[angle=0,width=\columnwidth]{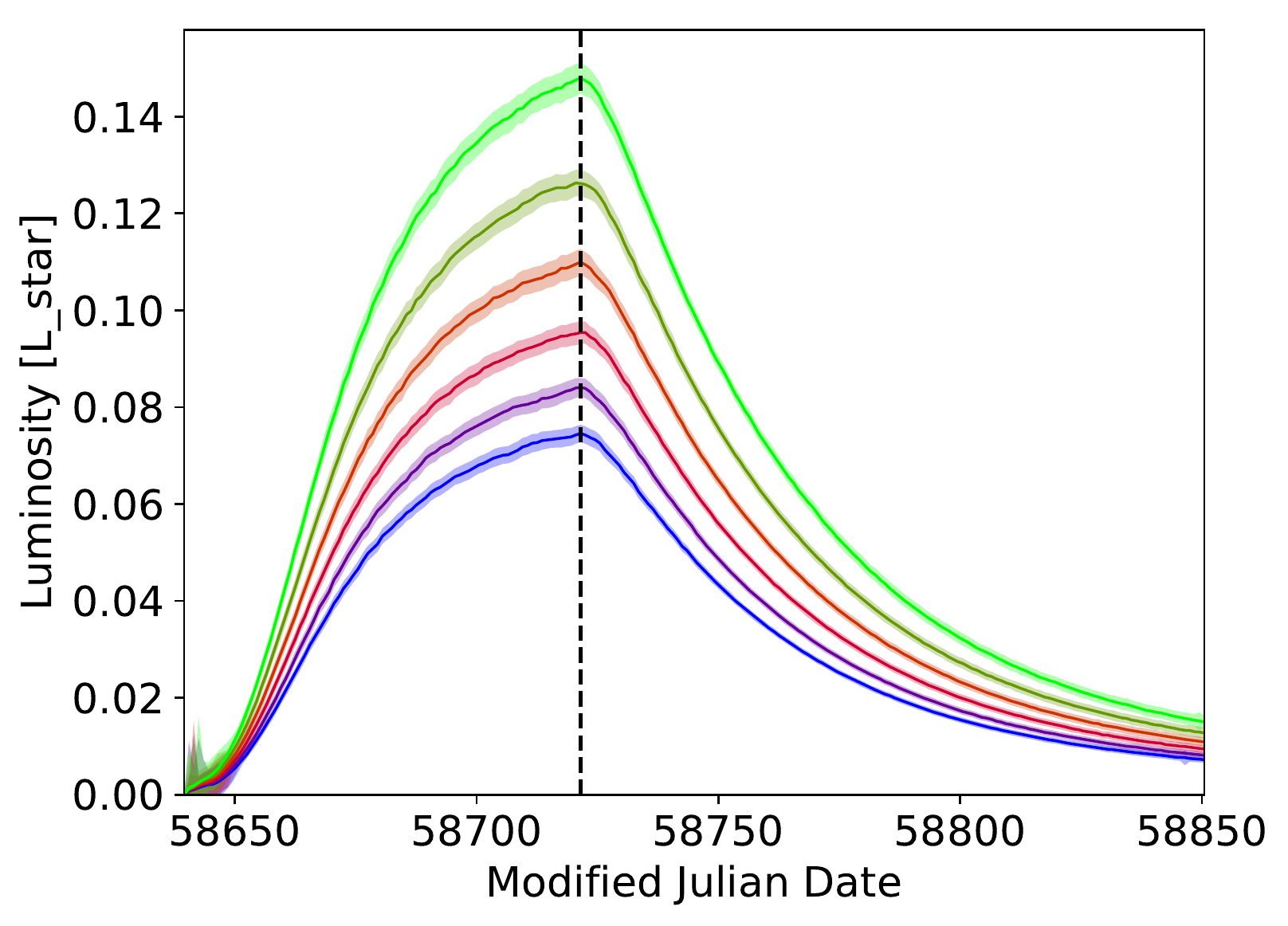} \\
\includegraphics[angle=0,width=\columnwidth]{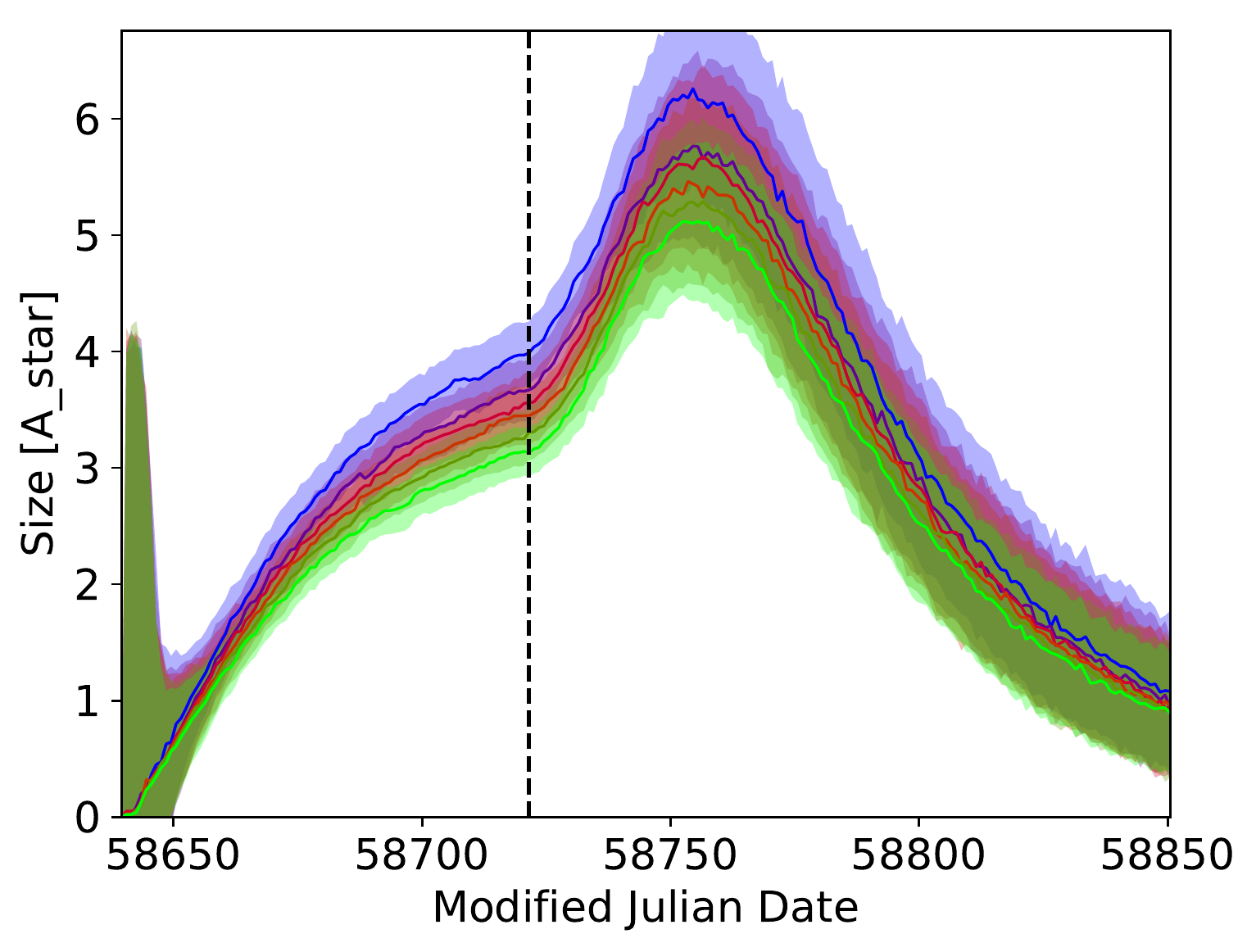} \hfill
\includegraphics[angle=0,width=\columnwidth]{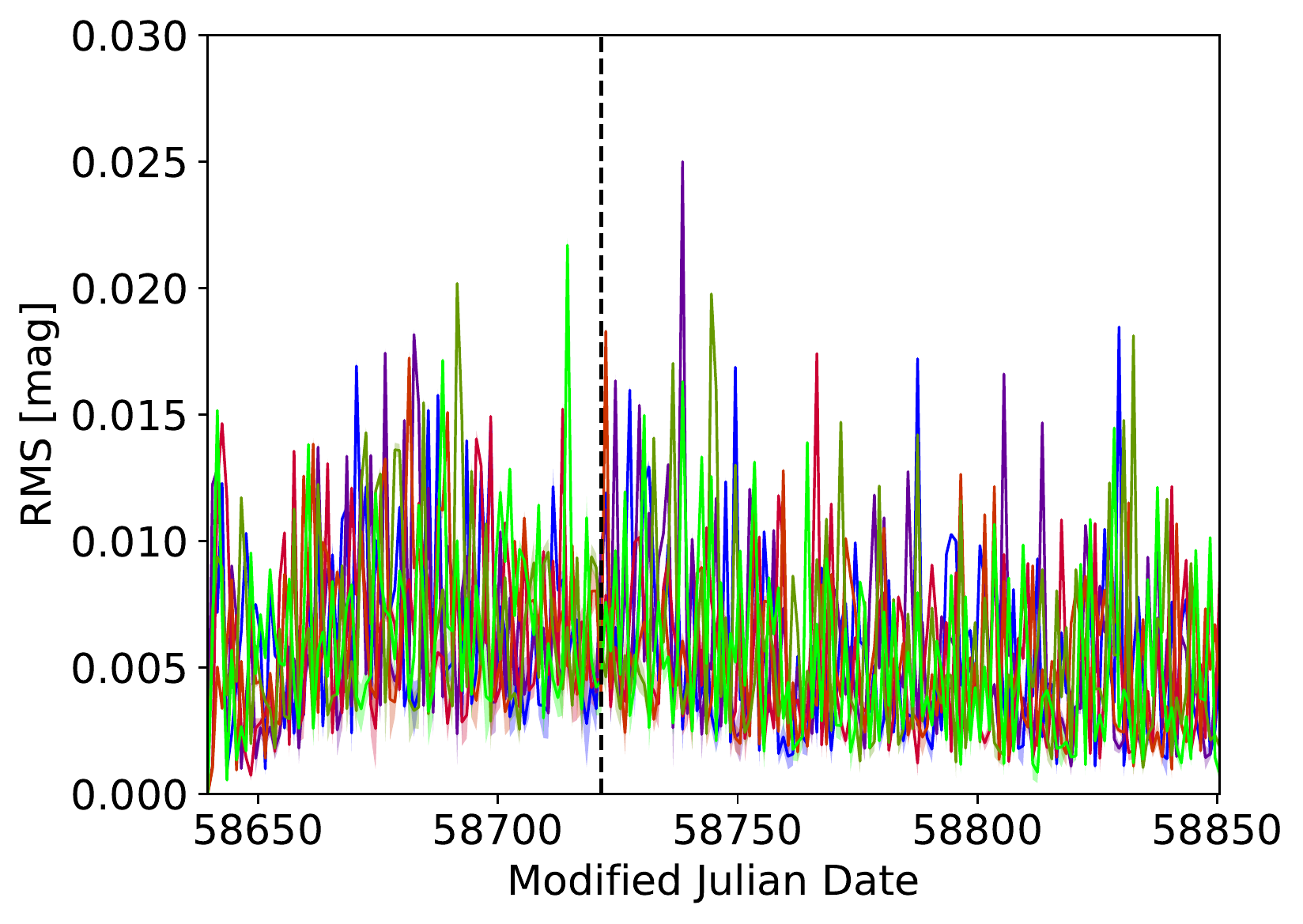} \\
\caption{Example of a burst fit using different stellar temperatures. We show the fit to burst B5 of NSW\,284, using the PHOENIX stellar atmosphere models and the $VRI$ filters. We use $\log(g) = 4.0$ and [M/H]\,=\,0.0. The colours indicate the adopted stellar temperature ranging from 15500\,K (green) to 20500\,K (blue) in steps of 1000\,K and the lightly shaded areas the uncertainties. \label{fit_test_temps}}
\end{figure*}

\bsp	
\label{lastpage}
\end{document}